\documentclass[a4paper,fleqn,usenatbib]{mnras}

\usepackage{newtxtext,newtxmath}
\usepackage[T1]{fontenc}
\usepackage{ae,aecompl}

\usepackage{amsmath}	% Advanced maths commands
\usepackage{amssymb}	% Extra maths symbols
\usepackage{graphicx,color,afterpage}
\usepackage{times,subcaption,rotating,tikz}
\usepackage{float}

\def\atlas{{{ATLAS}}$^{\rm 3D}$}

\def\kms{km s$^{-1}$}

\def\msun{M$_{\odot}$}

\def\arcsec{$^{\prime \prime}$}
\definecolor{Mygrey}{gray}{0.75}

\newcommand{\ltsimeq}{\raisebox{-0.6ex}{$\,\stackrel{\raisebox{-.2ex}{$\textstyle <$}}{\sim}\,$}}
\newcommand{\gtsimeq}{\raisebox{-0.6ex}{$\,\stackrel{\raisebox{-.2ex}{$\textstyle >$}}{\sim}\,$}}
\newcommand{\farc}{\mbox{\ensuremath{.\!\!^{\prime\prime}}}}

\mathchardef\mhyphen="2D

\usepackage[compact]{titlesec}
\titlespacing{\section}{0pt}{*2}{*1}
\title[Molecular gas in MASSIVE galaxies]{The MASSIVE survey -- XI. What drives the molecular gas properties of early-type galaxies} 
\author[Timothy A. Davis et al.]{\parbox{\textwidth}{Timothy A. Davis$^{1}$\thanks{E-mail: \texttt{DavisT@cardiff.ac.uk}}, Jenny E. Greene$^{2}$, Chung-Pei Ma$^{3}$, John P. Blakeslee$^{4,5}$, James M. Dawson$^{1}$, Viraj Pandya$^{6}$, Melanie Veale$^{3}$ and Nikki Zabel$^{1}$}
\vspace{0.4cm}\\
\parbox{\textwidth}{$^{1}$School of Physics \&\ Astronomy, Cardiff University, Queens Buildings, The Parade, Cardiff, CF24 3AA, UK\\
$^{2}$Department of Astrophysics, Princeton University, Princeton, NJ 08544, USA\\
$^{3}$Department of Astronomy, University of California, Berkeley, CA 94720, USA\\
$^{4}$Dominion Astrophysical Observatory, NRC Herzberg Astronomy and Astrophysics Research Centre, Victoria, BC V9E 2E7, Canada\\
$^{5}$Gemini Observatory, Casilla 603, La Serena, Chile\\
$^{6}$Department of Astronomy and Astrophysics, University of California, Santa Cruz, CA 95064, USA
}}
\begin{document}
\date{Accepted 2019 March 21. Received 2019 March 20; in original form 2018 September 26}

\pagerange{\pageref{firstpage}--\pageref{lastpage}} \pubyear{2015}

\maketitle

\label{firstpage}

\begin{abstract}
In this paper we study the molecular gas content of a representative sample of 67 of the most massive early-type galaxies in the local universe, drawn uniformly from the MASSIVE survey. We present new IRAM-30m telescope observations of 30 of these galaxies, allowing us to probe the molecular gas content of the entire sample to a fixed molecular-to-stellar mass fraction of 0.1\%. The total detection rate in this representative sample is 25$^{+5.9}_{-4.4}$\%, and by combining the MASSIVE and \atlas\ molecular gas surveys we find a joint detection rate of 22.4$^{+2.4}_{-2.1}$\%. This detection rate seems to be independent of galaxy mass, size, position on the fundamental plane, and local environment. We show here for the first time that true slow rotators can host molecular gas reservoirs, but the rate at which they do so is significantly lower than for fast-rotators. Objects with a higher velocity dispersion at fixed mass (a higher kinematic bulge fraction) are less likely to have detectable molecular gas, and where gas does exist, have lower molecular gas fractions. In addition, satellite galaxies in dense environments have $\approx$0.6 dex lower molecular gas-to-stellar mass ratios than isolated objects. In order to interpret these results we created a toy model, which we use to constrain the origin of the gas in these systems. We are able to derive an independent estimate of the gas-rich merger rate in the low-redshift universe. These gas rich mergers appear to dominate the supply of gas to ETGs, but stellar mass loss, hot halo cooling and transformation of spiral galaxies also play a secondary role. 

\end{abstract}


\section{Introduction}

Low redshift lenticular and elliptical galaxies have been shown in recent years to not be simple `red and dead' objects, as was often assumed in the past. Various studies have shown that these early-type galaxies (ETGs) often contain cold gas \citep[e.g.][]{1986A&A...164L..22W,1987ApJ...322L..73P,1996ApJ...460..271K} and associated star formation \citep{2000AJ....119.1645T,2005ApJ...619L.111Y}. {These new stars typically form dynamically cold (and often chemically distinct; e.g. \citealt{2010MNRAS.408...97K,2015MNRAS.448.3484M}) structures at the hearts of these systems.}  These objects provide an ideal laboratory in which to study the processes capable of regenerating galaxies. 

This molecular material is important because it provides the fuel for both future star formation, {and potentially for cold-mode active galactic nucleus (AGN) accretion \citep[e.g.][]{2007MNRAS.376.1849H,2019MNRAS.tmp..253R}}.  Molecular gas has been shown to be present in at least 22\% of local ETGs \citep{Welch:2003ev,2007MNRAS.377.1795C,Sage:2007jq,Welch:2010in,2011MNRAS.414..940Y}.
{These volume limited surveys have shown little difference in molecular gas properties between lenticular and elliptical galaxies, but that the specific stellar angular momentum (a {kinematic morphology} indicator) seems to be important. For instance \cite{2011MNRAS.414..940Y} show that the majority of the molecular gas in ETGs is present in fast-rotators. Such studies were, however, not able to probe significant numbers of slow-rotating ETGs.}

Exactly where the molecular gas that regenerates these objects comes from is debated. Internally, the stellar mass-loss rates of massive ellipticals can approach several solar masses a year \citep[e.g.][]{2001A&A...376...85J}, which, if it were able to cool, could easily maintain reservoirs like those observed. However, the detection rate of molecular gas seems to be independent of galaxy properties such as stellar mass, suggesting either that the majority of the gas in these systems has instead been accreted from external sources, {or that some mechanism (such as AGN) may suppress the cooling of this material in a manner that leaves no dependence on galaxy mass.}

Various observational studies (e.g. using the kinematic misalignment of gas reservoirs; \citealt{2006MNRAS.366.1151S, 2011MNRAS.417..882D, 2016MNRAS.457..272D}, motions of large scale atomic gas distributions; \citealt{2010MNRAS.409..500O,2012MNRAS.422.1835S}, gas metallicities and dust-to-gas ratios; \citealt{2012ApJ...748..123S,2015MNRAS.449.3503D} etc.) have suggested that minor mergers, and accretion of gas from close interactions/flybys dominate the supply of gas to {fast-rotating} ETGs. 
However, various open questions remain.  Some theoretical studies have suggested that cooling from the hot halo could still be important in supplying the gas to ETGs \citep{2014MNRAS.443.1002L,2014MNRAS.445.1351N,2015MNRAS.448.1271L}. In addition, the lack of a change in detection rates of cold gas in cluster ETGs, where mergers should be rare, has yet to be explained.

The most massive early type galaxies ($M_*>10^{11.5}$\msun) provide the most stringent tests of the origin of this regenerating gas. 
The majority of the stars in ETGs are old \citep{2015arXiv150402483G}, and thus the mass loss is dominated by low mass AGB stars. These most massive ETGs should thus have $>3$\,\msun\ a year of material returned to their interstellar medium (ISM). The star-formation rates of such objects are typically $\ltsimeq$0.1\,\msun\,yr$^{-1}$ \citep[e.g.][]{2016MNRAS.455..214D}, and thus their gas reservoirs should be building up. 
These most massive galaxies also host the largest hot halos \citep[e.g.][]{2016ApJ...826..167G}, and are typically slow-rotators \citep[e.g.][]{2017MNRAS.471.1428V}.

{Some of the brightest ETGs are the central galaxy in cluster/group environments, and thus have preferential access to cooling gas from their surrounding inter-group/cluster medium. Such objects are routinely found to have large molecular gas reservoirs \citep[e.g.][]{2001MNRAS.328..762E,2003A&A...412..657S,2016MNRAS.458.3134R,2017ApJ...848..101V,2018A&A...618A.126O,2018ApJ...865...13T,2019arXiv190209227R}, likely as a result of cooling from the hot reservoir \citep[e.g.][]{2018ApJ...854..167G,2018arXiv181011465B}}.
{However, not all massive ETGs live in such privileged positions within large halos. For instance, in a complete, volume limited sample of $M_* \gtsimeq10^{11.5}M_\odot$ galaxies nearly half are isolated, or satellites embedded in a larger halo \citep{2017MNRAS.471.1428V}. It is thus crucial to study the molecular gas content of representative samples of massive ETGs in a systematic way}.

We aim to address this problem in this work, as part of the MASSIVE project. MASSIVE is a survey of the  $\approx$100 most massive galaxies within 108 Mpc, using a combination of wide-field and adaptive-optics-assisted integral field spectroscopy (IFS). This volume-limited survey targets the distinct stellar mass range ($M_* \gtsimeq10^{11.5}M_\odot$; $M_K{\,<\,}{-}25.3$ mag) that has not been systematically studied to date. Full details of this survey are presented in \cite{2014ApJ...795..158M}. 

In \cite{2016MNRAS.455..214D}, paper III of the MASSIVE series, we presented a pilot study attempting to detect molecular gas in these high mass galaxies. We observed 15 MASSIVE galaxies with the IRAM-30m telescope and studied their gas masses, star-formation efficiencies and circular velocities. 
Within this sample, it seemed that the very most massive objects may have a lower probability of hosting cold gas. 
However, the sample of objects presented in Paper III were selected via various proxies of the presence of gas, and thus are not representative of the MASSIVE sample as a whole. In this paper we aim to address this, extending our survey to a representative subset of the MASSIVE galaxies. In this way we are able to draw more robust conclusions about the sources of gas in these objects, and how these may change with galaxy properties.

In Section \ref{sample} of this paper we present details of our sample selection and the properties of the target objects. Section \ref{data} details the observation parameters and reduction. In Section \ref{results} we present our observational results, and present a toy-model to help explain them in Section \ref{model}. Finally we discuss and conclude in Section \ref{conclude}.

\section{Sample selection}
\label{sample}

MASSIVE is a volume-limited, multi-wavelength,
spectroscopic and photometric survey of the most massive galaxies in the
local universe.  The full sample includes 116 galaxies in the northern
sky with distance $D < 108$ Mpc and absolute $K$-band magnitude
$M_K{\,<\,}{-}25.3$ mag, corresponding to stellar masses M$_* \gtsimeq
10^{11.5}M_\odot$.  
The MASSIVE survey volume is more than an order of magnitude larger than that probed by \atlas\ and there are only 6 overlapping galaxies in the two surveys.
Wide-field IFS data, taken using the Mitchell Spectrograph (formerly called
VIRUS-P; \citealt{2008SPIE.7014E..06H}) at McDonald Observatory, are being obtained for all objects in the MASSIVE survey, enabling us to derive stellar and gas parameters (and study the galaxies stellar populations; \citealt{2015arXiv150402483G}) out to beyond $\sim$2 effective radii. 

\begin{figure} 
\begin{center}
\includegraphics[width=0.48\textwidth,angle=0,clip,trim=0cm 0cm 0cm 0.0cm]{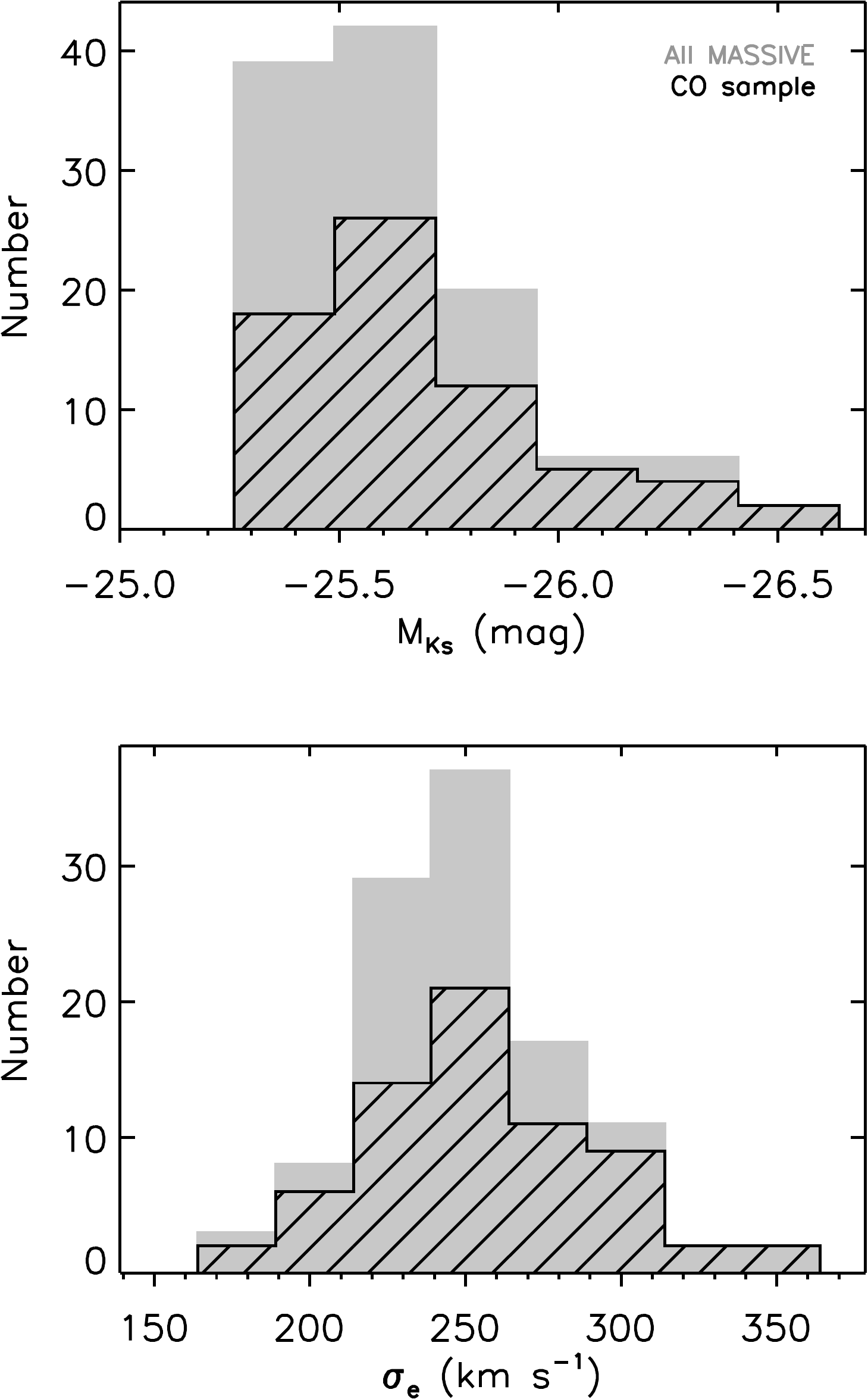}
\caption{Absolute $Ks$-band magnitude (top panel) and stellar velocity dispersion within the effective radius (bottom panel) for the entire MASSIVE galaxy sample (grey histograms) and the representative CO sub-sample discussed in this paper. The subset of objects discussed here is as representative as possible of the parent population.}
\label{fig:howselect}
 \end{center}
 \end{figure}

The objects observed in this paper were selected to form a \textit{representative} sample, which matches the full distribution of stellar masses and velocity dispersions found in the MASSIVE parent catalog. 
In order to select an unbiased sample we divided our parent sample (which is itself complete, mass-selected and volume-limited) into bins of $K_{\rm s}$-band magnitude and velocity dispersion (0.25 magnitudes wide, 40 km/s high). We rank the objects in each cell by their distance. We then choose the closest 50\% of objects in each cell, creating a fully representative sample of 57 objects, which match the full distribution of stellar masses and velocity dispersions found in the parent catalog by construction (see Figure \ref{fig:howselect}).  Twenty seven objects from this unbiased sample already have suitable observations (15 of these presented in \citealt{2016MNRAS.455..214D}), and in this paper we present new observations of 30 objects.

In what follows we also include data on an additional 10 objects, such that we utilise all data available from the literature for the MASSIVE sample. Including this data slightly over-represents high mass/velocity dispersion systems (see Figure \ref{fig:howselect}), but we found that their inclusion did not change any of our results.
We thus here study a total sample of 67 objects, with observations that allow us to detect gas (or set an upper limit to the gas content) down to a fixed molecular-to-stellar mass fraction of 0.1\%.

\subsection{Lower mass ETGs}
{In order to study how the molecular gas properties of ETGs change as a function of mass we also include information from \cite{2011MNRAS.414..940Y}, the molecular gas survey of the \atlas\ sample \citep{2011MNRAS.413..813C}  of ETGs. \atlas\ is a volume limited survey, covering all the morphologically classified ETGs with distances $<42$ Mpc,  $|\delta- 29^\circ | < 35^\circ$ and $|b| > 15^\circ$, that are  brighter than a $K_s$-band magnitude of $-21.5$ mag (stellar mass $M_*\gtsimeq 6\times10^9$ \msun).}

{\cite{2011MNRAS.414..940Y} observed all of these galaxies with the IRAM-30m telescope, to a fixed sensitivity of 3 mK Ta$^*$, and thus provides us with a complete, volume limited sample (which is thus dominated by lower mass, fast-rotating systems) without incompleteness biases. See \cite{2014ApJ...795..158M} for a more in depth comparison of these two samples.}

\begin{table*}
\centering
\caption{Properties of the MASSIVE representative sample objects.}
\begin{tabular*}{0.9\textwidth}{@{\extracolsep{\fill}}lrrrrrrr}
\hline
Galaxy  &  Distance & $M_{\rm K_S}$ & $L_{\rm K_S}$ & $\sigma_e$ & R$_e$ & log$_{10}$(M$_{\rm H_2})$& Ref\\
 &  (Mpc)  & (mag) & (L$_{\odot}$) &  (km/s) & (kpc) & (M$_{\odot}$) & \\
 (1) & (2) & (3) & (4) & (5) & (6) & (7) & (8)\\
 \hline
IC0310 &       77.5 &     -25.35 &      11.45 &       205. &        5.7 & =      9.02 &        1 \\
NGC0057 &       76.3 &     -25.75 &      11.61 &       251. &       10.0 & $<$      8.60 &        2 \\
NGC0227 &       75.9 &     -25.32 &      11.44 &       262. &        5.4 & $<$      8.52 &        2 \\
NGC0315 &       70.3 &     -26.30 &      11.83 &       341. &        8.5 & $<$      8.81 &        6 \\
NGC0383 &       71.3 &     -25.81 &      11.64 &       257. &        7.1 & =      9.23 &        1 \\
NGC0410 &       71.3 &     -25.90 &      11.67 &       247. &       10.9 & $<$      8.66 &        2 \\
NGC0467 &       75.8 &     -25.40 &      11.47 &       247. &        9.0 & =      8.77 &        2 \\
NGC0499 &       69.8 &     -25.50 &      11.51 &       266. &        5.3 & $<$      8.62 &        2 \\
NGC0507 &       69.8 &     -25.93 &      11.68 &       257. &       13.0 & $<$      8.68 &        2 \\
NGC0533 &       77.9 &     -26.05 &      11.73 &       258. &       15.3 & $<$      8.67 &        2 \\
NGC0547 &       74.0 &     -25.83 &      11.64 &       232. &        7.1 & $<$      8.76 &        2 \\
NGC0665 &       74.6 &     -25.51 &      11.52 &       164. &        4.9 & =      9.18 &        1 \\
NGC0708 &       69.0 &     -25.65 &      11.57 &       219. &       16.5 & =      8.83 &        3 \\
NGC0741 &       73.9 &     -26.06 &      11.74 &       289. &        9.6 & $<$      8.66 &        2 \\
NGC0890 &       55.6 &     -25.50 &      11.51 &       194. &        8.2 & $<$      8.79 &        2 \\
NGC0910 &       79.8 &     -25.33 &      11.44 &       219. &        9.9 & $<$      8.12 &        2 \\
NGC0997 &       90.4 &     -25.40 &      11.47 &       215. &       10.3 & =      9.26 &        1 \\
NGC1060 &       67.4 &     -26.00 &      11.71 &       271. &       12.0 & $<$      7.89 &        4 \\
NGC1129 &       73.9 &     -26.14 &      11.77 &       259. &       10.8 & $<$      8.81 &        6 \\
NGC1132 &       97.6 &     -25.70 &      11.59 &       218. &       14.6 & $<$      8.61 &        1 \\
NGC1167 &       70.2 &     -25.64 &      11.57 &       172. &       10.1 & =      8.52 &        4 \\
NGC1453 &       56.4 &     -25.67 &      11.58 &       272. &        7.9 & $<$      8.91 &        2 \\
NGC1497 &       87.8 &     -25.31 &      11.44 &       190. &        7.9 & =      9.10 &        1 \\
NGC1573 &       65.0 &     -25.55 &      11.53 &       264. &        7.9 & $<$      8.66 &        2 \\
NGC1600 &       63.8 &     -25.99 &      11.71 &       293. &       12.7 & $<$      8.41 &       7 \\
NGC1684 &       63.5 &     -25.34 &      11.45 &       262. &        9.0 & =      9.20 &        1 \\
NGC1700 &       54.4 &     -25.60 &      11.55 &       223. &        6.0 & $<$      8.31 &        7 \\
NGC2256 &       79.4 &     -25.87 &      11.66 &       259. &       16.8 & $<$      8.54 &        2 \\
NGC2258 &       59.0 &     -25.66 &      11.58 &       254. &       10.1 & $<$      8.81 &        1 \\
NGC2274 &       73.8 &     -25.69 &      11.59 &       259. &       10.1 & $<$      8.58 &        2 \\
NGC2320 &       89.4 &     -25.93 &      11.68 &       298. &        8.4 & =      8.41 &        5 \\
NGC2418 &       74.1 &     -25.42 &      11.48 &       247. &        7.1 & $<$      8.68 &        7 \\
NGC2513 &       70.8 &     -25.52 &      11.52 &       253. &        8.2 & $<$      8.49 &        2 \\
NGC2672 &       61.5 &     -25.60 &      11.55 &       262. &        4.3 & $<$      8.65 &        2 \\
NGC2693 &       74.4 &     -25.76 &      11.62 &       296. &        5.5 & $<$      8.40 &        2 \\
NGC2783 &      101.4 &     -25.72 &      11.60 &       264. &       18.7 & $<$      8.85 &        8 \\
NGC2832 &      105.2 &     -26.42 &      11.88 &       291. &       10.8 & $<$      8.59 &        2 \\
NGC2892 &      101.1 &     -25.70 &      11.59 &       234. &       11.4 & $<$      8.84 &        3 \\
NGC3158 &      103.4 &     -26.28 &      11.82 &       289. &        8.1 & $<$      8.81 &        2 \\
NGC3805 &       99.4 &     -25.69 &      11.59 &       225. &        7.9 & $<$      8.30 &        9 \\
NGC3816 &       99.4 &     -25.40 &      11.47 &       207. &        8.8 & $<$      8.67 &        2 \\
NGC3842 &       99.4 &     -25.91 &      11.68 &       231. &       11.6 & $<$      8.15 &        9 \\
NGC3862 &       99.4 &     -25.50 &      11.51 &       232. &       19.2 & =      8.49 &       10 \\
NGC4055 &      107.2 &     -25.40 &      11.47 &       270. &        7.1 & =      8.72 &        6 \\
NGC4073 &       91.5 &     -26.33 &      11.84 &       292. &       10.2 & $<$      8.90 &        2 \\
NGC4472 &       16.7 &     -25.72 &      11.60 &       258. &       14.3 & $<$      7.25 &       11 \\
NGC4486 &       16.7 &     -25.31 &      11.44 &       336. &        5.7 & =      6.70 &        1 \\
NGC4649 &       16.5 &     -25.36 &      11.46 &       340. &        5.4 & $<$      7.83 &        1 \\
NGC4839 &      102.0 &     -25.85 &      11.65 &       275. &       14.4 & $<$      8.88 &        2 \\
NGC4874 &      102.0 &     -26.18 &      11.78 &       258. &       15.8 & $<$      8.86 &        2 \\
NGC4889 &      102.0 &     -26.64 &      11.97 &       337. &       16.3 & $<$      8.93 &        2 \\
NGC4914 &       74.5 &     -25.72 &      11.60 &       225. &       11.3 & $<$      8.79 &        2 \\
NGC5208 &      105.0 &     -25.61 &      11.56 &       235. &        9.3 & =      9.49 &        1 \\
NGC5252 &      103.8 &     -25.32 &      11.44 &       196. &        7.9 & $<$      8.86 &        1 \\
NGC5322 &       34.2 &     -25.51 &      11.52 &       239. &        3.3 & $<$      7.76 &       11 \\
NGC5353 &       41.1 &     -25.45 &      11.49 &       290. &        4.8 & $=$      8.28 &       12 \\
NGC5490 &       78.6 &     -25.57 &      11.54 &       282. &        7.4 & $<$      8.73 &        3 \\
NGC5557 &       51.0 &     -25.46 &      11.50 &       223. &        3.6 & $<$      7.92 &       11 \\
NGC6482 &       61.4 &     -25.60 &      11.55 &       291. &        4.9 & $<$      8.62 &        1 \\
NGC7052 &       69.3 &     -25.67 &      11.58 &       266. &        9.2 & =      9.63 &        3 \\
NGC7265 &       82.8 &     -25.93 &      11.68 &       206. &       12.7 & $<$      8.65 &        2 \\
\hline
\end{tabular*}
\label{tab:info}
\end{table*}

\begin{table*}
\centering
\contcaption{}
\begin{tabular*}{0.9\textwidth}{@{\extracolsep{\fill}}lrrrrrrr}
\hline
Galaxy  &  Distance & $M_{\rm K_S}$ & $L_{\rm K_S}$ & $\sigma_e$ & R$_e$ & log$_{10}$(M$_{\rm H_2})$& Ref\\
 &  (Mpc)  & (mag) & (L$_{\odot}$) &  (km/s) & (kpc) & (M$_{\odot}$) & \\
 (1) & (2) & (3) & (4) & (5) & (6) & (7) & (8)\\
 \hline
NGC7274 &       82.8 &     -25.39 &      11.47 &       244. &        9.4 & $<$      8.55 &        2 \\
NGC7550 &       72.7 &     -25.43 &      11.48 &       224. &        9.9 & =      8.85 &        2 \\
NGC7556 &      103.0 &     -25.83 &      11.64 &       243. &       13.2 & $<$      8.56 &        1 \\
NGC7618 &       76.3 &     -25.44 &      11.49 &       265. &        6.2 & $<$      8.60 &        2 \\
NGC7619 &       54.0 &     -25.65 &      11.57 &       277. &        9.0 & $<$      7.52 &        1 \\
NGC7626 &       54.0 &     -25.65 &      11.57 &       250. &        7.0 & $<$      7.90 &       10 \\
\hline
\end{tabular*}
\parbox[t]{0.9\textwidth}{\textit{Notes:} The galaxy distances (Column 2) magnitudes/luminosities (Column 3 \& 4) and effective radii (Column 6) are reproduced here from \protect \cite{2014ApJ...795..158M}, while the velocity dispersions in Column 5 are effective velocity dispersions measured within one effective radius from \protect \cite{2017MNRAS.464..356V} where available, or \protect \cite{2014ApJ...795..158M} otherwise. Column 7 contains the measured molecular gas mass (or limit) for each object, taken from the paper referenced in Column 8, corrected to our assumed X$_{\rm CO}$. References: 1; \protect \cite{2016MNRAS.455..214D}, 2; this work, 3; \cite{2010A&A...518A...9O}, 4; \cite{2015A&A...573A.111O}, 5; \cite{2005ApJ...634..258Y}, 6; WISDOM project - private communication, 7; \cite{2001MNRAS.326.1431G}, 8; \cite{1995A&A...297..643W}, 9; \cite{1996ApJ...460..271K}, 10; \cite{2003ASPC..290..525L}, 11; \cite{2011MNRAS.414..940Y}, 12; \cite{2018A&A...618A.126O}} %A star indicates the objects that are not in MASSIVE, but were detected serendipitously during this survey.}
%\label{tab:info}
\end{table*}

\begin{figure*} 
\begin{center}

\begin{subfigure}[b]{0.95\textwidth}
\begin{center}
\begin{tikzpicture}\node[anchor=south west,inner sep=0,inner xsep=0mm,inner ysep=0mm] (image) at (0,0) {\includegraphics[height=3.6cm,angle=0,clip,trim=0cm 0cm 0.0cm 0.0cm]{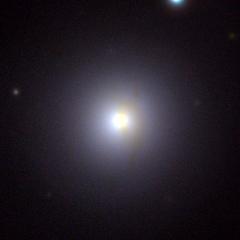}};
  \begin{scope}[x={(image.south east)},y={(image.north west)}]
		           \node[anchor=west, text=white] at (0.03,0.9) {NGC0467};	
           \end{scope}
\end{tikzpicture}
\includegraphics[height=3.6cm,angle=0,clip,trim=0cm 0cm 0.0cm 0.0cm]{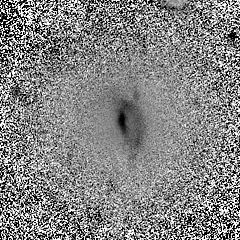}\hspace{0.2cm}
\includegraphics[height=3.6cm,angle=0,clip,trim=0cm 0cm 0.0cm 0.0cm]{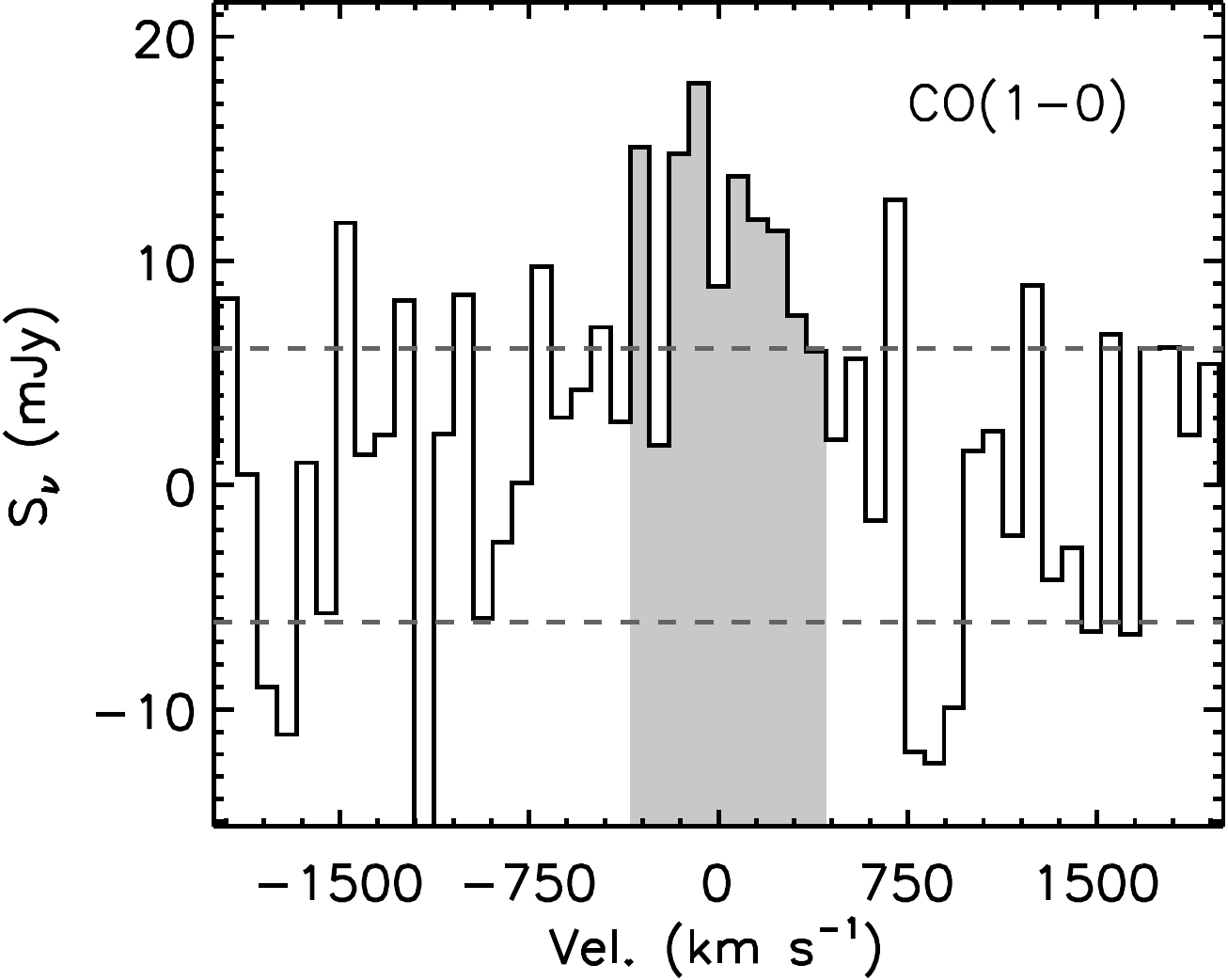}\hspace{0.0cm}
\includegraphics[height=3.6cm,angle=0,clip,trim=1.1cm 0cm 0cm 0.0cm]{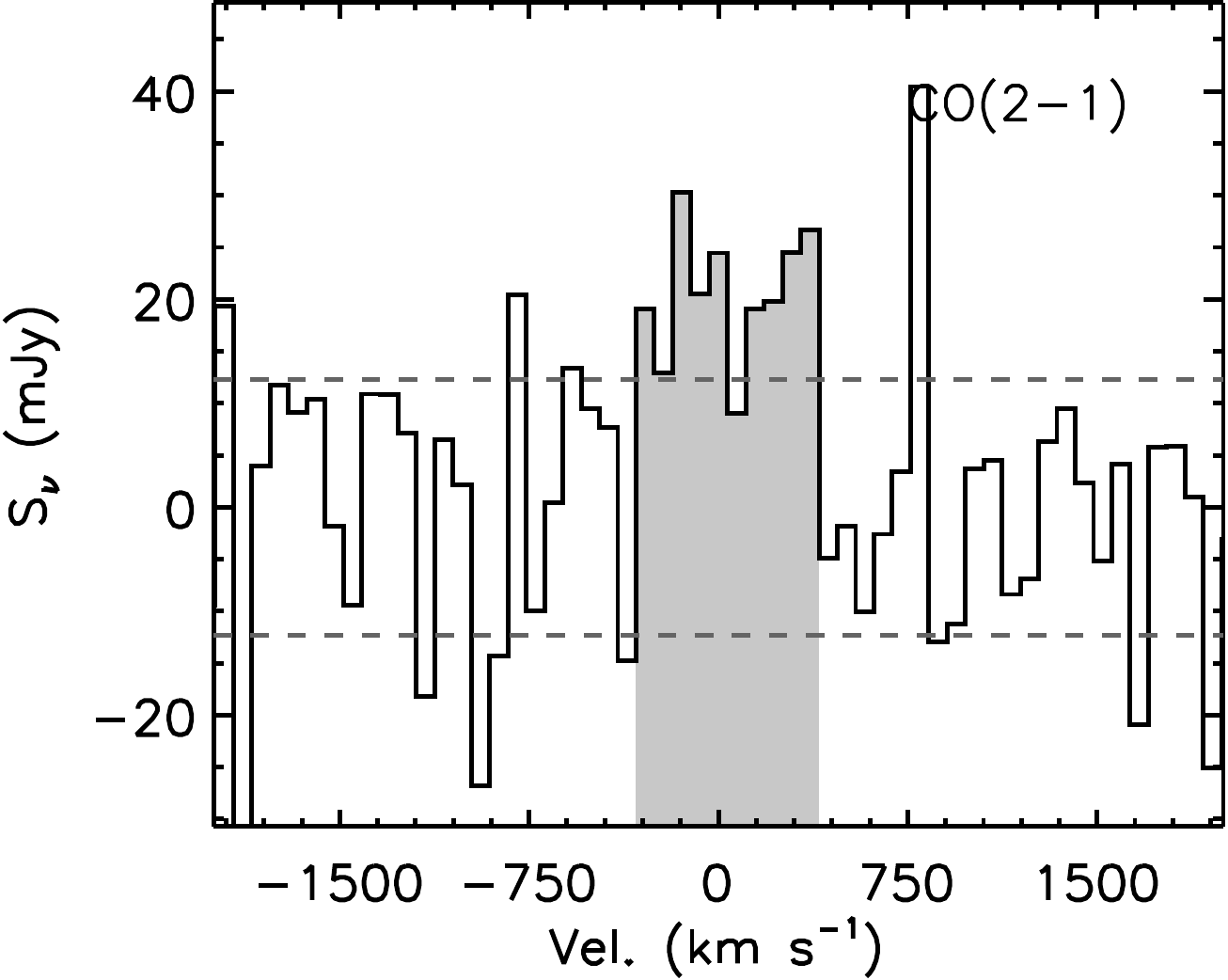}
%\caption{NGC0467}
\end{center}
\end{subfigure}\vspace{0.6cm}
\begin{subfigure}[b]{0.95\textwidth}
\begin{center}
\begin{tikzpicture}\node[anchor=south west,inner sep=0,inner xsep=0mm,inner ysep=0mm] (image) at (0,0) {\includegraphics[height=3.6cm,angle=0,clip,trim=0cm 0cm 0.0cm 0.0cm]{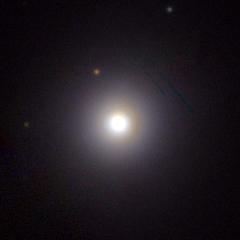}};
  \begin{scope}[x={(image.south east)},y={(image.north west)}]
		           \node[anchor=west, text=white] at (0.03,0.9) {NGC7550};	
           \end{scope}
\end{tikzpicture}
\includegraphics[height=3.6cm,angle=0,clip,trim=0cm 0cm 0.0cm 0.0cm]{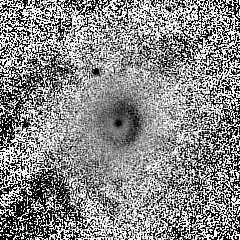}\hspace{0.2cm}
\includegraphics[height=3.6cm,angle=0,clip,trim=0cm 0cm 0cm 0.0cm]{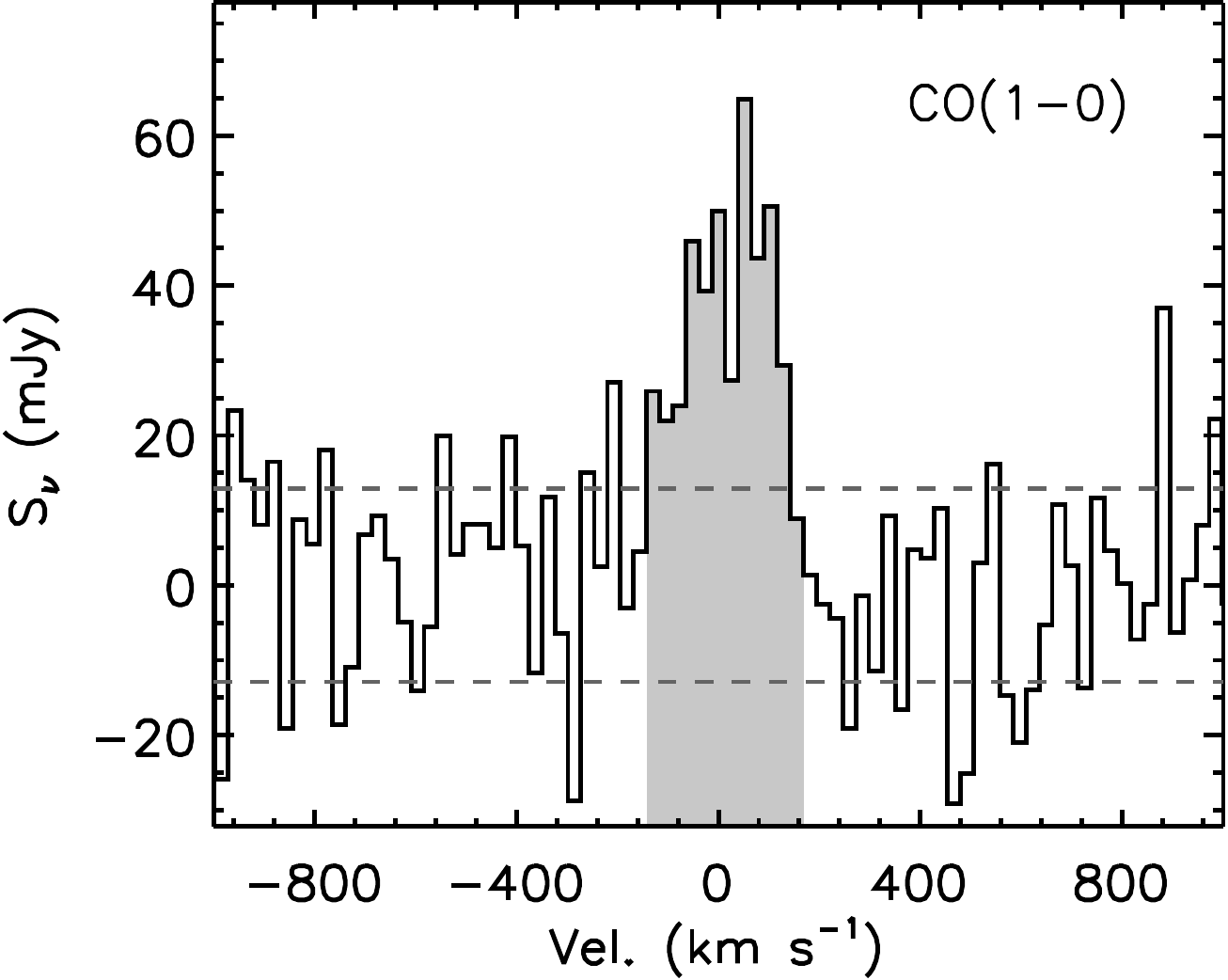}
\includegraphics[height=3.6cm,angle=0,clip,trim=1.1cm 0cm 0cm 0.0cm]{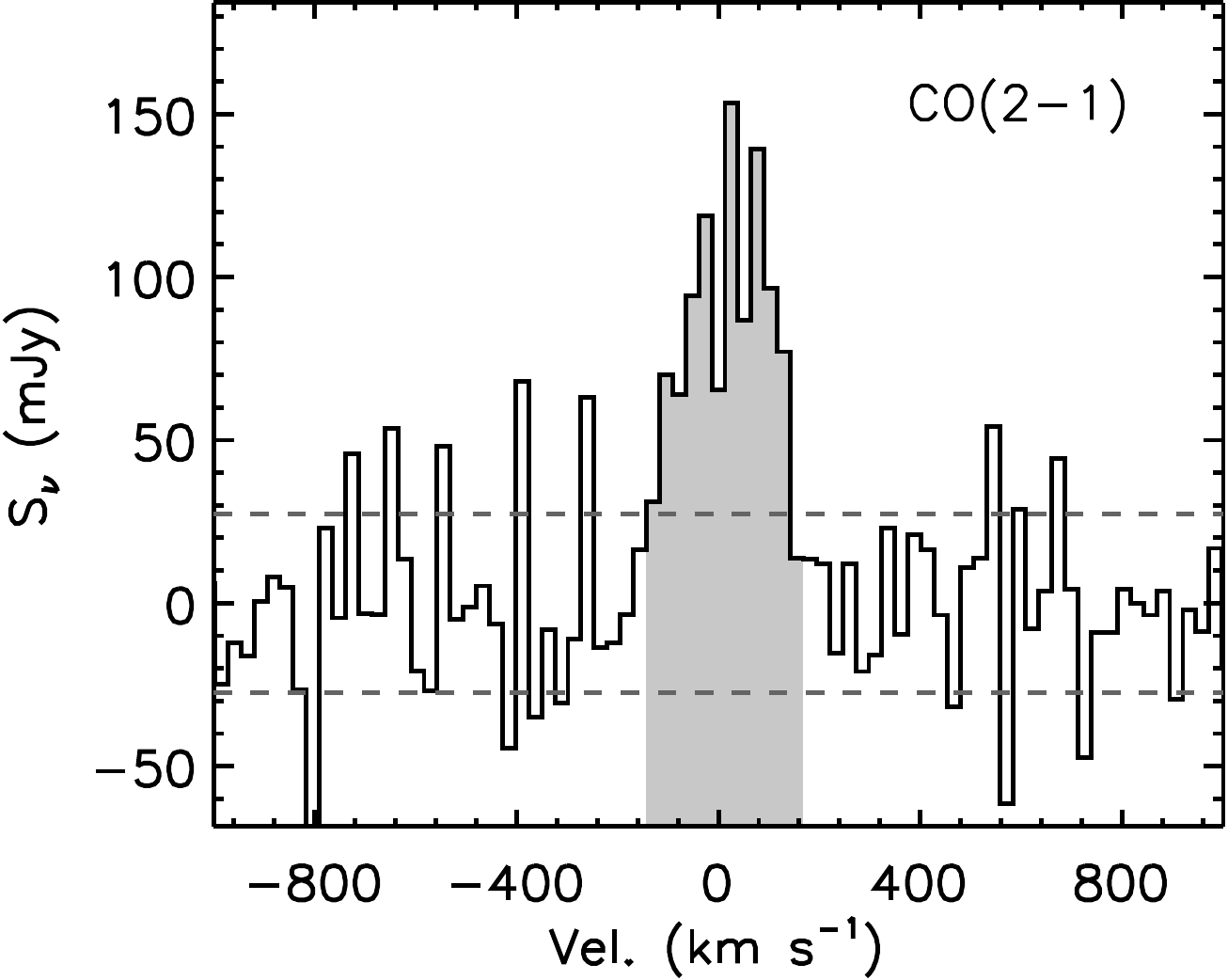}
%\caption{NGC7550}
\end{center}
\end{subfigure}
\caption{\textit{Far-Left:} Three-colour (\textit{y/i/g}) PanSTARRS images \protect \citep{2016arXiv161205560C} of our two CO detected targets, each with an angular size of 60$\times$60 arcsec$^2$. \textit{Centre-left:} $g$-$i$ colourmaps of our sources (from the same PanSTARRS imaging) that highlight the dust structures that likely host the detected molecular gas.  \textit{Centre-Right \& Right:} CO(1-0) and CO(2-1) spectra of these objects. The grey shaded region on the spectra denotes the detected line, and the dashed
lines show the $\pm$1$\sigma$ RMS level. The CO(1-0) detection in NGC0467 is marginal ($\approx$4.7$\sigma$), while NGC7550 is well detected ($\approx$10.4$\sigma$). }
\label{codetsfig}
 \end{center}
 \end{figure*}

 \begin{table*}
\caption{Observational parameters and derived molecular gas masses/limits for the sample ETGs}
\begin{tabular*}{\textwidth}{@{\extracolsep{\fill}}l r r r r r r r r r rr r r r}
\hline
Name & D$_{\rm beam}$$_{1\mhyphen0}$ &Peak$_{1\mhyphen0}$ &RMS$_{1\mhyphen0}$ & $\delta^{\rm chan}_{1\mhyphen0}$ & $\int S_{\rm \nu}\, \delta V$$_{1\mhyphen0}$ & Peak$_{2\mhyphen1}$ &  RMS$_{2\mhyphen1}$ & $\delta^{\rm chan}_{2\mhyphen1}$ & $\int S_{\rm \nu}\, \delta V$$_{2\mhyphen1}$ & $W_{20}$ & log$_{10}$(M$_{\rm H_2}$) \\ 
  & (kpc) & (mJy) & (mJy) & (\kms) & (Jy \kms) & (mJy) & (mJy) & (\kms) & (Jy \kms) & (\kms) & ($10^8$ M$_{\odot}$) \\%& &(\%) \\
 (1) & (2) & (3) & (4) & (5) & (6) & (7) & (8) & (9) & (10)& (11) & (12)\\
\hline
NGC0057 & 8.1 & -- &       7.66 &        52. & -- & -- &      19.73 &        52. & -- & -- & <      3.94\\
NGC0227 & 8.1 & -- &       6.50 &        52. & -- & -- &      14.39 &        52. & -- & -- & <      3.31\\
NGC0410 & 7.6 & -- &      10.27 &        52. & -- & -- &      33.61 &        52. & -- & -- & <      4.62\\
NGC0467 & 8.1 &      17.94 &       6.09 &        78. &       8.67 $\pm$       1.57 &      30.31 &      12.29 &        72. &       13.89 $\pm$       2.95 &  370.     &        5.87 $\pm$       1.06\\
NGC0499 & 7.4 & -- &       8.21 &        52. & -- & -- &      14.18 &        52. & -- & -- & <      4.17\\
NGC0507 & 7.4 & -- &     11.11 &        52. & -- & -- &      27.40 &        52. & -- & -- & <      4.79\\
NGC0533 & 8.3 & -- &      10.94 &        52. & -- & -- &      25.30 &        52. & -- & -- & <      4.71\\
NGC0547 & 7.9 & -- &      10.70 &        52. & -- & -- &      23.02 &        52. & -- & -- & <      5.74\\
NGC0741 & 7.9 & -- &       9.44 &        52. & -- & -- &      24.25 &        52. & -- & -- & <      4.57\\
NGC0890 & 5.9 & -- &      12.67 &        52. & -- & -- &      33.87 &        52. & -- & -- & <      6.12\\
NGC0910 & 8.5 & -- &       4.84 &        52. & -- & -- &      13.62 &        52. & -- & -- & <      1.32\\
NGC1453 & 6.0 & -- &      14.33 &        52. & --  & --&     5.83 &        50. & -- & -- & <      8.07\\
NGC1573 & 6.9 & -- &       8.14 &        52. & -- & -- &      17.03 &        52. & -- & -- & <      4.58\\
NGC2256 & 8.5 & -- &       9.37 &        52. & -- & -- &      20.35 &        52. & -- & -- & <      3.50\\
NGC2274 & 6.0 & -- &       6.83 &        52. & -- & -- &      15.78 &        52. & -- & -- & <      3.81\\
NGC2513 & 6.9 & -- &       6.41 &        52. & -- & -- &      16.89 &        52. & -- & -- & <      3.09\\
NGC2672 & 8.5 & -- &      10.06 &        52. & -- & -- &      30.66 &        52. & -- & -- & <      4.46\\
NGC2693 & 7.9 & -- &       7.60 &        52. & -- & -- &      20.11 &        52. & -- & -- & <      2.54\\
NGC2832 & 11.2 & -- &       7.86 &        52. & -- & -- &      19.90 &        52. & -- & -- & <      3.85\\
NGC3158 & 11.0 & -- &       6.56 &        52. & -- & -- &      16.90 &        52. & -- & -- & <      6.42\\
NGC3816 & 10.6 & -- &       4.96 &        52. & -- & -- &       9.19 &        52. & -- & -- & <      4.69\\
NGC4073 & 9.7 & -- &       9.06 &        52. & -- & -- &      14.84 &        52. & -- & -- & <      7.92\\
NGC4839 & 10.9 & -- &      10.14 &        52. & -- & -- &      38.80 &        52. & -- & -- & <      7.50\\
NGC4874 & 10.9 & -- &       7.94 &        52. & -- & -- &      19.91 &        52. & -- & -- & <      7.31\\
NGC4889 & 10.9 & -- &       9.31 &        52. & -- & -- &      15.90 &        52. & -- & -- & <      8.56\\
NGC4914 & 7.9 & -- &       6.77 &        52. & -- & -- &      12.77 &        52. & -- & -- & <      6.22\\
NGC7265 & 8.8 & -- &       9.00 &        52. & -- & -- &      24.06 &        52. & -- & -- & <      4.42\\
NGC7274 & 8.8 & -- &       5.90 &        52. & -- & -- &       7.07 &        52. & -- & -- & <      3.57\\
NGC7550& 7.7  &      64.85 &      12.89 &        26. &      11.28 $\pm$       1.20 &     153.54 &      27.41 &        26. &        26.58 $\pm$       2.56 & 180.   &        7.03 $\pm$       0.75\\
NGC7618 & 8.1 & -- &       6.62 &        52. & -- & -- &      16.79 &        52. & -- & -- & <      4.02\\
         \hline
\end{tabular*}
\parbox[t]{1\textwidth}{ \textit{Notes:}  Column 1 lists the name of each source. Column 2 lists the diameter of the 22\arcsec\ CO(1-0) beam of the IRAM-30m telescope in kiloparsecs, given the distance to the object as listed in Table \ref{tab:info}. The beam at the frequency of CO(2-1) is half this size. Column 3 lists the CO(1-0) peak flux, while Column 4 lists the RMS noise reached in the CO(1-0) observation with a channel size as shown in Column 5. Column 6 lists the integrated intensity of the CO(1-0) line and its error. Column 7 --9 are as Column 3 --6, but for the CO(2-1) line observations.
Column 11 shows the CO linewidth at 20\% of the peak intensity, which was found to be the same for both of the CO lines. Dashes indicate sources where the lines were not detected. Column 12 shows the H$_2$ mass for each source in units of $10^8$ M$_{\odot}$, derived using Equation \protect \ref{co2h2}. For the non detected sources Column 12 lists the 3$\sigma$ upper limit to the H$_2$ mass derived assuming a 250 \kms\ velocity width for the line.}
\label{obstable}
\end{table*}

\section{Data and Derived Quantities}
\label{data}

The IRAM 30-m telescope at Pico Veleta, Spain, was used between both the 28th July -- 9th September 2015 and the 1st June -- 16th July 2018 to observe CO emission in our sample galaxies (proposals 067-15 and 064-18, PI Davis). 
 We aimed to simultaneously detect CO(1-0) and CO(2-1), at rest-frequencies of 115.27 and 230.54 GHz respectively, in the 3\,mm and 1\,mm atmospheric windows.
The beam full width at half-maximum (FWHM) of the IRAM-30m at the frequency of these lines is 21\farc3 and 10\farc7, corresponding to physical scales of between 5.7 and 10.8 kpc at the frequency of CO(1-0), and between 2.9 and 5.4 kpc at the frequency of CO(2-1), given the varying distance to these sources. 

The Eight MIxer Receiver (EMIR) was used for observations in the wobbler switching mode, with reference positions offset by $\pm$100\arcsec\ in azimuth. The Fourier Transform Spectrograph (FTS) back-end gave an effective total bandwidth of $\approx$4 GHz per window, and a raw spectral resolution of 200 kHz ($\approx$0.6 \kms\ at 3\,mm, $\approx$0.3 \kms\  at 1\,mm). The Wideband Line Multiple Autocorrelator (WILMA) back-end was used simultaneously with the FTS.

The system temperatures ranged between 154 and 493 K at 3~mm and between 200 and 2160 K at 1~mm. The time on source ranged from 24 to 120 min, being weather-dependent, and was interactively adjusted by the observers to reach our required sensitivity (where we could detect objects with a molecular-to-stellar mass fraction of 0.1\%).
 
The individual $\approx$6 minute scans were inspected, and the baseline removed, using a zeroth-, first- or second-order polynomial, depending on the scan. Scans with poor quality baselines or other problems were discarded.
The good scans were averaged together, weighted by the inverse square of the system temperature. We consider emission lines where the integrated intensity has greater than a 3$\sigma$ significance (including the baseline uncertainty;  \citealt{2012MNRAS.421.1298C}) to be detected.

We convert the spectra from the observed antenna temperature (T$_a^*$) into units of Janskys, utilising the point-source conversion as tabulated on the IRAM website\footnote{http://www.iram.es/IRAMES/mainWiki/EmirforAstronomers} (5.9 Jy/K at 3mm, 8.0 Jy/K at 1mm). 

\subsection{Line detections}

We detected line emission in two out of the 30 galaxies observed in this work (see Figure \ref{codetsfig}). This low detection rate reflects the fact that the selection criteria used in the previous biased survey did a good job in selecting the objects which were most likely to be detected. Table \ref{obstable} lists the properties of the observed spectra, and the detected lines. 
The CO(1-0) detection in NGC0467 is somewhat marginal ($\approx$4.7$\sigma$), while NGC7550 is well detected ($\approx$10.4$\sigma$).

The ratio of the CO(1--0) and CO(2--1) lines we observe is affected by both the gas excitation temperature and the spatial distribution of the gas.
 If the observed CO emission were to fill the telescope beam at both frequencies (and the CO is not sub-thermally excited) we would expect a line ratio of one (measured in main beam temperature units). However, if the CO emission is compact compared to the beam then the measured intensity in the CO(2--1) line should be larger by up to a factor of 4 (as the beam at such frequencies covers a 4 times smaller area). In our two detected sources we find a line ratio (in beam temperature units) of 1.2 for NGC0467, and 1.75 in NGC7550. Thus, all else being equal, we expect these gas reservoirs to be somewhat extended, but not to completely fill the IRAM-30m beam. This is consistent with the molecular gas being co-spatial with the dust-discs present in these two sources, which are seen in absorption in optical imaging (see Fig \ref{codetsfig}).

\subsubsection{H$_2$ masses}
We estimate H$_2$ gas masses and limits from our CO observations in the standard manner, using the following equation

\begin{equation}
M_{H_2} = 2m_H \frac{\lambda^2}{2 k_b} X_{\rm CO} D_L^2 \int{S_v \delta V},
\label{co2h2}
\end{equation}
\noindent where $m_H$ is the mass of a hydrogen atom, $\lambda$ is the wavelength, $k_b$ is Boltzmann's constant, $D_L$ is the luminosity distance, $\int{S_v \delta V}$ is the integrated CO flux density and  X$_{\rm CO}$ is your CO-to-H$_2$ conversion factor of choice in units of K \kms.

As in  \cite{2016MNRAS.455..214D} we here use a Galactic X$_{\rm CO}$ factor of 3$\times$10$^{20}$ cm$^{-2}$ (K \kms)$^{-1}$ \citep{Dickman:1986jz}. As these ETGs are massive, and mass correlates positively with metallicity, such a value seems reasonable. It is possible, however, that the gas in these galaxies has been accreted from a low-metallicity source. In such a case we would be underestimating the total gas mass. 

In order to set upper-limits on the H$_2$ mass of our undetected sources we use the 3$\sigma$ RMS on the CO(1-0) spectrum with 52 \kms\ channels, assuming a total velocity width of 250 \kms, to allow direct comparison to \cite{2011MNRAS.414..940Y}. The detected MASSIVE objects all have larger velocity widths than this, and if we assumed a velocity width of 500 \kms\ and 100 \kms\ channels our upper-limits would increase by a factor of $\sqrt{2}$. The gas masses and limits derived are presented in Table \ref{obstable}.

\section{Results}
\label{results}
Combining this work, and the objects from our pilot sample (and the literature) we now have access to the gas properties of a representative sample, drawn uniformly from the MASSIVE parent volume. Sixty seven of the MASSIVE ETGs have been observed in molecular gas to date (from this work, \citealt{2016MNRAS.455..214D} and other literature sources; see Table \ref{obstable}). Crucially these objects were not selected on any proxy for the presence of cold gas, and we use this fact in what follows to allow us to probe the average gas properties of this galaxy population.  

\subsection{Detection rate}

From the 67 observed MASSIVE galaxies we detect line emission in 17; a detection rate of 25$^{+5.9}_{-4.4}$\%. This rate is very similar to the detection rate of of 22$\pm3$\% (56/259) found by \cite{2011MNRAS.414..940Y} from a survey of all the lower mass \atlas\ ETGs, and the 26\% reported by in the smaller volume limited studies of \cite{1996ApJ...460..271K} and \cite{Welch:2010in}. Combining the MASSIVE and \atlas\ surveys yields a joint detection rate of 22.4$^{+2.4}_{-2.1}$\% for ETGs of all masses.

Below we combine the MASSIVE and \atlas\ surveys to explore if (and how) the molecular gas detection rate varies with the physical properties of ETGs. The detection rate is an interesting quantity, as it tells us the fraction of objects with molecular masses greater than $\approx$0.1\% of the stellar mass (the detection threshold in this work). We move on in a later section to consider how the molecular masses themselves correlate with these quantities. 

In what follows we utilise the $K_s$-band luminosity as a measure of the stellar mass of our objects. Dynamical masses are available for the \atlas\ objects, but not (to date) for the MASSIVE galaxies. None of the results reported in this paper change if dynamical masses are used instead of $K_s$-band luminosities for the \atlas\ objects.

We also note that the molecular gas observations of MASSIVE and \atlas\ followed a different rationale. In this work, as described above, we observed each ETG such that we were sensitive to a molecular-to-stellar mass fraction of $\approx$0.1\% (at 3$\sigma$). In \atlas, however, each object was observed to the same RMS noise level, meaning that objects nearer to us have deeper observations than those further away. Here we combine these two surveys directly, in order to maximise our sample size, but we note that essentially none of the results we present would change if we instead removed the eleven objects from \atlas\ with a molecular-to-stellar mass fraction of $<$0.1\%. In the few cases where results depend on the choice to include these objects we highlight this explicitly in the text.

\begin{figure*}
%\begin{flushleft}
\includegraphics[width=13cm,angle=0,clip,trim=0cm 1.5cm 0cm 0.0cm]{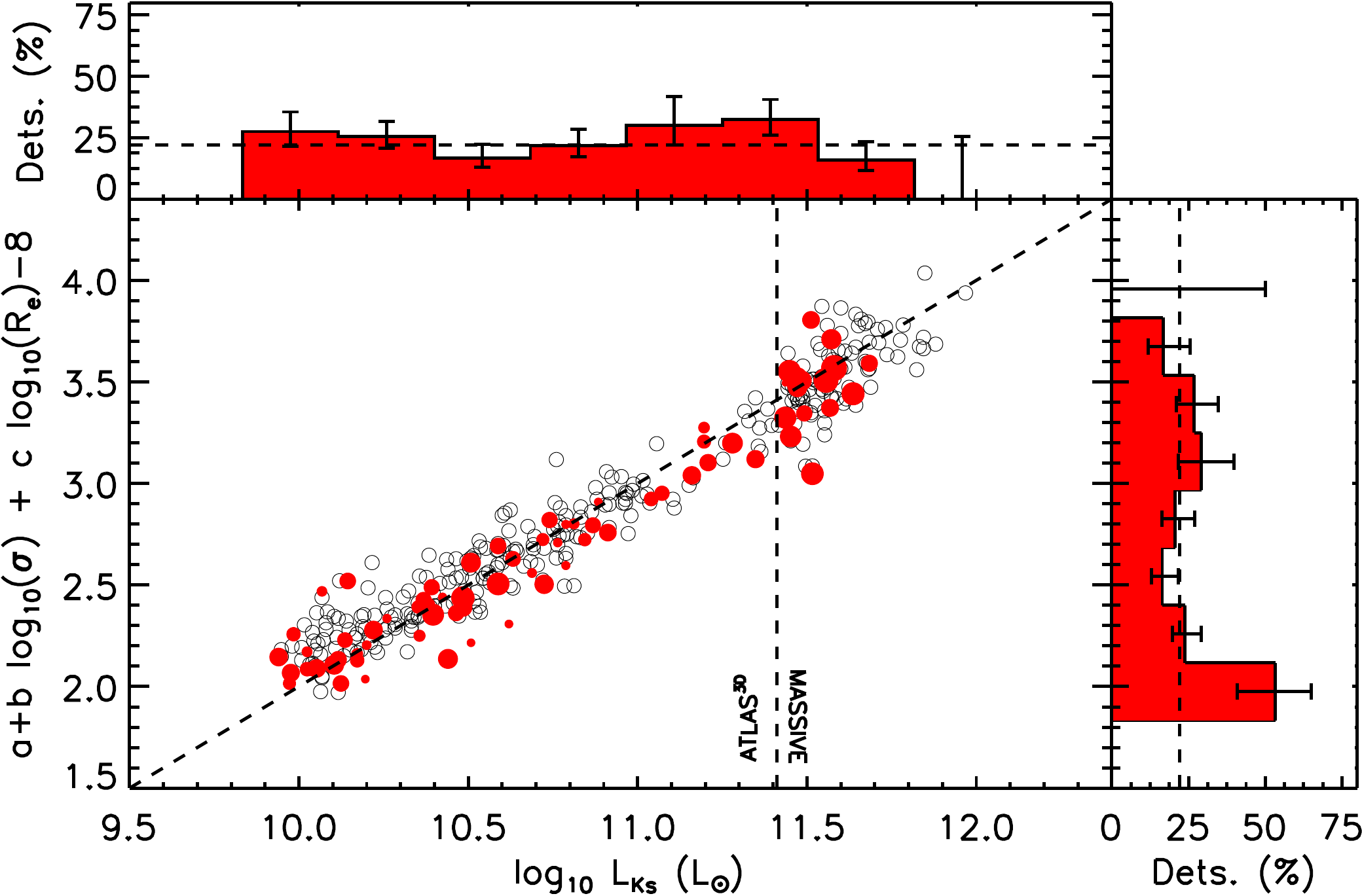}
\includegraphics[width=13cm,angle=0,clip,trim=0cm 1.5cm 0cm 0.0cm]{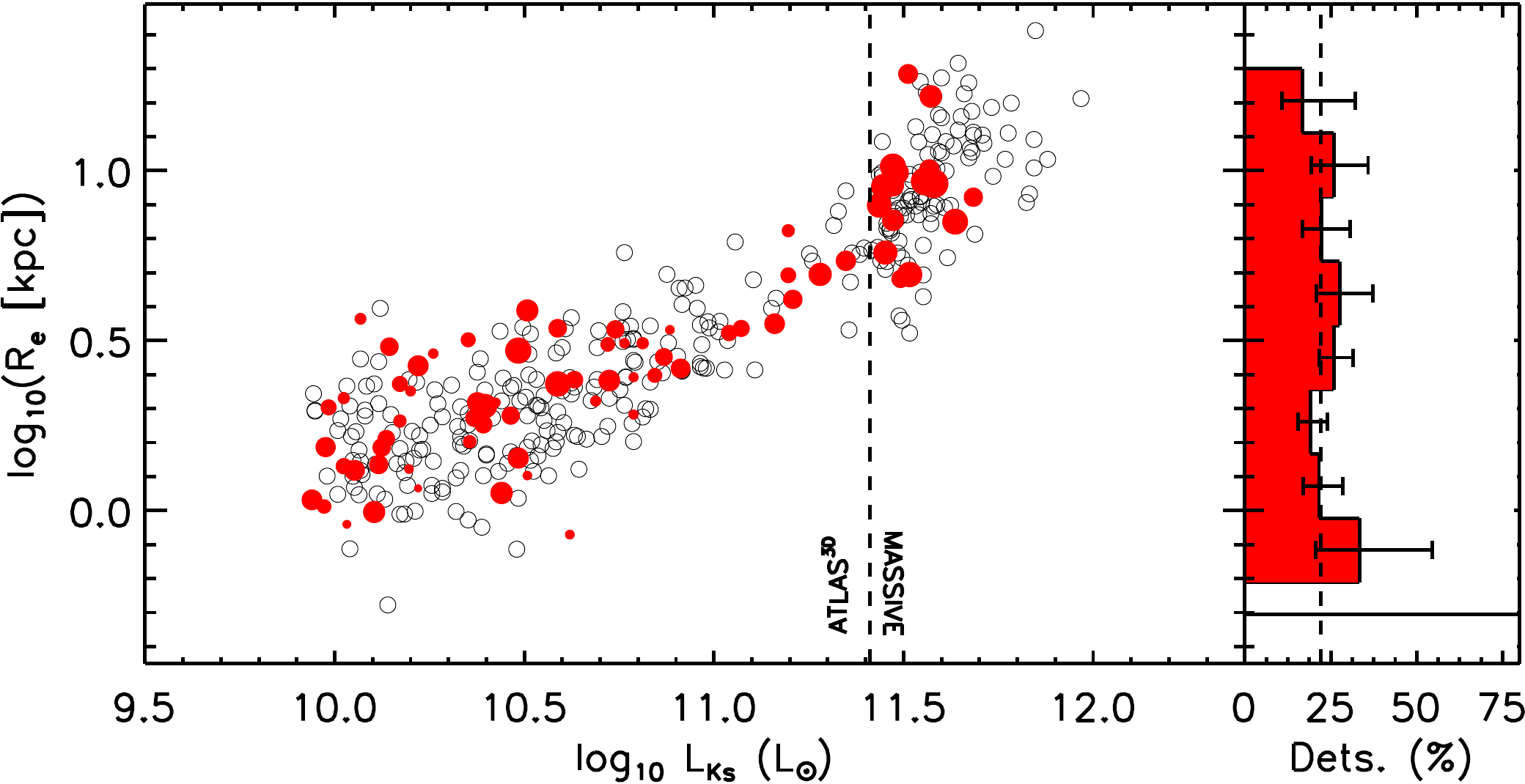}\\
\includegraphics[width=13cm,angle=0,clip,trim=0.07cm 0.0cm 0cm 0.0cm]{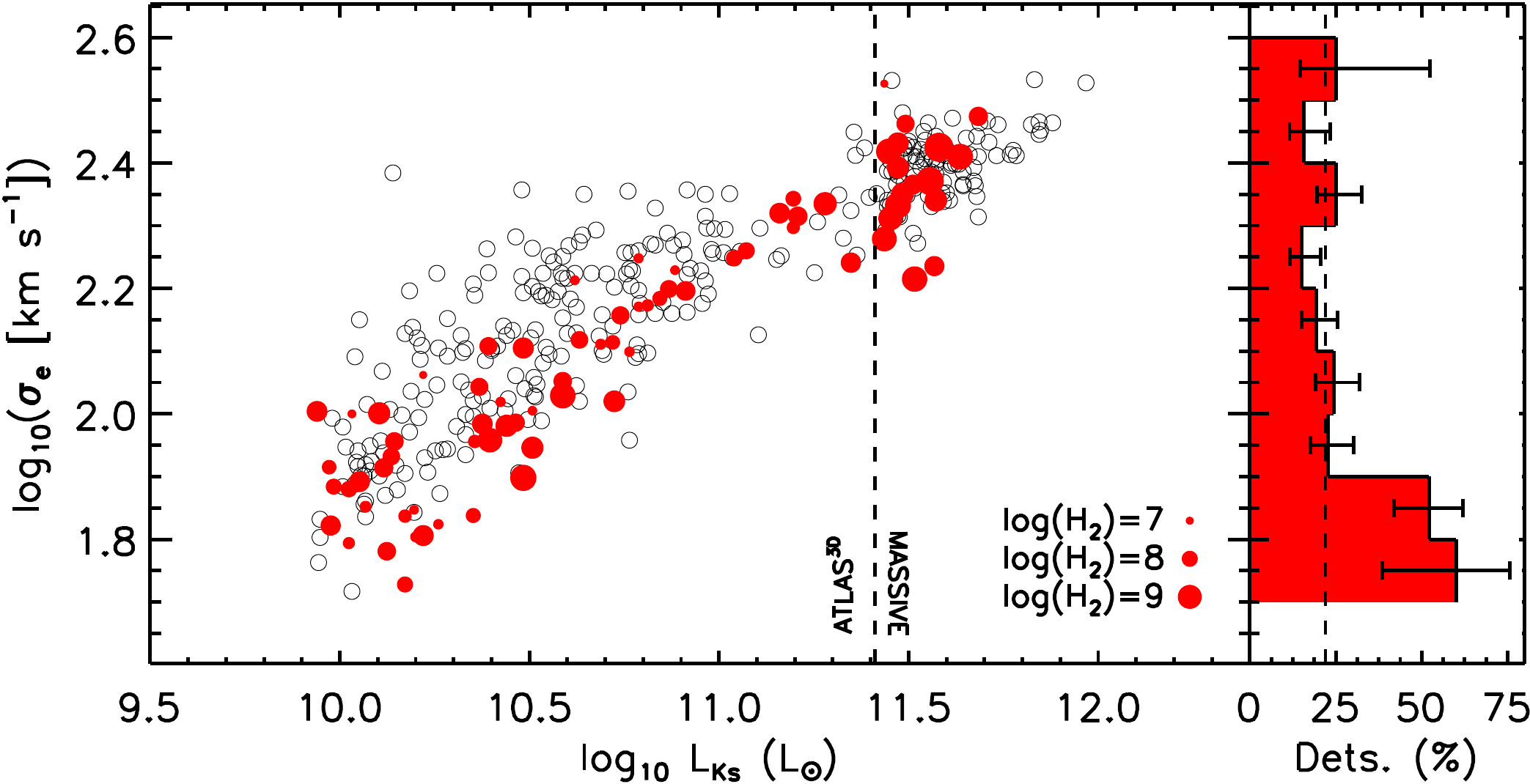}\\
\caption{\textit{Main panels:} $K_s$-band luminosity of the combined MASSIVE and \atlas\ CO survey galaxies, plotted against a combination of the effective radius (the radius containing half of the stellar light) and velocity dispersion (measured within that effective radius) which together form an edge-on view of the fundamental plane of ETGs (top panel), the effective radius of the galaxy (middle panel) and the velocity dispersion measured within the effective radius of the galaxy (bottom panel). The approximate dividing line between the MASSIVE and \atlas\ surveys is shown as a black dashed line. Objects which were not detected in molecular gas are shown as open circles, while detections are shown with red filled symbols, whose size scales with the total H$_2$ mass (between $10^7$ and $10^{9.5}$ \msun), as shown in the legend.
\textit{Top and right panels:} Histograms, whose bins denote the fraction of ETGs detected within that bin when the distribution is projected down onto this axis, with error bars denoting the binomial 1$\sigma$ confidence limits. The black dashed line represents the total detection rate of 22\%. We are unable to reject the null hypothesis that the detection rate is the same for galaxies in all parts of the parameter space, apart from at very low $\sigma_e$.} 
\label{fig:fundplane}
%\end{flushleft}
 \end{figure*}

\subsubsection{Fundamental plane}

As highlighted in detail in \cite{2013MNRAS.432.1862C} many of the properties of ETGs vary over the mass-size, and mass-velocity dispersion planes. These surfaces are projections of the `Fundamental Plane' of early-type galaxies \citep{1987ApJ...313...59D,1987ApJ...313...42D}. We here consider the detection rate (and mass fraction) of molecular gas as a function of the various fundamental plane parameters.

On the top and right of Figure \ref{fig:fundplane} are histograms, whose bins denote the fraction of ETGs detected by either survey within that bin when the distribution is projected down onto this axis. The error bars shown on each bin are the binomial 1$\sigma$ confidence limits. The black dashed line represents the total detection rate of 22\%. This allows us to show that the detection rate is the same for galaxies in all parts of the parameter space.

In the top panel of Figure \ref{fig:fundplane} we show the $K_s$-band luminosity of the combined MASSIVE and \atlas\ CO survey galaxies, plotted against a combination of the effective radius (the radius containing half of the stellar light) and velocity dispersion (measured within that effective radius) which together form an edge-on view of the fundamental plane of ETGs. The observational quantities are all reproduced from \\ \cite{2011MNRAS.413..813C,2013MNRAS.432.1709C,2014ApJ...795..158M,2017MNRAS.464..356V}, and the constants a, b and c (which are used to produce the dashed fit line) are those which produce an edge on view of the plane \citep{2013MNRAS.432.1709C}. The approximate dividing line between the MASSIVE and \atlas\ surveys is shown on the figure. Objects which were not detected in molecular gas by either survey are shown as open circles, while detections are shown with red filled symbols, whose size scales with the total H$_2$ mass (between $10^7$ and $10^{9.5}$ \msun).

We note that the luminosities and radii of massive ETGs are hard to measure in shallow infrared imaging, and may be somewhat underestimated in the 2MASS survey data we use here \citep[e.g.][]{Lauer2007}. This issue will be addressed in detail in Quenneville et al., in prep, who will present details of a dedicated deep infrared imaging campaign. For consistency with \atlas\ we use the 2MASS photometry here, but note that none of our conclusions would change if we instead used these new data.

The top panel of Figure \ref{fig:fundplane} clearly shows that molecular gas is detected in ETGs in all parts of the fundamental plane. The detection rate of objects appears to be constant both as a function of galaxy mass and fundamental plane parameters, apart from at very low values of the fundamental plane axis. This signature is likely an artefact, as discussed in Section \ref{massveldisp_present} below.
No molecular gas is detected in the very most massive objects, but the low number statistics mean that we cannot reject the null hypothesis of no correlation between the detection rate and stellar mass. The amount of gas present (as indicated by the size of the symbols) also appears to not correlate with either axis (see Section \ref{sec:gasmass_frac}).

\begin{figure*} 
\begin{center}
\includegraphics[width=0.75\textwidth,angle=0,clip,trim=0cm 0cm 0cm 0.0cm]{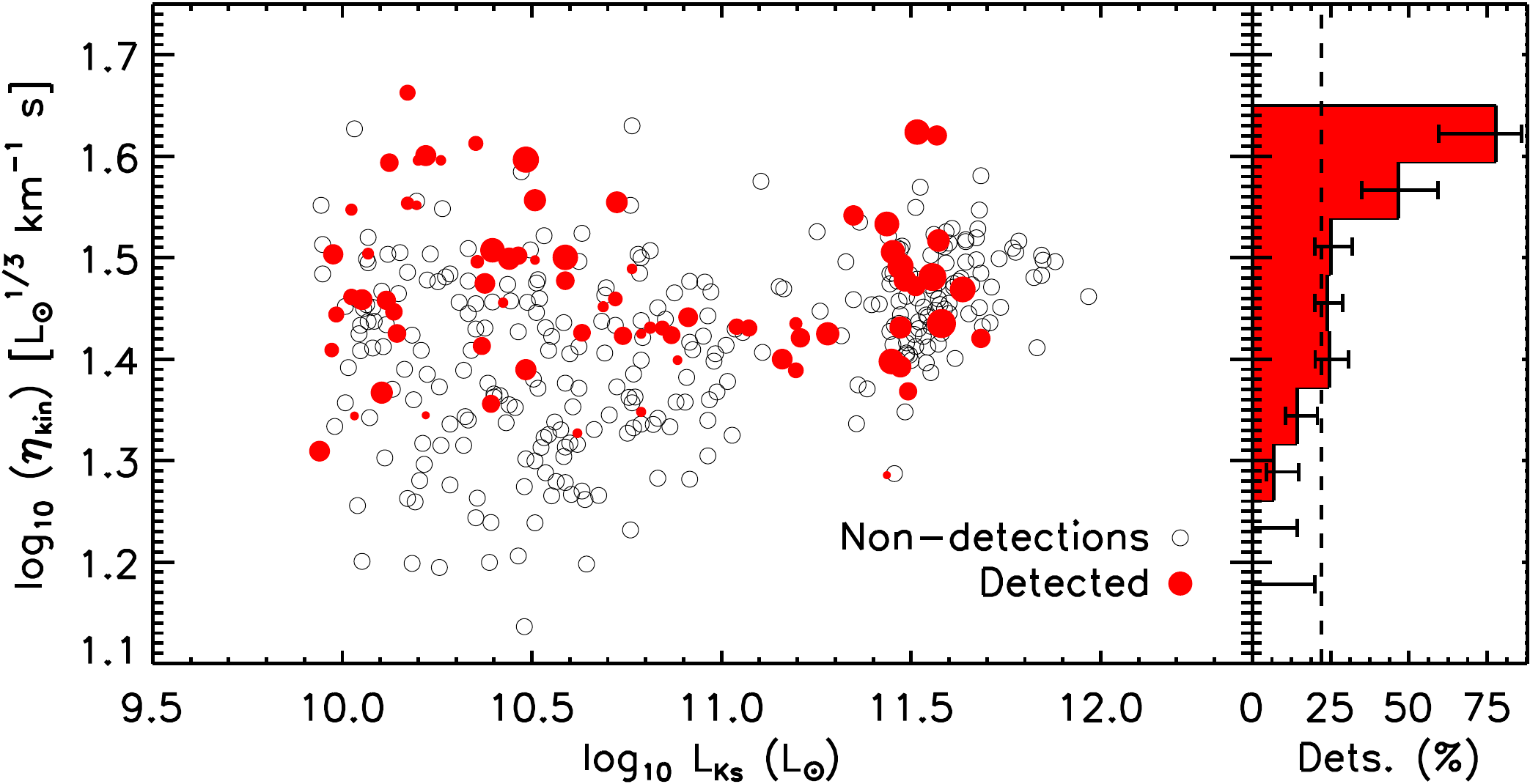}
\caption{\textit{Left panel:} As Figure \ref{fig:fundplane}, but showing the $K_s$-band luminosity of the combined MASSIVE and \atlas\ CO survey galaxies against $\eta_{\rm kin}\equiv\frac{\sqrt[3]{\mathrm{L}_{Ks}}}{\sigma_e}$ (where $L_{Ks}$ is the galaxies $Ks$-band luminosity and $\sigma_e$ is the velocity dispersion). This parameter can be interpreted as a kinematic bulge fraction, and is formulated as a rotation of the mass -- velocity dispersion plane to remove its mass-dependant slope. As shown in the histogram on the right panel $\eta_{\rm kin}$ clearly correlates with detection fraction. }
\label{fig:gaspredictor}
 \end{center}
 \end{figure*}

\subsubsection{Mass -- Size plane}

In the middle panel of Figure \ref{fig:fundplane} we show the mass -- size projection of the fundamental plane, plotting the $K_s$-band luminosity of the combined MASSIVE and \atlas\ CO survey galaxies against their effective radius (the radius containing half of the stellar light).

Molecular gas is detected in ETGs in all parts of the mass -- size plane. The detection rate of objects appears to be constant both as a function of galaxy mass and size. Once again, the amount of gas present (as indicated by the size of the symbols) also appears to not correlate with either axis (see Section \ref{gasmass_frac}).

 \subsubsection{Mass -- Velocity Dispersion plane}
 \label{massveldisp_present}

In the bottom panel of Figure \ref{fig:fundplane} we show the mass -- velocity dispersion projection of the fundamental plane, plotting the $K_s$-band luminosity of the combined MASSIVE and \atlas\ CO survey galaxies against their velocity dispersion measured within one effective radius. 

This panel shows clearly that the excess at low fundamental plane parameters is caused by an excess of molecular gas detections at very low velocity dispersions, below $\approx$80 \kms, where the detection rate of molecular gas increases to $\approx$50\%. This signature comes entirely from the \atlas\ survey, and was already reported in \cite{2011MNRAS.414..940Y}. It is likely due to misclassification of a small number of low-mass spiral/dwarf galaxies for which no good imaging exists, and thus contaminate the sample. 

Despite the constant detection fraction as a function of velocity dispersion shown across most of the combined MASSIVE+\atlas\ sample, it is clear that the molecular gas detections do not fill the mass -- velocity dispersion plane uniformly (with a lack of detections in the top left of the figure). At fixed mass the detection rate of cold molecular gas is dependent on velocity dispersion, at least in the lower mass \atlas\ sample. The dynamic range covered by MASSIVE is lower, so it is harder to tell if the same is true in the more massive objects. The molecular gas fraction at fixed mass also seems to vary across this plane (as discussed previously by \citealt{2013MNRAS.432.1862C}). 

In order to show this trend more clearly in Figure \ref{fig:gaspredictor} we plot the $K_s$-band luminosity of the combined sample objects versus ${\eta_{\rm kin}\equiv\frac{\sqrt[3]{M}}{\sigma_e}}$, a combination of variables that allows us to rotate the mass -- velocity dispersion plane to remove its mass-dependent slope.
$\eta_{\rm kin}$ correlates with various measures of galaxy morphology (such as Sersic index, concentration index, and stellar mass surface density), and can be interpreted as a kinematic bulge fraction at fixed stellar mass \citep[e.g][]{2013MNRAS.432.1862C}.
 This Figure clearly shows that the gas detection rate varies significantly across this plane. At low $\eta_{\rm kin}$ the detection rate of cold gas is $<$10\%, while it increases to $\approx$60\% at high $\eta_{\rm kin}$ (low kinematic bulge fraction at fixed mass). The low $\sigma_{e}$ outliers discussed above do not drive this trend, as it persists when all objects with $\sigma_e<$80 \kms are removed. 

It is less clear if the detection rate in the MASSIVE sample alone correlates with $\eta_{\rm kin}$ in the same way as for the lower mass \atlas\ sample. A Kolmogorov-Smirnov test provides some evidence that the distribution of $\eta_{\rm kin}$ values for MASSIVE detections differs significantly from that for \atlas\ detections (at $>5\sigma$). This difference seems driven by the slow-rotators; when these objects are removed the $\eta_{\rm kin}$ distributions become statistically indistinguishable. 
We will discuss this correlation further in Section \ref{discuss}.

\subsubsection{Colour}
\begin{figure*}
%\begin{flushleft}
\includegraphics[width=13cm,angle=0,clip,trim=0cm 0cm 0cm 0.0cm]{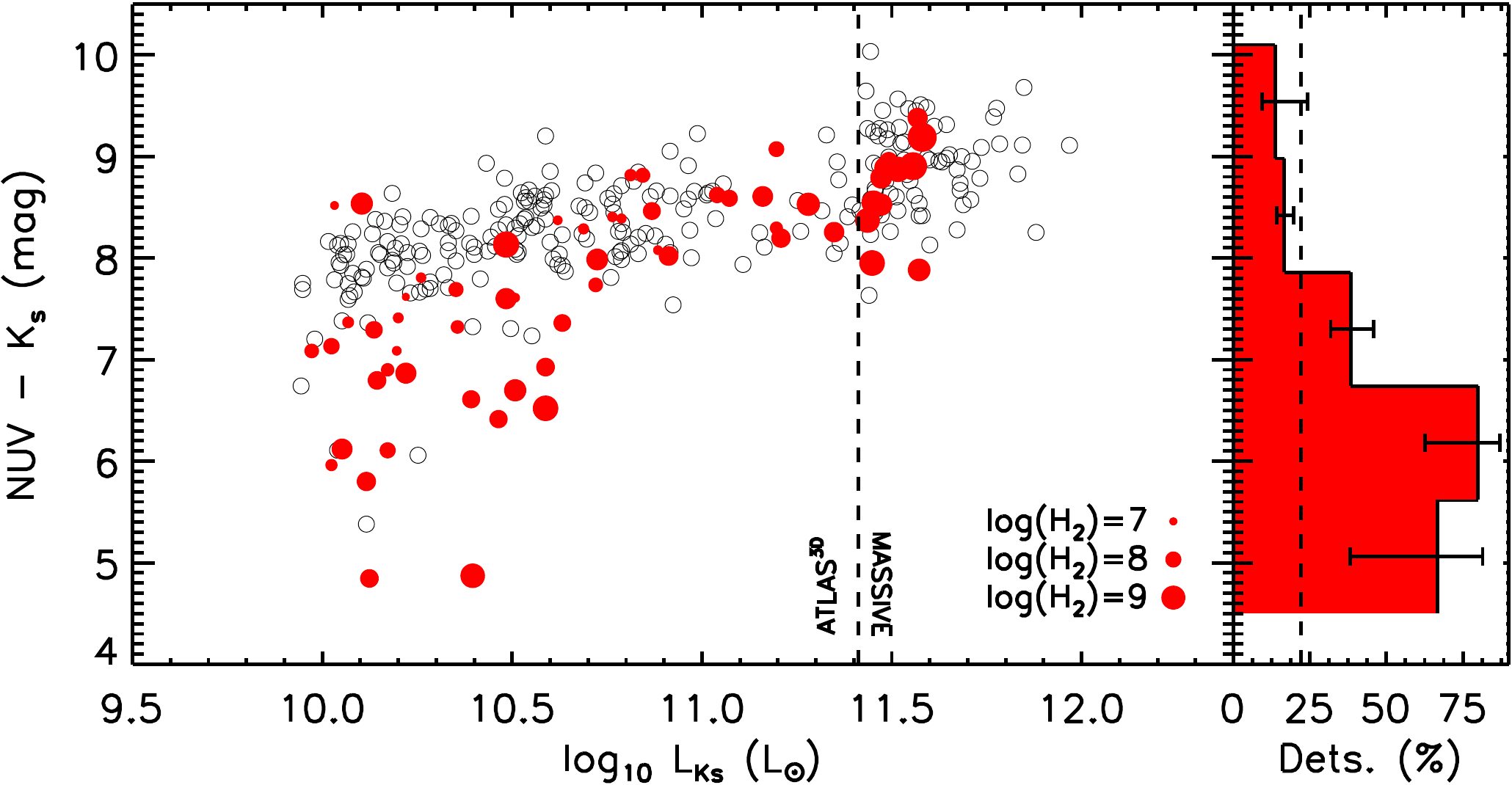}
\caption{As Figure \ref{fig:fundplane}, but showing the $K_s$-band luminosity of the combined MASSIVE and \atlas\ CO survey galaxies against their $NUV$ to $Ks$-band colour. While the detection rate of cold gas correlates with colour over the whole sample, at high stellar mass this is no longer true. }
\label{fig:cmd}
%\end{flushleft}
 \end{figure*}

One would expect to find a correlation between the presence of gas in an ETG and its colour, especially in the Ultraviolet bands, which are sensitive to the presence of young stars. 
In Figure \ref{fig:cmd} we plot such a figure, showing the $K_s$-band luminosity of the combined MASSIVE and \atlas\ CO survey galaxies against their Near Ultraviolet (NUV) minus $K_{s}$ colours (from GALEX and 2MASS, respectively). 
NUV magnitudes for the galaxies used in this work were obtained from the GALEX catalogue server, {release GR7}. Where multiple observations of the same target exist, we always used the deepest observation.
{Not all of our objects lie within the GALEX survey area, but all that do are detected in the NUV band. In what follows we have assumed that ignoring those ETGs that fall outside the GALEX survey footprint does not bias our results.}

UV emission in ETGs can typically arise from two sources. The first is OB stars indicating ongoing star formation. The second is emission from evolved stars (the UV upturn effect; e.g. \citealt{2005ApJ...619L.111Y}). 
The UV-upturn is, however, important mostly in the FUV, and if even a small amount of young stars are present they will dominate the UV emission. In Figure \ref{fig:cmd} clearly the tail of low-mass galaxies with blue $NUV$-$K$ colours are much more likely to be molecular gas rich, as already shown in \cite{2014MNRAS.444.3408Y}. {We note that excluding galaxies with molecular-to-stellar mass fraction below 0.1\% in \atlas\ (where our combined survey is not complete) strengthens the colour trend, as these systems are preferentially found on the $NUV$-$Ks$ red sequence.}
The addition of the MASSIVE sample, however, shows how insensitive even ultraviolet colours are to the presence of gas (and star formation) in the most massive galaxies.

 \subsubsection{Specific angular momentum}
 \label{lr_matters}
A variety of authors have shown that classifying ETGs by their specific angular momentum allows one to predict their properties (and constrain their evolutionary histories) in a physically motivated way. 

\cite{2011MNRAS.414..940Y} were able to study the molecular gas detection rate of the \atlas\ sample as a function of $\lambda_{Re}$, a simplified and dimensionless version of the luminosity-weighted specific angular momentum. They found that there was no dependence of the detection rate with $\lambda_{Re}$ amongst the fast-rotators ($\lambda_{Re}$/$\sqrt{\epsilon}$>0.31), but that the detection rate was very much lower amongst the slow-rotators. Indeed, none of the CO detected objects classified as slow-rotators were true bulge/dispersion dominated systems. NGC1222, is an ongoing merger (\citealt{2007AJ....134.1237B}),  NGC4550 is an object with two counter-rotating stellar discs (denoted 2CR objects; \citealt{2009MNRAS.393.1255C, 2011MNRAS.414.2923K}), and NGC4476 is a complex case with a large kinematically decoupled core \citep{2002AJ....124..788Y,2011MNRAS.417..882D}. This lead to speculation that the mechanisms that supply gas to ETGs were suppressed in true slow-rotators. This study was hampered, however, by a lack of true bulge/dispersion dominated systems within the survey volume. With MASSIVE we are in a prime position to investigate this further. 

\begin{figure*} 
\begin{center}
\includegraphics[width=0.75\textwidth,angle=0,clip,trim=0cm 0cm 0cm 0.0cm]{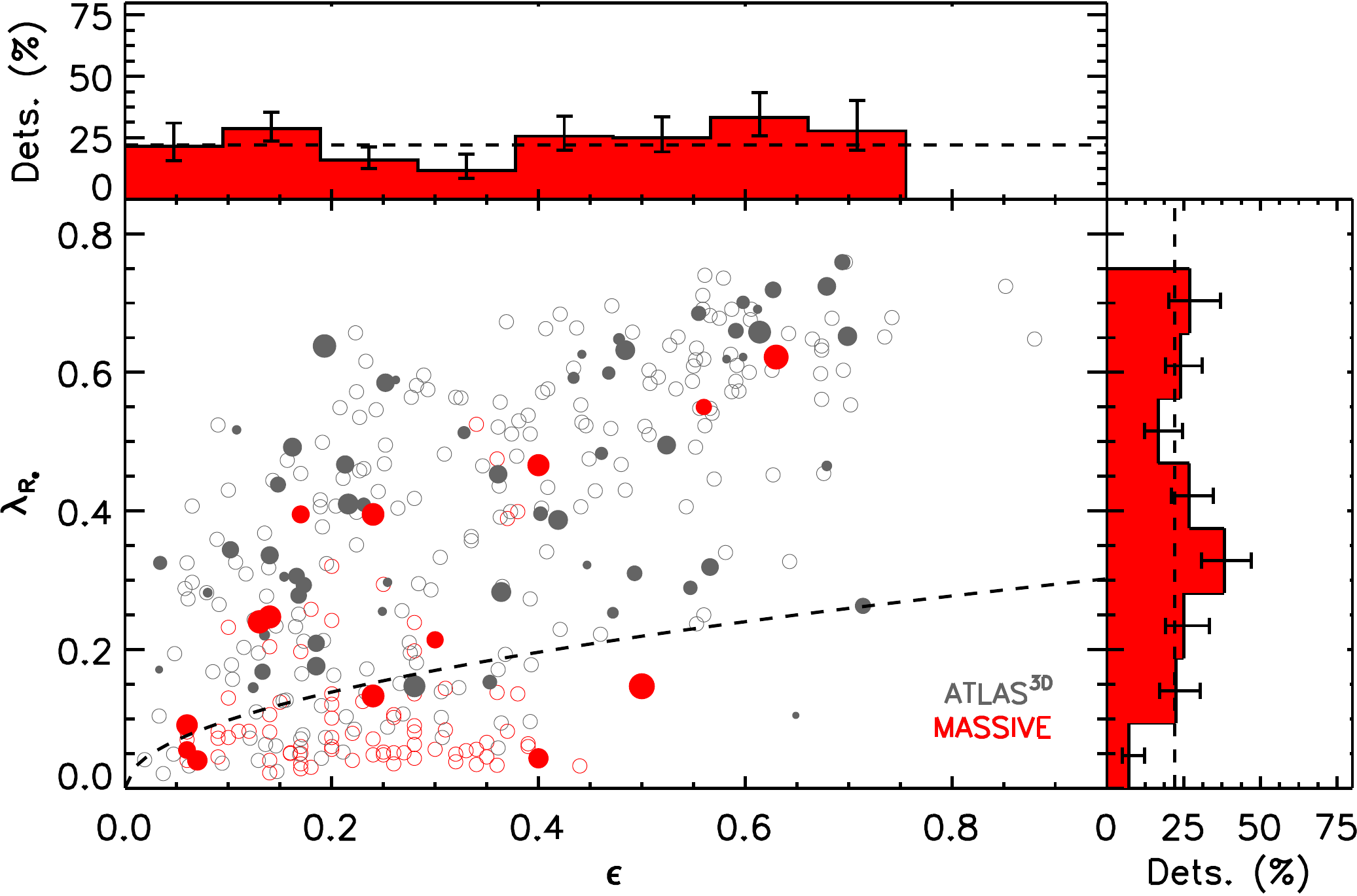}
\includegraphics[width=0.75\textwidth,angle=0,clip,trim=0cm 0cm 0cm 0.0cm]{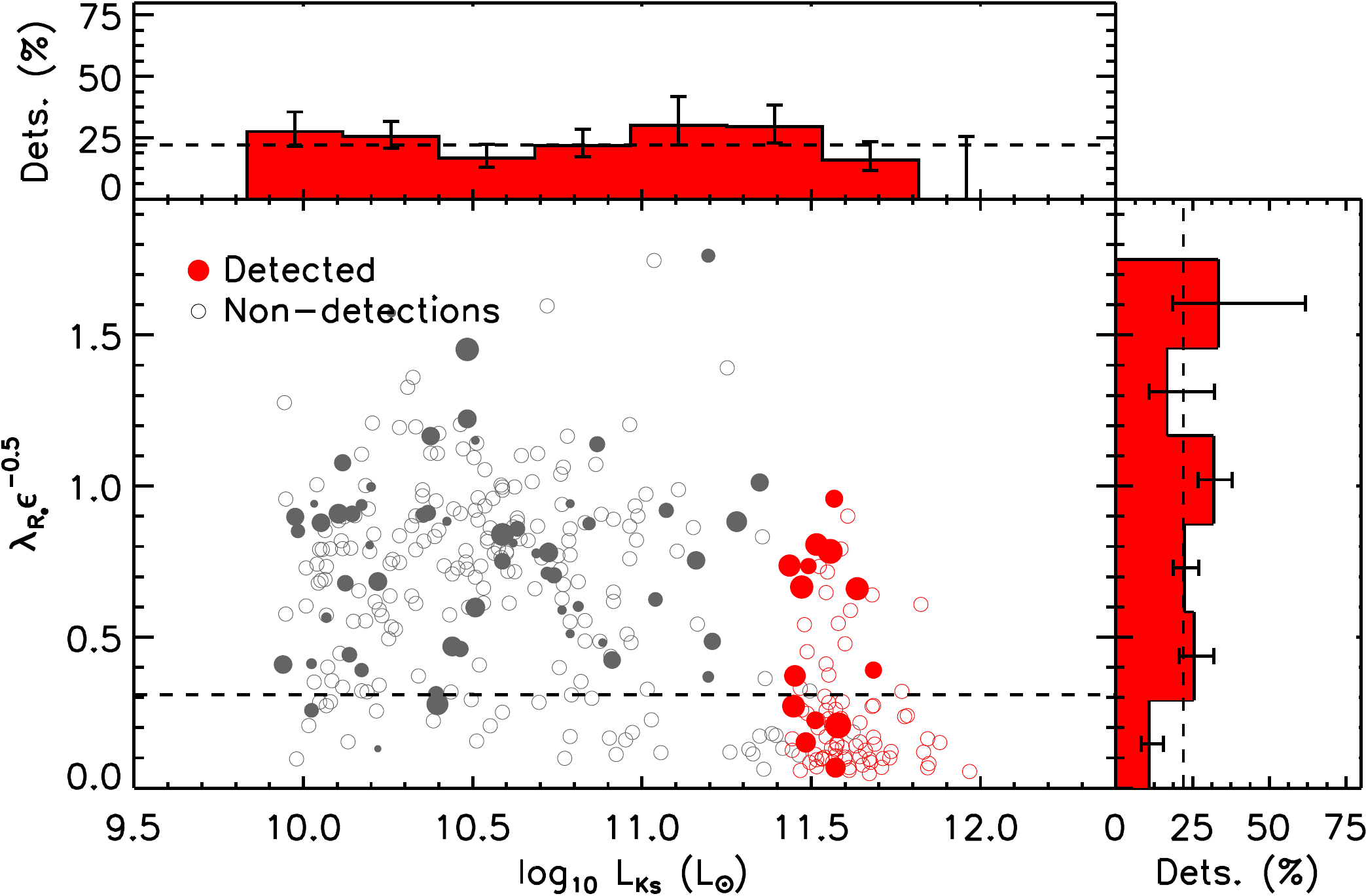}
\caption{\textit{Top plot}: $\lambda_{\rm R_e}$ vs $\epsilon$ plane.  \textit{Bottom plot}: $\lambda_{\rm R_e}/\sqrt{\epsilon}$ vs L$_{\rm K_s}$ plane. Both are shown as in Figure \ref{fig:fundplane}, but with MASSIVE and \atlas\ galaxies indicated by grey and red points, respectively. The empirical divide between slow- and fast-rotators from \protect \cite{2011MNRAS.414..888E} is shown as a black dashed line in both plots. Both plots show that while fast-rotators have a similar detection rate at all $\lambda_{\rm R_e}$ values, slow-rotators are detected less often. }
\label{fig:lambdar_mass}
\label{fig:lambdar_eps}
 \end{center}
 \end{figure*}

In the top panel of Figure \ref{fig:lambdar_eps} we show the $\lambda_{Re}$ vs $\epsilon$ diagram  \citep{2007MNRAS.379..401E,2011MNRAS.414..888E} for the MASSIVE galaxies in red, and \atlas\ in grey.  As before, objects which were not detected in molecular gas are shown as open circles, while detections are shown with filled circles, whose size scales with the total H$_2$ mass (between $10^7$ and $10^{9.5}$ \msun). $\lambda_{Re}$ and $\epsilon$ are taken from \cite{2011MNRAS.414..888E} for \atlas\ and \cite{2018MNRAS.479.2810E} for MASSIVE. The empirical divide between slow- and fast-rotators from \cite{2011MNRAS.414..888E} is shown as a dashed line. 
As in Figure \ref{fig:fundplane} we show histograms at the top and right of the plot, in order {to show if} the detection rate is the same for galaxies in all parts of the parameter space.
For clarity we also present $\lambda_{Re}$/$\sqrt{\epsilon}$ vs $K_s$-band luminosity for the sample objects in the bottom panel of Figure \ref{fig:lambdar_mass}, which allows a clearer separation of fast- and slow-rotators.

It is clear from Figure \ref{fig:lambdar_eps} that with MASSIVE we significantly increase the number of slow-rotators with limits on their molecular gas content. We detect cold gas both in MASSIVE fast-rotators, and in some of the slow-rotators {(NGC0708, NGC1684, NGC3862, NGC7052, NGC7550). All of these systems appear to be true bulge/dispersion dominated systems based on the data available, apart from one which has a kinematic twist \citep[NGC1684;][]{2018MNRAS.479.2810E}. } The detection rate of cold gas is consistent with being constant as a function of both $\lambda_{Re}$ and $\lambda_{Re}$/$\sqrt{\epsilon}$ in the fast-rotating population, but drops to $8.6_{-2.6}^{+3.0}$\% in the slow-rotators. If one were to remove the merging/2CR systems from \atlas\ this drops to $5.2_{-1.7}^{+2.9}$\%.
Overall it seems that slow-rotators are able to host cold molecular gas reservoirs, but the frequency with which they do so is significantly reduced. We will discuss the mechanisms potentially driving this in Section \ref{discuss}.

{Slow rotators dominate at the highest masses, so given their lower detection rate one might expect the molecular gas detection rate to drop in the most massive galaxies. The histograms in the bottom panel of Figure \ref{fig:lambdar_eps} show that there is a drop in the detection rate at the highest masses, that could well be due to the dominance of slow rotators, but this is not significant given our sample size. }

\begin{figure*} 
\begin{center}
\includegraphics[width=0.75\textwidth,angle=0,clip,trim=0cm 1.5cm 0cm 0.0cm]{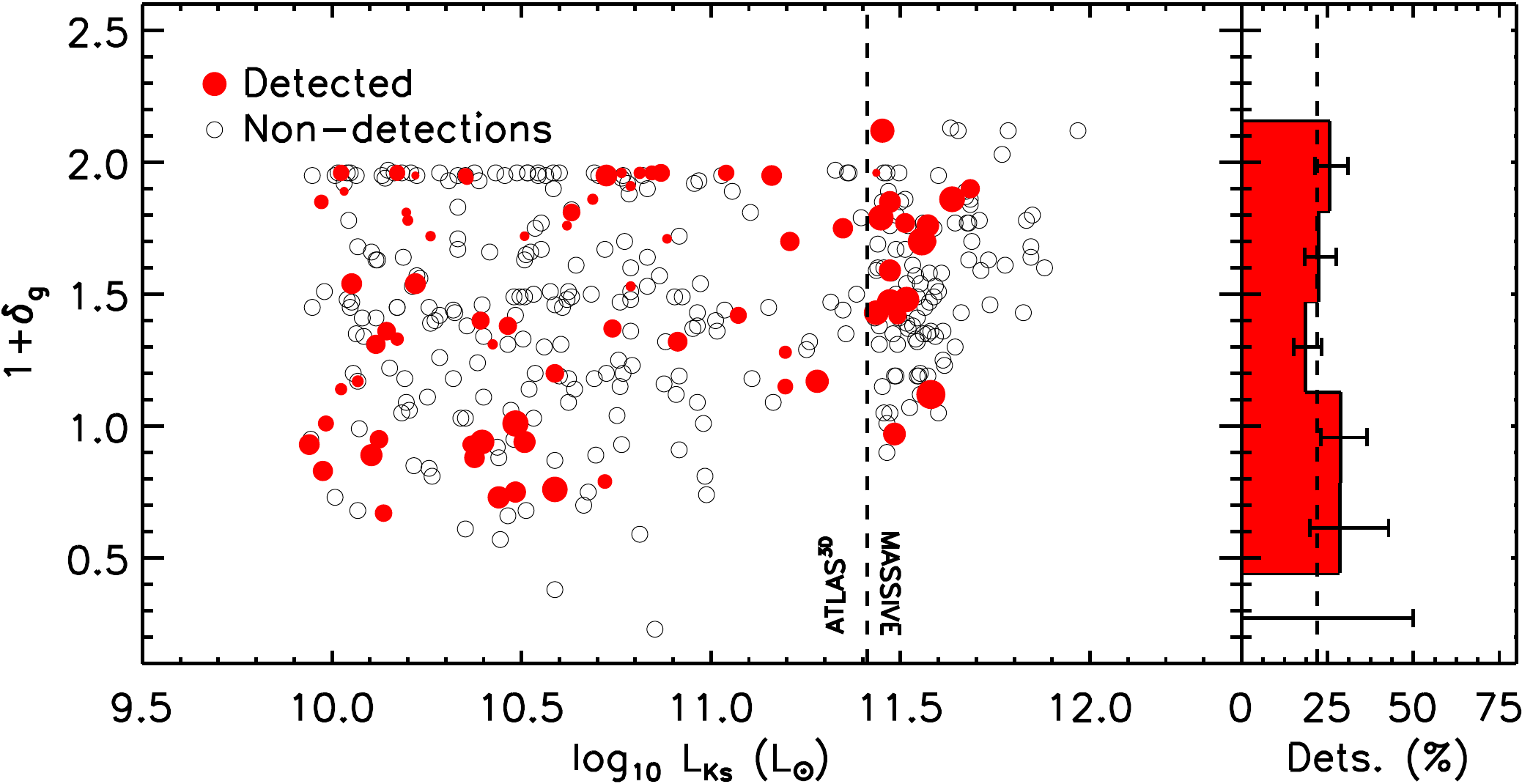}
\includegraphics[width=0.75\textwidth,angle=0,clip,trim=0cm 0cm 0cm 0.0cm]{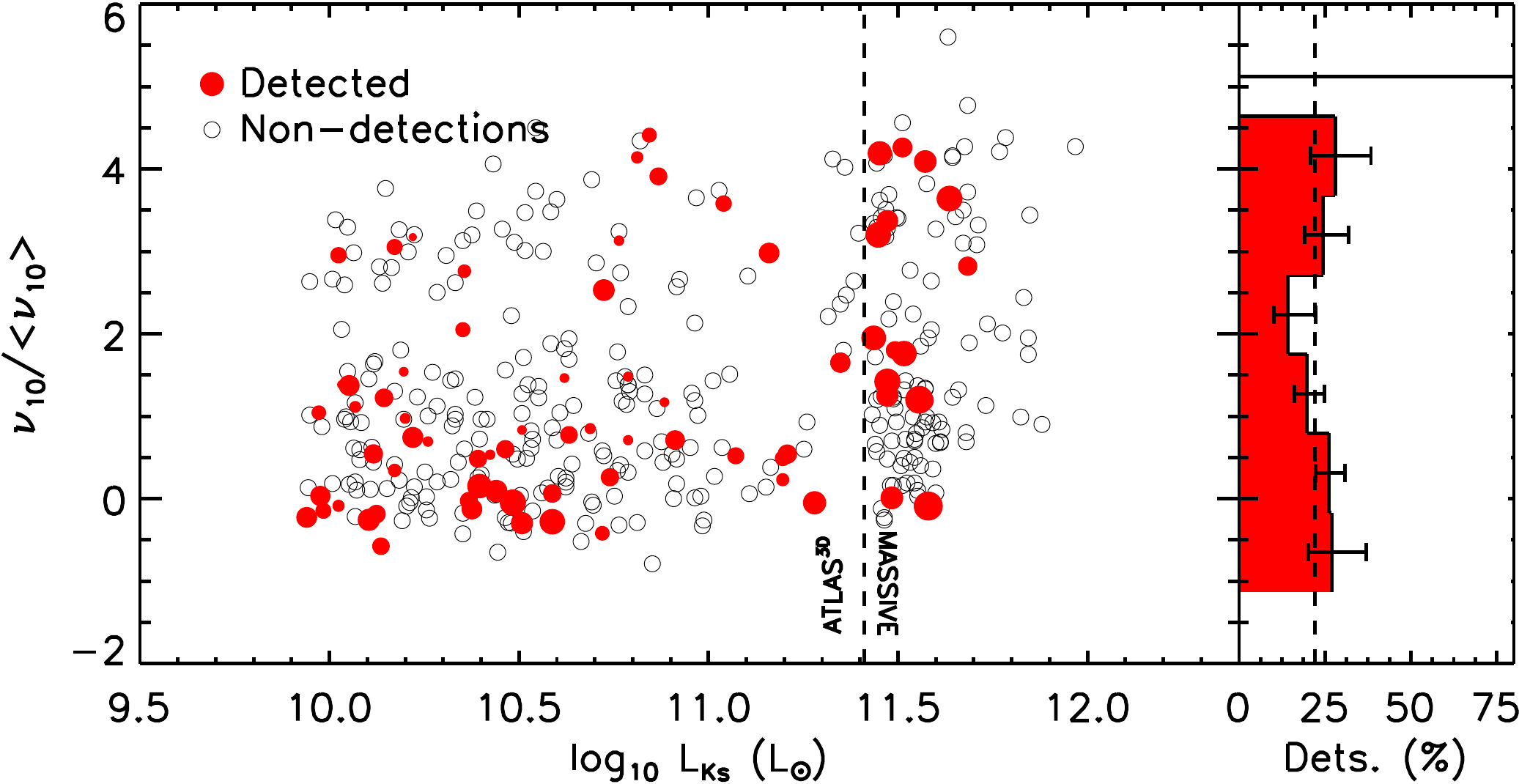}
\caption{As Figure \ref{fig:fundplane}, but showing the $K_s$-band luminosity of the combined MASSIVE and \atlas\ CO survey galaxies against large-scale (top panel) and small scale (bottom panel) environment indicators from \protect \cite{2017MNRAS.471.1428V}. No clear correlation is seen between detection rate of cold gas and environment. }
\label{fig:nu_delta_env}
 \end{center}
 \end{figure*}

\begin{figure} 
\begin{center}
\includegraphics[width=0.45\textwidth,angle=0,clip,trim=0cm 0cm 0cm 0.0cm]{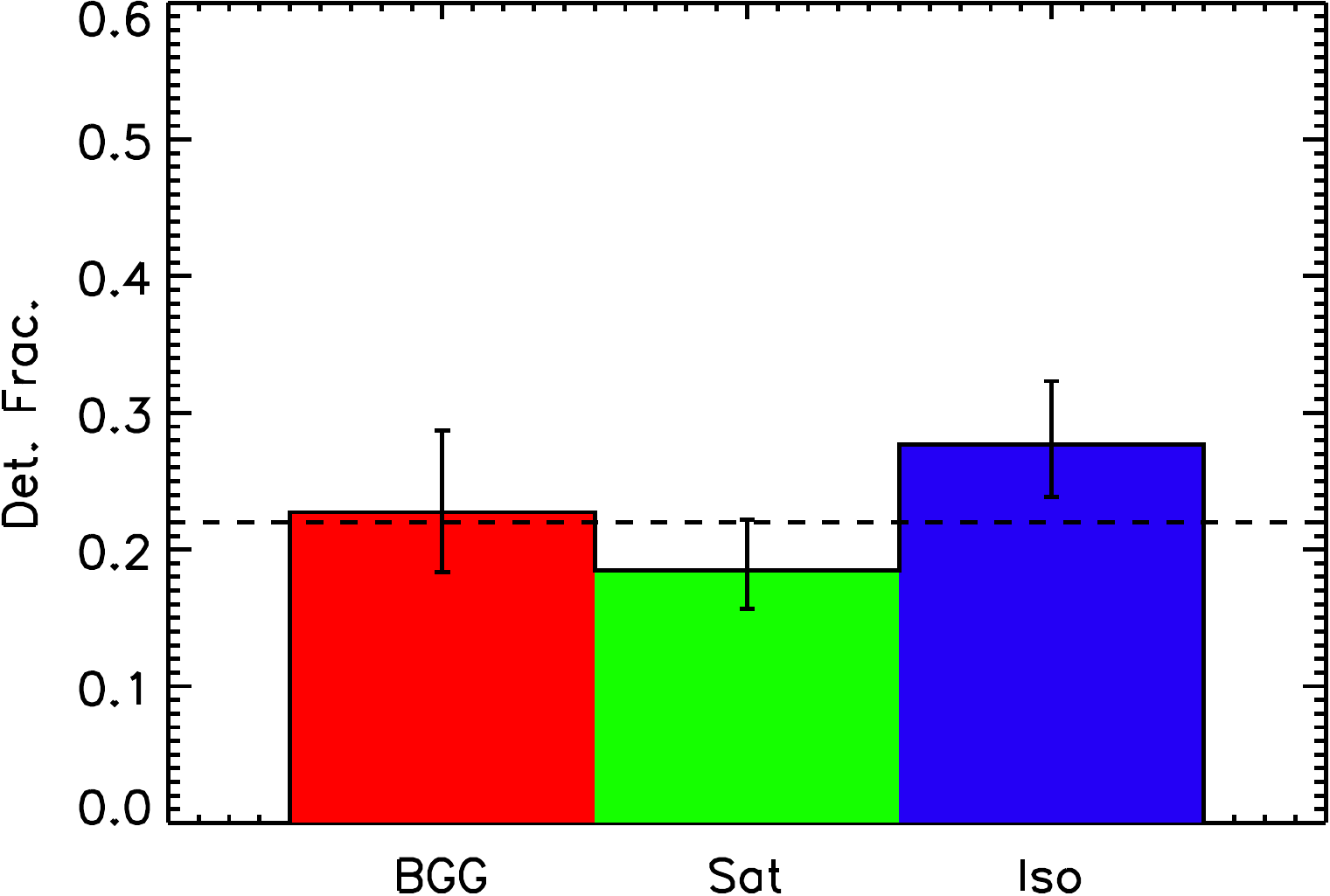}
\caption{Molecular gas detection fraction in brightest group, satellite and isolated galaxies in MASSIVE and \atlas (as classified by \protect \cite{2017MNRAS.471.1428V}). Although isolated systems have a slightly higher detection rate, this is {not statistically significant, and with this analysis} we cannot reject the null hypothesis that detection rates are independent of halo position. }
\label{fig:bgg_sat_iso_env}
 \end{center}
 \end{figure}

\subsubsection{Environment}

Environment is another potential driver of the molecular gas content in galaxies. As mentioned above, many of the mechanisms thought to be fuelling the cold gas in ETGs should be shut down in cluster environments {(at least for satellites, which cannot get gas from cooling flows)}. In \atlas\ \cite{2011MNRAS.414..940Y} did not find any dependance of the molecular gas detection rate on environment in the single cluster with their volume. In MASSIVE we probe systems with dark matter halo masses from $\approx 6\times10^{11}$ to $2\times10^{15}$\,\msun\ \citep{2017MNRAS.471.1428V}, giving us the ability to test this once more. MASSIVE includes the largest ETGs that lie in several different clusters, and a larger number of central brightest group galaxies (BGGs) which may have other mechanisms fuelling their gas reservoirs.  Our sample does not, however, include many brightest cluster galaxies, and we refer the reader to other works on these distinct objects (e.g. \citealt{2001MNRAS.328..762E,2003A&A...412..657S}). 

In Figure \ref{fig:nu_delta_env} we show the global overdensity indicator $\delta_g$ (the luminosity-weighted galaxy density contrast calculated over a 5.7 Mpc Gaussian kernel), and the local overdensity indicator $\nu_{10}/<\nu_{10}>$ (the normalised luminosity density within a region enclosing the 10 nearest neighbours of each galaxy), both reproduced here from \cite{2017MNRAS.471.1428V}, plotted against the $K_s$-band luminosity of the combined MASSIVE and \atlas\ CO survey galaxies.

 The plots are laid out as above, with round coloured symbols denoting molecular gas detections, and the size of those symbols encoding the molecular gas mass. 

In both panels there does seem to be a slight drop in the molecular gas detection rate at intermediate densities, but this is not significant. Overall we thus conclude that there is no strong dependance of the molecular gas detection rate on environment density.

The position a galaxy occupies within its halo can be equally as important as the density of galaxies. For example an object that is the brightest in its vicinity is likely to lie close to the centre of the halo, where gas cooling from the hot gas halo can accumulate. Satellite galaxies, however, do not have this advantage. 
To test if the molecular gas detection rate depends on the position of an ETG in its halo, we plot in Figure \ref{fig:bgg_sat_iso_env} the molecular gas detection rate for objects classified as brightest group galaxies, satellites and isolated by \cite{2017MNRAS.471.1428V}. Once again we show the global detection rate as a dashed line, in order to show that the detection rate is the same for galaxies in all environments.

As Figure \ref{fig:bgg_sat_iso_env} shows overall we do not find any significant trend in molecular gas detection rate with the position of each object within its halo. We do see a marginal enhancement of the detection rate in isolated galaxies, and a corresponding drop in the detection rate of satellite objects, but this is not hugely significant. These differences are detected at less than the $<2\sigma$ level, so we would need a significant enhancement in the sample size to determine if these effects are real.

\subsubsection{Ionised gas emission}

Ionised gas emission is present in the majority of early-type galaxies \citep[e.g. in $\approx$75\% of \atlas\ objects;][]{2006MNRAS.366.1151S,2011MNRAS.417..882D}. Within MASSIVE at least 38\% of objects show clear [OII] emission with an equivalent width $>$2 {\AA} \citep{2017ApJ...837...40P}. 

{Insufficient overlap between the MASSIVE and \atlas\ samples (in terms of emission line selection/sensitivity) exists to compare the properties of the molecular emission with those of the ionised medium systematically.} However, all of the molecular gas detected objects presented here that are also studied in \cite{2017ApJ...837...40P} show detectable [OII] emission. The inverse is not true, however, as some objects without detectable molecular gas do have detectable ionised gas emission. The same trend was found in the lower mass \atlas\ sample \citep{2006MNRAS.366.1151S,2011MNRAS.417..882D}. 
This is consistent with the expectation that an ionised gas layer would form in the outer regions of molecular clouds (where self-shielding is not yet efficient) and in {\mbox{H\sc{ii}}} regions within the clouds, in addition to the presence of a dynamically hotter \citep{2011MNRAS.417..882D,2018ApJ...860...92L}, more diffuse component which can be maintained by ionisation from the older stellar population \citep[e.g.][]{2010MNRAS.402.2187S}.

\subsubsection{X-ray emission}

Given the possible importance of the hot halo in both quenching and refuelling ETGs, we searched for a correlation between the properties of X-ray emission (that arises from cooling hot-halos) and the detection rate of cold gas. 
{Suitable observations unfortunately are not available for all sample members.} 
We combined the X-ray observations presented in \cite{2013MNRAS.432.1845S} and \cite{2016ApJ...826..167G} for the available sample objects. We do not find any correlation between hot halo properties (X-ray luminosity, X-ray temperature, difference between the X-ray temperature and the Virial expectation) and the presence/absence of molecular gas, or the amount of gas present in this limited sub-sample of galaxies {(see Figure \ref{fig:lk_xray} in Appendix B)}. This lack of correlation will be discussed further in Section \ref{discuss}.

\subsection{Gas mass and gas fraction}
\label{sec:gasmass_frac}
In the previous section we attempted to determine the physical parameters of galaxies that might be important in controlling the acquisition of cold gas. Now we move on to understand if any of these processes control the amount of gas present, either in an absolute sense or as a function of the mass of the stars within each system. 

\subsubsection{Stellar mass}

As argued above, most of the internal processes that can return material to the ISM of ETGs are related in some way to the mass of stars present. In addition many other gas supply/depletion processes in galaxy evolution also correlate with the stellar (or halo) mass of the system (e.g. the formation of virial shocks, the masses of supermassive black holes, etc). It is thus crucial to know if the mass of ISM material present in ETGs correlates directly with the stellar mass (and thus gas fractions are fairly constant across the ETG population). Previous studies have found no such dependance of the gas mass on stellar mass \citep[e.g.][]{2011MNRAS.414..940Y}, suggesting the gas is dominantly supplied by external processes.
We adopt this as our null hypothesis, and will here determine if there is any significant evidence of variation from this within our new observations.

\begin{figure} 
\begin{center}
\includegraphics[width=0.48\textwidth,angle=0,clip,trim=0cm 0cm 0cm 0.0cm]{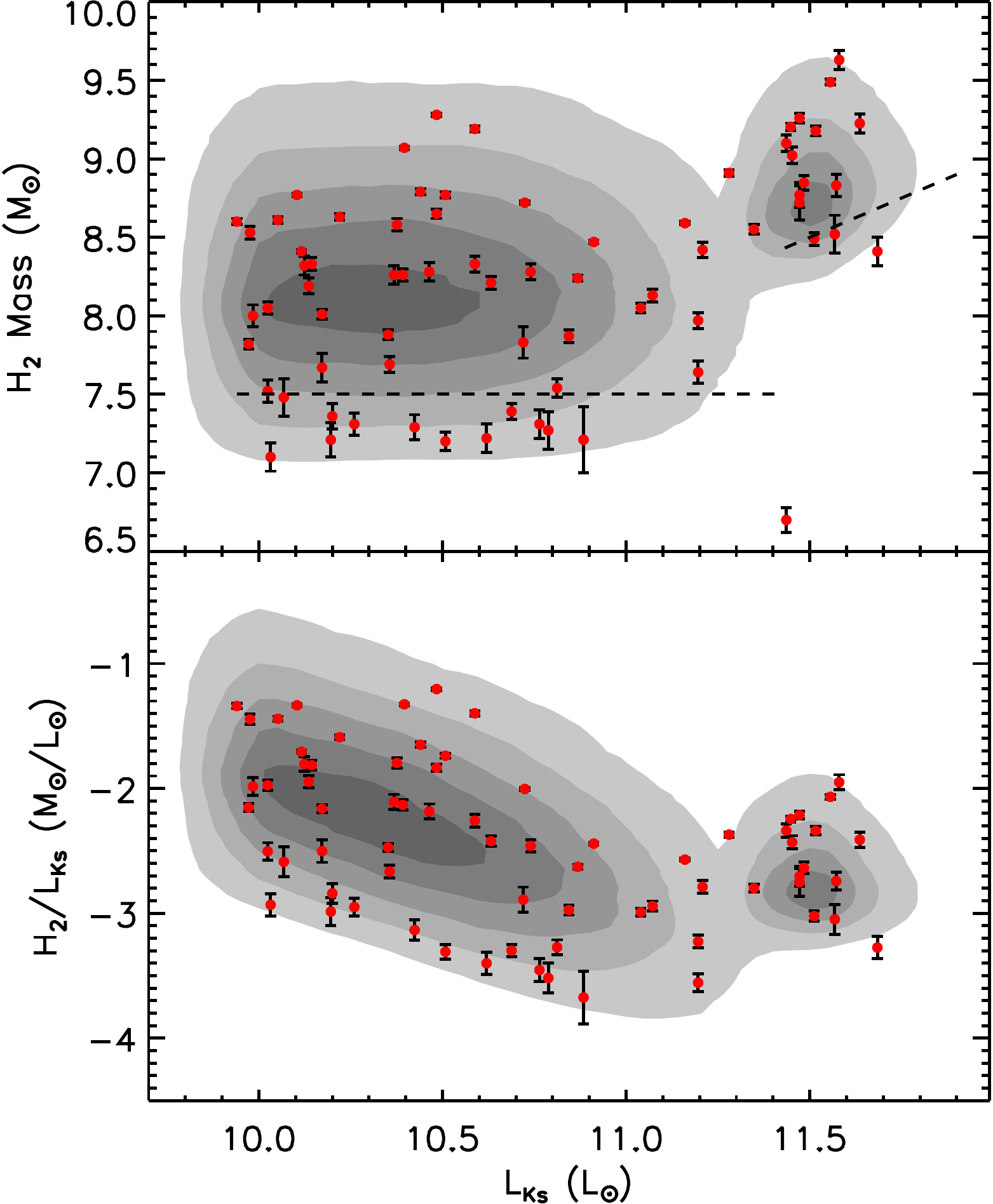}
\caption{Molecular gas mass (top panel) and molecular gas mass to stellar luminosity ratio (bottom panel) plotted against $K_s$-band luminosity of the detected MASSIVE and \atlas\ CO survey galaxies. The approximate detection limits of each survey are shown as dashed black lines in the top panel. Shown as grey contours, which well match the observed distribution, is a Monte-Carlo realisation of the distribution expected if the molecular gas mass is independent of stellar mass.}
\label{gasmass_frac}
 \end{center}
 \end{figure}

The combination of two different volume limited samples, which have observations that reach different sensitivity levels complicates our analysis. In the top panel of Figure \ref{gasmass_frac} we show the H$_2$ masses of {detected} \atlas and MASSIVE galaxies, plotted against $K_s$-band luminosity, while in the bottom panel we show the molecular gas fractions (here defined as $M_{\rm H_2}/L_{\rm Ks}$). The two surveys cover disjoint regions of the parameter space, and it seems at first glance as if the more massive objects have much higher H$_2$ masses (and thus gas fractions that do not follow the induced trend with stellar mass). 
This effect is, however, entirely caused by the different sensitivity limits of the two surveys, as indicated by a dashed lines in the top panel of Figure \ref{gasmass_frac}. {We note that in each case a small fraction of objects had deeper observations available, which we include for completeness. Removing these galaxies would not change any of the results of this paper.}

In order to determine if there is any true deviation away from our null hypothesis we create a {Monte-Carlo} simulation. This simulation incorporates these detection limits in order to produce a sample of model data points which correspond to our null hypothesis (no dependance of the gas mass on stellar mass). In order to do so we start by drawing galaxy masses randomly from the observed mass distributions of the MASSIVE and \atlas\ surveys. We then assign a random amount of gas to be present in each object, by drawing from the H$_2$ mass function of \cite{2011MNRAS.414..940Y}, that was
fitted with a Gaussian at the high mass end, and was assumed constant below the completeness limit down to an H$_2$ mass of $10^6$~\msun, as in \cite{2016MNRAS.457..272D}. Removing the low-mass
cutoff does not affect our results. In the subsample representing \atlas\ we assign each object a random distance between 10 and 42 Mpc, and consider the object as detected if the CO flux from that gas reservoir would be detectable at that distance. We note that the true distribution of distances is not uniform (e.g. $\approx$25\% of \atlas\ galaxies are in the Virgo cluster), but this is a second order effect. 
For the subset of galaxies representing MASSIVE we impose a detection limit of 0.1\% of the stellar mass (with a small scatter of 0.15 dex to match the observations, some of which are slightly deeper due to weather variations, etc). 

In this way we are able to generate a model of the distribution of galaxies within our parameter space (shown with grey shading representing galaxy density on Figure \ref{gasmass_frac}), which we can compare to the observed distribution to determine if we can reject the null hypothesis of no correlation between the variables. Both Kolmogorov-Smirnov and Mann--Whitney U tests are unable to reject the possibility that the observed and model distributions are drawn from the same parent distribution, with no evidence for variation greater than $\approx1\sigma$. This suggests, once again, that the amount of H$_2$ present in ETGs is not correlated with the stellar mass of the host galaxy, and thus that stellar mass loss is not the dominant source of this material.

Given the lack of correlation between gas mass and stellar mass, it is clear that we would expect to find a weak correlation between the gas-to-stellar mass \textit{fraction} and stellar mass, as seen in the bottom panel of Figure \ref{gasmass_frac}. This correlation arises simply because the typical amount of gas in each object stays the same, but the mass of the galaxy changes \citep[see e.g.][]{2011MNRAS.414..940Y,2015MNRAS.451.3815A}. We repeated the test described above for the gas-to-stellar mass fractions, and again show the results as grey contours. Again Kolmogorov-Smirnov and Mann--Whitney U tests are unable to reject the possibility that the observed and model distributions are drawn from the same parent distribution, suggesting no additional correlation between these variables is detectable, beyond that induced by the correlated axes.

 \subsubsection{Specific angular momentum}
  
 We showed in Section \ref{lr_matters} above that the specific angular momentum of galaxies correlates with the detection rate of cold gas. Slow rotating ETGs seem to be detected far less often than their fast-rotating counterparts. These detected slow rotating systems have mean gas masses of (6.5$\pm1.6)\times10^8$\,\msun, while detected fast-rotating galaxies have an average molecular gas mass of (4.5$\pm1.2)\times10^8$\,\msun. Slow rotators are systematically more massive than fast rotators, so despite this higher average gas content, the median gas fraction of detected slow rotators is actually lower than fast rotators (0.23$\pm$0.08\% vs 0.3$\pm$0.1\%). However, given the significant scatter seen, our low number statistics, and the different selection functions of our two surveys these differences are not significant at more than $\approx1\sigma$. We thus conclude that there is no strong evidence of the specific angular momentum itself correlating with the amount of cold gas present in ETGs. 

\subsubsection{Environment}

 \begin{figure} 
\begin{center}
\includegraphics[width=0.48\textwidth,angle=0,clip,trim=0cm 0cm 0cm 0.0cm]{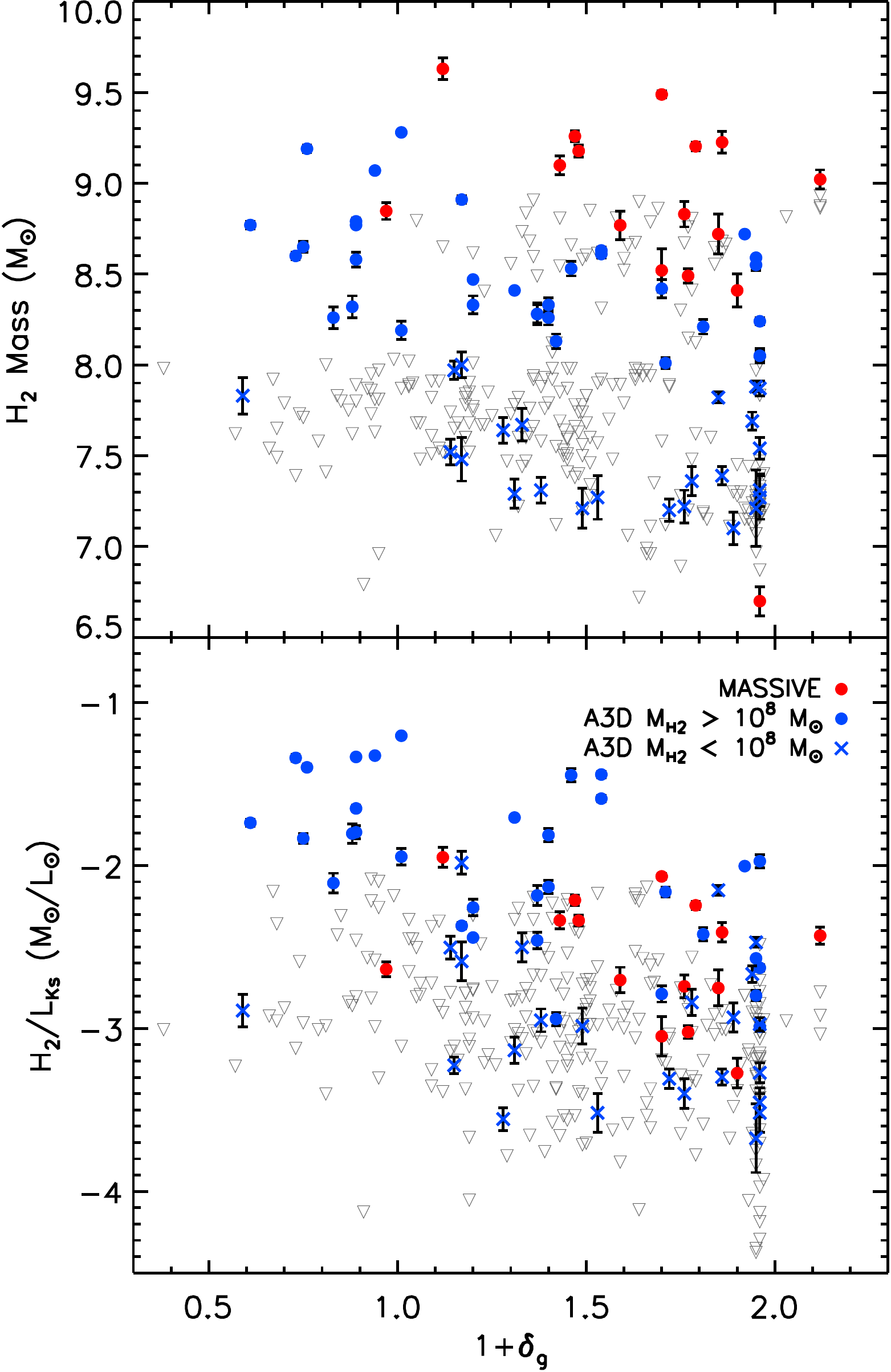}
\caption{Molecular gas mass (top panel) and molecular gas mass to stellar luminosity ratio (bottom panel) plotted against the large scale environment of the MASSIVE and \atlas\ CO survey galaxies. MASSIVE detections are shown as red points, and \atlas\ detections with gas masses above and below $10^8$ \msun with blue points and crosses, respectively. Non-detections are shown as upper-limits with an open grey triangle. The molecular gas mass to stellar luminosity ratio does appear to correlate with large scale environment. }
\label{fig:gasmass_env_frac}
 \end{center}
 \end{figure}

As discussed above, environment is another potential driver of the molecular gas content of galaxies.
In Figure \ref{fig:gasmass_env_frac} we show the global overdensity indicator $1+\delta_g$ (from \citealt{2017MNRAS.471.1428V}) for our objects plotted against their H$_2$ masses (top panel), and H$_2$ mass fractions (bottom panel). The MASSIVE detections are shown as red points,  \atlas\ detections as blue symbols, with the non-detections shown as upper-limits with an open grey triangle. 

Both of these panels appear to show strong trends, with the most isolated objects (with low values of $1+\delta_g$) having larger H$_2$ masses, and thus higher molecular-to-stellar mass fractions.
We stress, however, that one does need to exercise caution in drawing strong conclusions from this. The galaxies in dense regions of the \atlas\ volume are typically closer than those in the void regions, due to the presence of the Virgo Cluster. This means that the \atlas\ observations were more sensitive in this region. If one were to remove all \atlas\ detections with H$_2$ masses less than $10^8$ \msun (indicated with blue crosses in Figure \ref{fig:gasmass_env_frac}), to ensure that the sample is complete, the trend in the top panel appears less significant.
{We tested this using a generalised Kendall's $\tau$ correlation test (implemented using the \textsc{bhkmethod} task in \textsc{iraf}), which properly accounts for the significant number of upper-limits in our study. This test was unable to reject the null-hypothesis, and thus suggests that the correlation between environment and the $H_2$ mass is not statistically significant ($\tau$=-0.062, probability=0.16).}

In the bottom plot, showing molecular-to-stellar mass fractions, however, the trend that exists is more robust. 
{A generalised Kendall's $\tau$ correlation test that properly takes into account upper-limits (as discussed above) finds a correlation strength of $\tau$=-0.11, with a probability of 0.01, suggesting the result is significant at $\approx$3$\sigma$.} This still does not mean that there is a true correlation between environment and molecular-to-stellar mass fraction, however, as the most massive ETGs ($>10^{11}$ \msun) are preferentially found in dense environments (see e.g. Fig \ref{fig:nu_delta_env}).  We are able to test if this affects our conclusions by only including ETGs with $<10^{11}$ \msun, as in our sample there is no strong correlation between mass and environment at these lower masses. When restricting ourselves to this lower mass range we find no sign of a weakening of the correlation {(in fact it becomes stronger; $\tau$=-0.1422, P=0.008)}, suggesting there is some true correlation present between the molecular-to-stellar mass fractions and environment. This correlation is not as apparent when comparing with the local over-density indicator $\nu_{10}/<\nu_{10}>$, suggesting whatever is driving this correlation acts over large scales. 

\begin{figure} 
\begin{center}
\includegraphics[width=0.5\textwidth,angle=0,clip,trim=0cm 0cm 0cm 0.0cm]{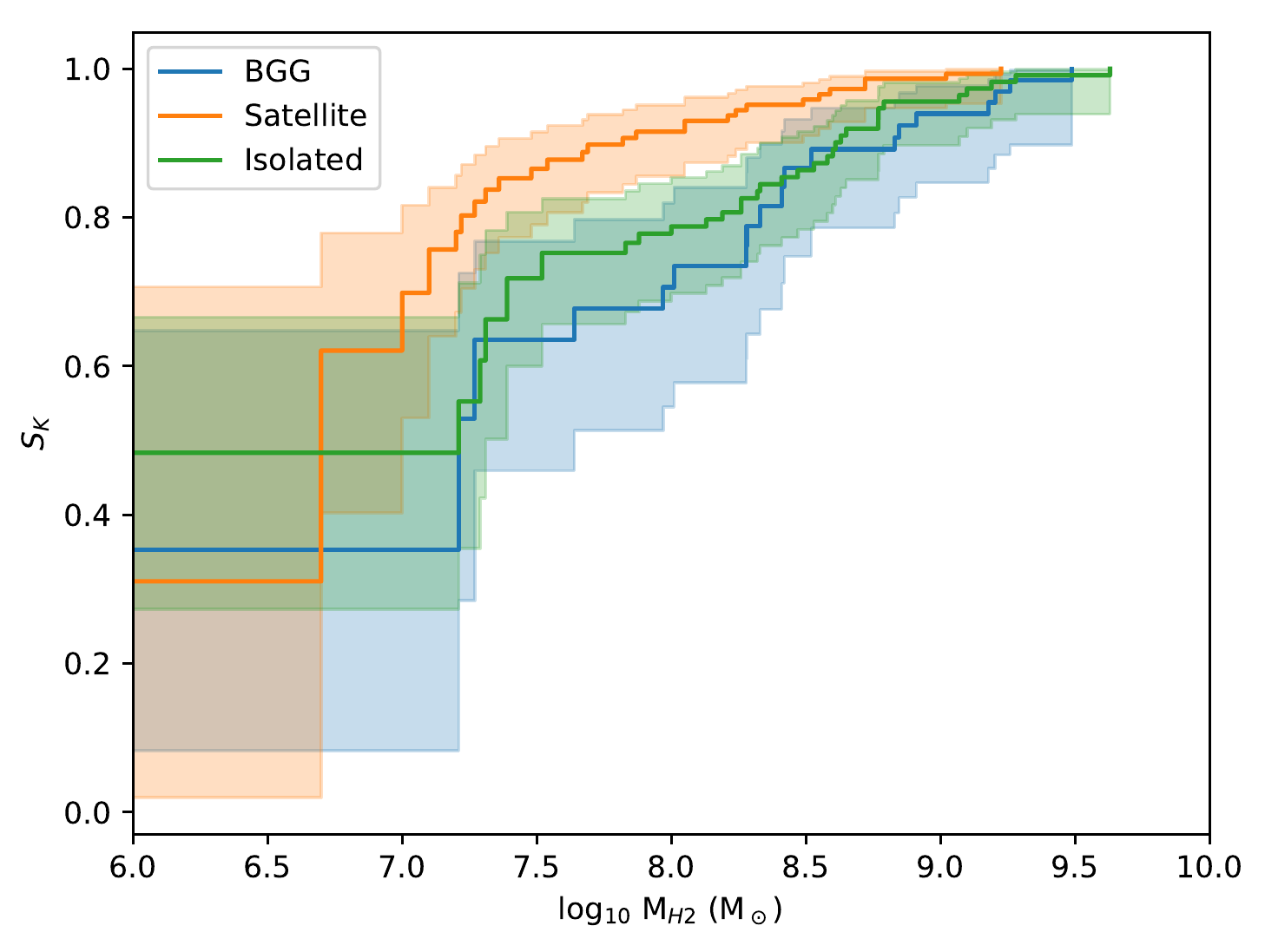}
\includegraphics[width=0.5\textwidth,angle=0,clip,trim=0cm 0cm 0cm 0.0cm]{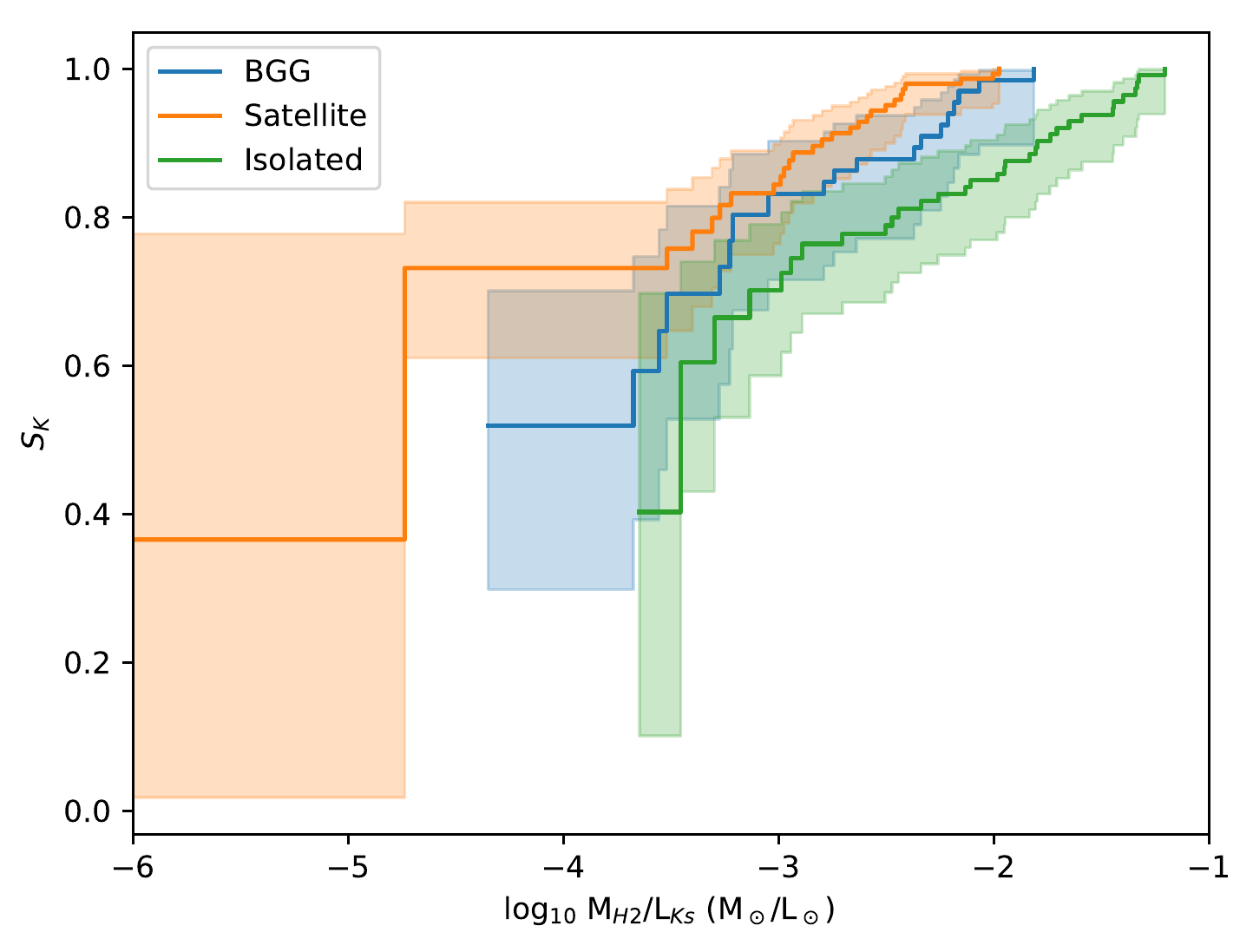}
\caption{{The Kaplan--Meier estimators for the distribution functions ($S_K$) of the cumulative molecular gas mass (top panel), and molecular gas mass to stellar luminosity ratio (bottom panel). Each plot shows estimates for our combined sample galaxies split by environment, with isolated galaxies shown in green,  satellite galaxies in orange and brightest group galaxies in blue. These estimators include the effects of censored data (upper-limits) in constructing the cumulative distribution function. Satellite galaxies have lower average molecular gas masses than BGG/isolated systems, and their gas fraction is also lower.}}
\label{fig:gasfrac_env}
 \end{center}
 \end{figure}

{A clue to the likely processes driving this correlation can be found by splitting our galaxies by their positions within their dark matter halos. Figure \ref{fig:gasfrac_env} shows the Kaplan--Meier estimators for the gas properties of the combined sample galaxies split by environment, with isolated galaxies shown in green, satellite galaxies in orange and brightest group galaxies in blue (as classified by \citealt{2017MNRAS.471.1428V}). The estimated distribution functions ($S_K$) of the cumulative molecular gas masses are shown in the top panel, and for the molecular gas mass to stellar luminosity ratio in the bottom panel. These estimators include the effects of upper-limits (censored data).} 

{In the top panel BGG and isolated systems are found to have similar molecular gas distribution functions, and within the errors cannot be shown to be significantly different. Satellite galaxies, however, have lower average molecular gas masses than BGG/isolated galaxies. The bottom panel of Figure \ref{fig:gasfrac_env} shows that isolated galaxies have the highest molecular gas fractions on average (as they are typically lower stellar mass systems), while BGG and satellites have fractionally less gas. The molecular gas to stellar mass fractions of isolated objects are $\approx$0.6 dex higher on average than satellite and BGG's. This suggests that group/cluster environments do affect the amount of gas present in ETGs, if not the total detection rates.}

 \subsubsection{Kinematic bulge fraction}
 
Given the strong dependance of the gas detection rate on the kinematic bulge fraction at fixed stellar mass (which we parameterised with $\eta_{\rm kin}\equiv\frac{\sqrt[3]{M}}{\sigma_e}$), one might expect to find a correlation between the molecular gas mass (or molecular gas-to-stellar mass fraction) and this parameter. Indeed \cite{2013MNRAS.432.1862C} report a correlation between molecular-to-stellar gas mass fraction and the position of ETGs on the mass-velocity dispersion plane, after adaptive smoothing of the gas mass fractions of the lower mass \atlas\ ETGs. 
{Using our combined sample a generalised Kendall's $\tau$ correlation test (which includes upper-limits) confirms that both molecular gas mass and molecular-to-stellar gas mass fraction correlate with $\eta_{\rm kin}$ at high significance (correlation strength $\tau$=0.21, P$< 0.001$).}

{It seems that the ability to host a cold gas reservoir correlates with $\eta_{\rm kin}$, and as this kinematic bulge fraction changes so does the mean amount of gas present.} Despite this, the scatter in H$_2$ masses (and mass fractions) at fixed $\eta_{\rm kin}$ is very large, so this parameter does not give us the ability to predict with confidence the amount of gas present in any individual object. This is discussed further in Section \ref{discuss}. 

\section{Analytic toy model}
 \label{model}
 
 In this section we attempt to synthesise the knowledge gained from the correlations discussed above via a simple analytic toy model. This model is based on that presented in \cite{2016MNRAS.457..272D}, but extended to include the effect of environment, transformation of spiral galaxies, and stellar mass loss. 
  The aim of this exercise is to determine in a simplistic way the relative importance of the various processes we suspect of being involved in supplying molecular gas to ETGs, and is not meant to reproduce every property of these systems. For instance, we completely ignore atomic gas in this analysis, where clearly it may be important, especially in field environments (e.g. \citealt{2012MNRAS.422.1835S}). Placing such models within the correct cosmological context is better done within semi-analytic or full-hydrodynamic simulations, and is beyond the scope of this work. 

\subsection{Model details}
 
 Our aim is to create a model that can match the overall molecular gas detection rate in ETGs, the molecular gas detection rates in different environments, as well as the mass and mass-fraction functions taken from this work. In addition, we aim to reproduce the gas kinematic misalignment statistics as a function of environment from \cite{2011MNRAS.417..882D}. {Our model includes a range of physical processes, and each is parameterised using simple analytic expressions.}
 We summarise below the salient details of the model, and highlight how our implementation differs from \cite{2016MNRAS.457..272D}.

The model begins at a lookback time of 7.7 Gyr ($z$=1), and follows the evolution of an arbitrary number of galaxies.
These systems have masses drawn uniformly from the observed mass distribution of the combined \atlas\ and MASSIVE sample. 
From the assigned stellar masses we calculate the maximum
circular velocity at which the gas can rotate by assuming the galaxies
follow the broken $K$-band CO Tully-Fisher relation of
\cite{2016MNRAS.455..214D}, and determine the radial extent of the gas, and hence the dynamical
time at its outer edge, by drawing uniformly from a log-normal fit to
the observed distribution of molecular disc radii in \cite{2013MNRAS.429..534D}. 
Molecular gas reaches beyond the turnover
of the rotation curve in $\approx70\%$ of ATLAS$^{\rm 3D}$ ETGs
\citep{2011MNRAS.414..968D,2013MNRAS.429..534D}, and all of the interferometrically mapped MASSIVE ETGs (Davis et al., in prep) so this should
provide a good estimate of the rotation velocity of the gas at the
edges of the gas discs for the bulk of the population. {We note that using the un-broken Tully-Fisher relation from \cite{2011MNRAS.414..968D} would not change our conclusions. }

Initially all of these systems are classed as isolated, and 27\% of them are designated as ETGs (as found for ETGs of all masses at $z$=1 in \citealt{2014A&A...570A.102T}). Although this is somewhat unrealistic, as the red-sequence in clusters is already in place by these redshifts \citep[e.g.][]{2007ApJ...665..265F} the majority of the satellite galaxies in this study reside in groups, which are still assembling. Our results do not strongly depend on the redshift range we choose to probe, and extending the model to start at $z=2$--3 (to better match the mean light weighted ages of our MASSIVE ETGs; \citealt{2015arXiv150402483G}) would not change our results. Initially all the ETGs are gas free.
 
The evolution of these systems is governed by a linked set of equations {(described in more detail below)}, which allow us to roughly approximate the action of gas rich mergers, star formation, stellar mass loss, environmental quenching and strangulation. These allow us to track the gas content, misalignment of the gas reservoir from the stellar kinematic axis and morphology of each system as a function of time. 

In each time-step we randomly select galaxies that will accrete gas, such that the average number of gas-rich mergers per galaxy per unit time is equal to $R_{\rm acc}$ (which we here treat as a free parameter). In this simple model, $R_{\rm acc}$ is independent of redshift, galaxy mass and morphology, which may not be true in reality. 

Each isolated galaxy has a fixed probability of entering a dense environment during any given time-step, which is constrained such that the correct fraction of satellite and BGGs (based on \citealt{2017MNRAS.471.1428V}) is reached at $z$\,$=$\,$0$. Once an object enters a dense environment mergers are prevented entirely.

For those systems that do experience mergers, the amount of gas accreted is chosen randomly from the H$_2$ mass function presented in \cite{2016MNRAS.457..272D}. As discussed in that work we assume that after coalescence each one of our mergers results in the creation of a disc of H$_2$ in the centre of the remnant, with a mass that is drawn from the observed distribution, and that this final state is obtained instantaneously. 
The angle at which gas is accreted onto galaxies is also chosen
uniformly, sampling the full range $0$-$180^{\circ}$. If gas already
exists in a galaxy, the interaction between the two reservoirs is
roughly included by mass-weighting the resulting position angle of the
gas (thus lying between the newly accreted and old material). 

The relaxation of the gas is tracked at each time-step. The gas relaxes
at a rate that has a cosine dependance on the current misalignment
angle (i.e.\ the rate increases as the disk approaches the galaxy
plane, where the torque is higher; e.g. \citealt{1982ApJ...252...92T}).

\begin{equation}
\Delta\Psi = 60^{\circ}\frac{|\cos(\Psi)|}{\cos{60^{\circ}}}\frac{\Delta T}{t_{\rm relax}} ,
\end{equation}
{where $\Delta\Psi$ is the change in angle in any given time-step, $\Psi$ is the current misalignment angle, $\Delta T$ is the duration of each time-step, and $t_{\rm relax}$ is a free parameter. }
The total time taken to relax is calculated such that the gas disc relaxes into the plane from a
misalignment of $60\degr$ in a time $t_{\rm relax}$, which we here fix as 25$t_{\rm dyn}$, following the results of \cite{2015MNRAS.451.3269V} and \cite{2016MNRAS.457..272D}.

{Star formation in our model is parameterised as follows:}
\begin{equation}
\Delta {M}_{H_2} = \epsilon_{sf}\frac{\mathrm{{M}_{H_2}}}{t_{\rm dyn}} \Delta T,
\end{equation}
{where M$_{\rm H_2}$ is the current gas mass, $\Delta {M}_{H_2}$ the change in the gas mass at any given time-step, $t_{\rm dyn}$ is the dynamical time, and $\Delta T$ is the duration of the time-step.
$\epsilon_{sf}$ defines the efficiency of star formation per dynamical time,} which we set to 10\%, equivalent to the standard star formation law derived from local star-forming galaxies \citep{1998ApJ...498..541K}. 
We note that various authors have suggested that ETGs actually have low star formation efficiencies \citep[e.g.][]{2011MNRAS.415...61S,2013MNRAS.432.1914M,2014MNRAS.444.3427D}. This was explored in detail in \cite{2016MNRAS.457..272D}, but for simplicity we do not explore this effect here.

{We include AGN feedback in our model in a very simplistic way. Each system has a fixed probability of experiencing an AGN outburst, set such that the average galaxy will experience such an episode once every $t_{\rm AGN}$ years. When this happens, we simply remove 99\% of the gas in the galaxy instantaneously. This treatment cannot hope to capture the true complexity of AGN feedback processes, but at least roughly may help us estimate the duty cycle of AGN driven quenching events.} 

In this work we have added two additional pathways by which ETGs can increase their gas reservoir. The first of these is stellar mass loss. As discussed above, the old low-mass AGB stars, which dominate the evolved stellar populations in ETGs, return $\approx10^{-11}$\,\msun\,yr$^{-1}$ (per solar mass of stars in the galaxy) of gas back into the ISM. This material likely shocks and enters the hot halo of the galaxy initially, but being metal loaded and in the dense region of the halo close to disk, it would be expected to cool back onto the galaxy. We here include this in our model by allowing a fraction of this material (f$_{\rm ML,iso}$ - which we here treat as a free parameter) to cool back during each time-step.

\begin{equation}
\Delta {M}_{H_2} = f_{\rm ML,iso} \frac{\mathrm{{M}_{*}}}{10^{11}} \Delta T,
\end{equation}
{where $\mathrm{M}_{*}$ is the stellar mass of the galaxy at that time, and the other parameters are as defined above.}

 In satellite galaxies, however, this cooling is likely to be suppressed as the hot halo of the system is ram-pressure stripped. Once an object becomes a satellite we reduce the fraction of the mass loss which cools to zero over a timescale ($t_{\rm strip}$), which here we fix to 1 Gyr following the results of \cite{2016A&A...591A..51S}. The results we derive are not strongly dependant on the exact value we assume for this parameter. In contrast, for BGGs we allow cooling to continue with a higher efficiency (f$_{\rm ML,BGG}$, which we again leave as a free parameter) due to the higher likelihood of cooling-flows in massive groups/clusters. 
This material is always assumed to enter the galaxy in a direction which is aligned with the stellar rotation, and if any misaligned gas exists in the system it interacts and torques it towards co-rotation (again via mass-weighting the two resulting position angles). 

Due to the shutdown of mergers, and the reduction of cooling mass loss in satellite galaxies due to halo stripping, a second mechanism of producing gas rich satellite ETGs is clearly needed. We know from the morphology-density relation \citep{1980ApJ...236..351D} that environmental effects can morphologically transform and quench galaxies. We include this here by giving each spiral galaxy in our model a chance to transition into a ETG when it is in a cluster environment. {Note that we do this in a very simple way, that is agnostic to the physical mechanism that is actually causing the quenching}. This probability is set such that we reproduce the relative fraction of spiral and ETGs in dense environments at $z=0$ from \cite{2011MNRAS.413..813C}.  The gas fraction of these spirals as they transition is treated as a free parameter, and parameterised as M$_{\rm H_2}$/M$_{\rm spiral}$. This gas is assumed always to be co-rotating within its host galaxy. 

At the end of the simulation, we are thus left with the molecular gas mass and kinematic position angle of each galaxy. We convolve the
kinematic misalignment angles with a Gaussian kernel to match the observational errors, and apply the observational limits of the \atlas\ and MASSIVE molecular gas surveys to define objects that are detectable.

We can then compare the output to the observational constraints at $z=0$. The constraints we use are the overall molecular gas detection rate, the molecular gas detection rates in different environments, the mass and mass-fraction functions (all taken from this work), and the gas kinematic misalignment statistics as a function of environment from \cite{2011MNRAS.417..882D}. We combine these constraints and generate a log-likelihood for each model.  We were then able to run a Markov-Chain Monte Carlo (MCMC) fitting process (using the Gibb's sampler with adaptive stepping from \citealt{2017MNRAS.468.4675D}) in order to find the best fit values for the free parameters (the gas rich merger rate, the fraction of stellar mass loss that cools in BGG and isolated galaxies, and the initial gas fraction of morphologically transformed spiral galaxies), and their uncertainties. 

\begin{figure*} 
\begin{center}
\includegraphics[width=0.9\textwidth,angle=0,clip,trim=0cm 0cm 0cm 0.0cm]{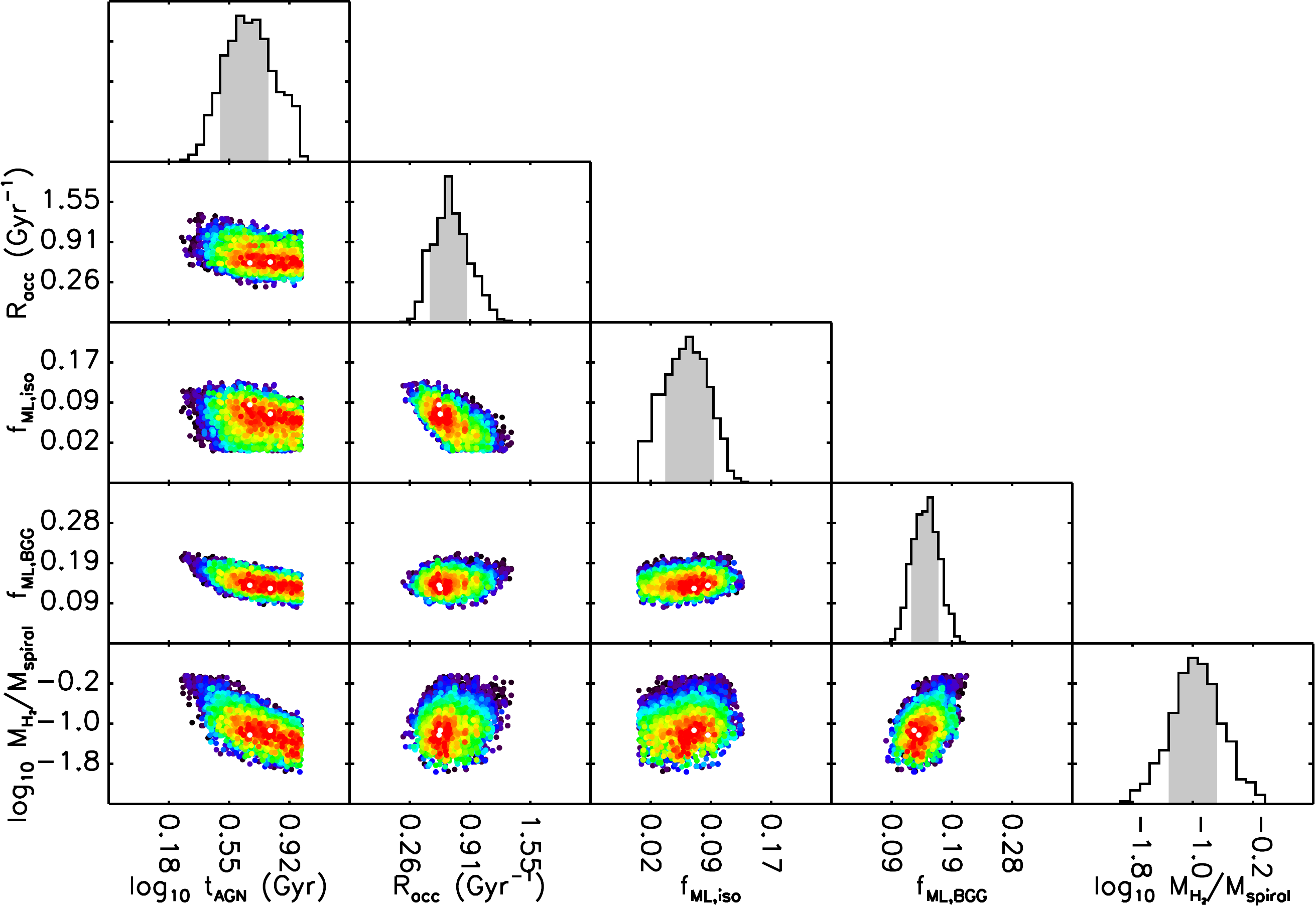}
\caption{Visualisation of the multidimensional parameter space explored by our toy model, fitting to the observational data for the gas in ETGs as described in Section \ref{model}. {The fitted parameters are the mean time between AGN outbursts ($t_{\rm AGN}$), the rate of gas rich mergers ($R_{\rm acc}$), the fraction of stellar mass loss material that can cool in isolated systems/brightest group galaxies (f$_{\rm ML,iso}$ and f$_{\rm ML,BGG}$ respectively), and the gas fraction of spiral galaxies that transform into ETGs in cluster environments (M$_{\rm H_2}$/M$_{\rm spiral}$).}
 In the top panel of each column a one-dimensional histogram shows the marginalised posterior distribution of that given parameter, with the 68\% (1$\sigma$) confidence interval shaded in pale grey. In the panels below, the points show the two-dimensional marginalisations of those fitted parameters, colour-coded to show the relative log-likelihood of that realisation, with white points the most likely.  Mergers appear to dominate the supply of gas to ETGs in this simple model.}
\label{model_results}
 \end{center}
 \end{figure*}

\subsection{Model results}
The results of the modelling process describe above are shown in Figure \ref{model_results}. While such a simple model is never going to be able to capture all the complexity of galaxy evolution, we do obtain a reasonable match to the observational quantities, with the best model having a reduced $\chi^2$ of 1.03.

One of the key parameters of our model is the rate of gas rich mergers/accretion events. As Figure \ref{model_results} shows our constraint on this parameter is degenerate with the fraction of stellar mass loss that cools in isolated galaxies (and somewhat with the typical AGN duty cycle) but only accretion rates in the range $\approx$0.3-1.2 Gyr$^{-1}$ are allowed by the model. The 1$\sigma$ estimate of the gas rich merger rate is 0.69$_{-0.18}^{+0.14}$ Gyr$^{-1}$, which is in reasonable agreement with other studies. For instance, \cite{2011ApJ...742..103L} presented a range of estimates from different techniques, and estimated a total merger rate of $0.2$ to $0.5$~Gyr$^{-1}$ with $75\%$ of these mergers being minor (mass ratios of 1:4 to 1:10). Theoretical studies such as that of \citet{2010ApJ...715..202H}
find a major+minor merger rate of $\approx0.2$~Gyr$^{-1}$ for objects
of $10^{11}$~\msun, but predict that this merger rate scales
positively with galaxy mass. The more massive galaxies in our sample
(with stellar masses $\approx10^{12}$~\msun) are predicted to have
merger rates of up to $0.9$~Gyr$^{-1}$. However, not all of these mergers are likely to be gas rich, especially at low $z$.

As mentioned above, the fraction of expected stellar mass loss material that cools is somewhat degenerate in our model with the merger rate. Our best estimate is 6.2$\pm$2\% of the total old-star stellar mass cools to the molecular phase, equivalent to a return rate per solar mass of stars of $\approx(6\pm2)\times10^{-13}$ yr$^{-1}$. If gas rich mergers/accretion events are rare {(or AGN outbursts happen frequently)} then this rate could be as high as $\approx2\times10^{-12}$ yr$^{-1}$, which is still an order of magnitude below the known rate of mass loss from AGB stars. 
This suggests the majority of this material is unable to cool (perhaps due to AGN maintenance mode feedback which is not modelled here; e.g. \citealt{2006MNRAS.370..645B,2006MNRAS.365...11C}), and the hot halos of these systems are building up over time. We note that the material that cools here doesn't have to be stellar mass loss, as this formalism would apply equally to any cooling from a kinematically aligned hot halo. If the hot halo were kinematically misaligned from the galaxy in some cases, as suggested in \cite{2015MNRAS.448.1271L}, then this fraction could increase. 
In brightest cluster/group galaxies, the model suggests a return of $\approx$14$\pm$2\% of the total old-star stellar mass material, or $\approx1.4\times10^{-12}$ yr$^{-1}$. This suggests even where cooling flows may be present, the majority of the stellar mass loss material must stay in the halo.

In our model the spiral galaxies that transition into ETGs in clusters ({due to any quenching mechanism}) have a mean gas fraction of M$_{\rm H_2}$/M$_{\rm spiral}\approx$10\% at the moment they enter the ETG population. This is consistent with the gas fraction of typical sprial galaxies (e.g. \citealt{1991ARA&A..29..581Y}), suggesting only a minimal reduction in the gas fraction of these galaxies is required prior to morphological transformation.

{Another parameter in the model is the typical duty cycle for galaxy quenching via an AGN. This is fairly poorly constrained by the model itself, and Figure \ref{model_results} shows that clear covariances exist between this timescale and the other parameters. If AGN driven quenching events are common (occurring $<$2 Gyrs apart on average), then the gas rich merger rate can be higher, a larger fraction of the stellar mass loss of isolated and BGG galaxies can cool, and spiral galaxies can transition with sizeable gas reservoirs, all without violating our observational constraints. If however the timescale is longer (similar to or greater than our best estimate of $\approx4.7^{+1.1}_{-1.35}$ Gyr) then the constraints on our other parameters become tighter. We note that the timescale we derive is very much longer than the typical AGN duty cycle ($\approx10^6$-$10^8$~yr: e.g. \citealt{2001ApJ...549..832S};\\ \citealt{2002ApJ...567L..37M}; \citealt{2006ApJS..163....1H,2008MNRAS.388..625S}). This suggests that only a small fraction of AGN outbursts go on to significantly affect the gas reservoir of their host galaxies.}

\section{Discussion \& Conclusions}
\label{discuss}
\label{conclude}

In this paper we studied a representative sample of 67 of the most massive ETGs in the universe, drawn uniformly from the MASSIVE survey. We presented new observations of the molecular gas content of 30 of these galaxies, which combined with measurements from the literature allows us to probe the molecular gas content of the entire sample down to a fixed molecular-to-stellar mass fraction of 0.1\%. Two of these objects observed in this work were detected, hosting between 5 and 7$\times$10$^{8}$\,\msun\ of molecular gas.  

The total detection rate of cold molecular gas in the MASSIVE representative sample we study here is 25$^{+5.9}_{-4.4}$\%.
This rate is very similar to the detection rate of of 22$\pm3$\% found by \cite{2011MNRAS.414..940Y} for lower mass \atlas\ ETGs.  By combining the MASSIVE and \atlas\ molecular gas surveys we find a joint detection rate of 22.4$^{+2.4}_{-2.1}$\%. This detection rate seems to be independent of mass, radius, position of the galaxy on the fundamental plane, and (both local and global) environment. The detection rate of cold gas does increase in galaxies with blue ultraviolet-to-infrared colours, reflecting the presence of underlying residual star-formation. In the MASSIVE galaxies alone, however, little effect is seen, showing how insensitive even ultraviolet colours are to the presence of gas in the most massive galaxies. As such, red sequence selections should be treated with caution, as they do not guarantee a purely passive galaxy sample.
 
 The position of a galaxy in the mass-velocity plane does seem to affect the molecular gas detection rate. Objects with a higher velocity dispersion at fixed mass (a higher kinematic bulge fraction at fixed stellar mass) are less likely to be detected in molecular gas. We showed that the parameter $\eta_{\rm kin}\equiv\frac{\sqrt[3]{M}}{\sigma_e}$ correlates strongly with the molecular gas detection rate, at least in fast rotating ETGs. Fast-rotating objects with high $\eta_{\rm kin}$ have a detection rate of $\approx$60\%, while those with low $\eta_{\rm kin}$ are detected $<10$\% of the time. Slow rotators do not seem to follow the same trend, suggesting the kinematic bulge fraction is less important in these objects. Assuming it holds, this correlation between $\eta_{\rm kin}$ and molecular gas could act as a predictor for future surveys which hope to find molecular gas in fast-rotating ETGs (e.g. at higher redshift; \citealt{2018ApJ...860..103S}). 
 
 This strong correlation between kinematic bulge fraction (at fixed mass) and the ability to host cold gas must encode information about the mechanisms (re-)fuelling massive galaxies. For instance, AGN feedback is invoked in galaxy models in order to stop cooling from the halo. Galaxies with higher kinematic bulge fraction at fixed stellar mass also likely have a higher SMBH mass (given the strong black hole mass -- velocity dispersion correlation;\\ e.g. \citealt{2013ApJ...764..184M}; \citealt{2016ApJ...831..134V}). This past black hole growth will have injected significant energy/heat into the surroundings, which in turn could have acted to suppress halo cooling \citep[see e.g.][]{2016ApJ...832L..11M,2017ApJ...844..170T}. The growth of a bulge through the action of bars or dry mergers, for instance, may also have helped evacuate or use up any gas remaining in these systems. Future works will be required to separate out, and determine the balance between these varying competing mechanisms.
    
 The other parameter that appears to play a role in the ability of ETGs to host gas is the stellar specific angular momentum. Fast rotating ETGs appear be be detected at approximately the same rate across the $\lambda_{\rm R_e}/\sqrt{\epsilon}$ vs L$_{\rm K_s}$ plane. While we show here (for the first time) that true slow rotators can host molecular gas reservoirs, the rate at which they do so is significantly lower than for fast rotators ($\ltsimeq8$\% in SR vs $\approx$22\% in FR).
 
The amount of molecular gas in ETGs does not appear to correlate with galaxy mass, which induces a weak correlation between gas-to-stellar mass ratio and galaxy mass. Environment does, however, appear to make a larger difference, with galaxies in denser environments having $\approx$0.5 dex lower molecular gas-to-stellar mass ratios. Brightest group galaxies appear to be somewhat less affected, likely because they are able to access additional gas from cooling flows. The cause of this lower gas fraction in satellite galaxies, at least in our toy model, is strangulation \citep{1980ApJ...237..692L}, as the gas accretion rate is strongly reduced for satellite galaxies \citep[e.g.][]{2017MNRAS.466.3460V}. Ram-pressure stripping \citep{1972ApJ...176....1G} may also play a role in the low gas fractions of the observed systems. For instance, it has been shown that the gas reservoirs in ETGs are less extended in dense environments \citep{2013MNRAS.429..534D}. Despite this, it is expected that ram-pressure affects the molecular gas in massive galaxies to a lesser degree than strangulation, as the gas is tightly bound at the centre of a deep potential well.

 Three possible scenarios are often discussed to explain the presence of molecular gas in ETGs:
 \begin{enumerate}
\item  The gas could be a remnant of the star forming reservoir of the galaxy before it was morphologically transformed. 
\item The gas could have been regenerated from an internal source, such as stellar mass loss or cooling of the hot halo.
\item The gas could be acquired externally, from minor/major mergers and cold accretion. 
\end{enumerate}

The first of these possibilities is explored somewhat in our toy model, where in cluster environments we include the transformation of spirals into ETGs. In our model around a quarter of the low redshift ETG population in clusters are the remnants of transformed spiral galaxies, which have maintained a detectable gas reservoir since they were morphologically transformed. While these systems are present, they do not dominate the population of gas rich ETGs, and thus other mechanisms also seem to be important.
 
The constant detection rate of molecular gas as a function of stellar mass we find in this work once again suggests that the stellar mass loss is not a dominant source of ISM in ETGs.  As mentioned above the gas return rate should correlate linearly with stellar mass at old ages, leading more massive galaxies to be detected more frequently, which is at odds with our data. In addition, if mass loss were the dominant process then it is much harder to explain the \textit{lack} of gas in 75\% of ETGs {without another process (such as AGN feedback) conspiring to keep the detection rate of cold gas constant}. Various other studies have come to the same conclusion, but we can now extend this result to even the most massive systems.
Indeed, the toy model outlined above suggests $<15\%$ of AGB star mass loss can cool into the ISM of isolated/satellite ETGs, and $<20\%$ in BGGs, in order to not violate the observational constraints. In our models the gas rich merger rate is $\approx$0.6 Gyr$^{-1}$, and the typical accreted gas mass is $\approx$3$\times10^7$\,\msun\,Gyr$^{-1}$, and thus accreted material dominates over in-situ production of gas by a factor of at least 10,000.
 
The only other internal reservoir of gas that could cool to form a molecular reservoir is the galaxy's hot halo. {This is known to be an important mechanism for the central galaxies in galaxy clusters and groups, and ALMA observations have begun to resolve this process in a variety of systems in recent years \citep[e.g.][]{2016MNRAS.458.3134R,2017ApJ...848..101V,2018A&A...618A.126O,2018ApJ...865...13T}. However, the importance of halo cooling in isolated systems is less well understood}. The correlations we find between the molecular gas detection rate and both kinematic bulge fraction and stellar specific angular momentum raise the possibility that the hot halo is also important in these systems, as cooling rates are predicted to be enhanced in more disky (kinematically colder) galaxies (\citealt{2014MNRAS.445.1351N}; \citealt{2018arXiv180805761J}). If the hot-halo is a dominant source of cold gas then one requires a large number of  halos to be kinematically misaligned from their host galaxy in order to match the fraction of kinematically misaligned gas discs in ETGs \citep[e.g.][]{2015MNRAS.448.1271L}. A misaligned hot halo could potentially show increased cooling rates due to collisions between stellar mass loss material and the halo.
Given the above, one might expect to find a correlation between these parameters and the X-ray emission that arises from cooling hot-halos, but no such correlation was found in this work. This could be due to a real lack of correlation, to the low number of probed systems, or to the difficulty of probing the colder gas $\approx10^5$\,K gas in galaxy halos. This issue will be probed further in a future work of this series (Goulding et al. in prep).

In summary, in this work we show that molecular gas reservoirs are common across the ETG population. The kinematic bulge fraction and specific angular momentum of galaxies affect their ability to host such reservoirs, while the amount of gas present is also affected by environment. Mergers appear to dominate the supply of gas, but stellar mass loss, hot halo cooling and transformation of spiral galaxies may also play a role,
Further observational and theoretical work will be required to fully disentangle the importance of each of these effects. For instance theoretical works may be needed to predict the observational signatures of hot halo cooling {in isolated} ETGs. {Resolved observations (with e.g. ALMA) will reveal the angular momentum of the molecular gas, and its spatial distribution. These observations can also reveal if this molecular material is important in fuelling AGN, which are very common in massive ETGs, but are often thought to be fuelled by hot-mode accretion.} A high-resolution X-ray spectrometer (to replace \textit{Hitomi} after its unfortunate destruction) would also reveal the rotation of hot halos for the first time, and thus their importance in refuelling galaxies across cosmic time.

 \vspace{0.5cm}
\noindent \textbf{Acknowledgments}

TAD acknowledges support from a Science and Technology Facilities Council Ernest Rutherford Fellowship, and thanks the referee for helpful comments which improved this paper.

The MASSIVE survey is supported in part by NSF AST-1411945 and AST- 1411642. This paper is based on observations carried out with the IRAM Thirty Meter Telescope, and we thank Claudia Marka and Wonju Kim, the pool supervisors for observational assistance. IRAM is supported by INSU/CNRS (France), MPG (Germany) and IGN (Spain). This research also made use of the NASA/IPAC Extragalactic Database (NED) which is operated by the Jet Propulsion Laboratory, California Institute of Technology, under contract with the National Aeronautics and Space Administration.

The Pan-STARRS1 Surveys (PS1) and the PS1 public science archive have been made possible through contributions by the Institute for Astronomy, the University of Hawaii, the Pan-STARRS Project Office, the Max-Planck Society and its participating institutes, the Max Planck Institute for Astronomy, Heidelberg and the Max Planck Institute for Extraterrestrial Physics, Garching, The Johns Hopkins University, Durham University, the University of Edinburgh, the Queen's University Belfast, the Harvard-Smithsonian Center for Astrophysics, the Las Cumbres Observatory Global Telescope Network Incorporated, the National Central University of Taiwan, the Space Telescope Science Institute, the National Aeronautics and Space Administration under Grant No. NNX08AR22G issued through the Planetary Science Division of the NASA Science Mission Directorate, the National Science Foundation Grant No. AST-1238877, the University of Maryland, Eotvos Lorand University (ELTE), the Los Alamos National Laboratory, and the Gordon and Betty Moore Foundation.
 
\bsp
\bibliographystyle{mnras}
\bibliography{bibMASSIVE_rep.bib}
\bibdata{bibMASSIVE_rep.bib}
\bibstyle{mnras}

\label{lastpage}
\clearpage
\appendix

\begin{figure*} 
\section{Observed Spectra}
\begin{center}
\begin{subfigure}[b]{0.48\textwidth}
\includegraphics[width=0.45\textwidth,angle=0,clip,trim=0cm 0cm 0cm 0.0cm]{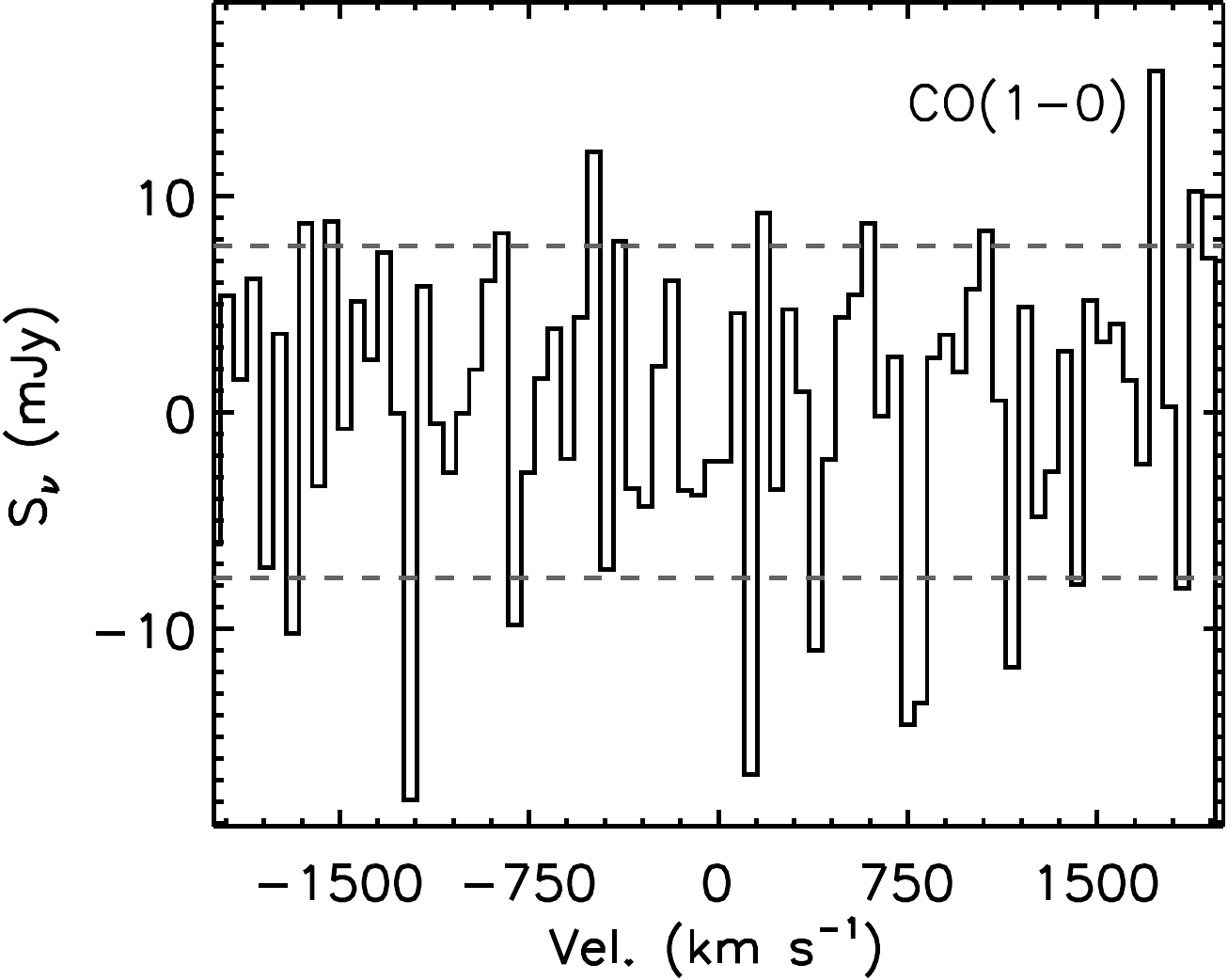}\hspace{0.25cm}
\includegraphics[width=0.45\textwidth,angle=0,clip,trim=0cm 0cm 0cm 0.0cm]{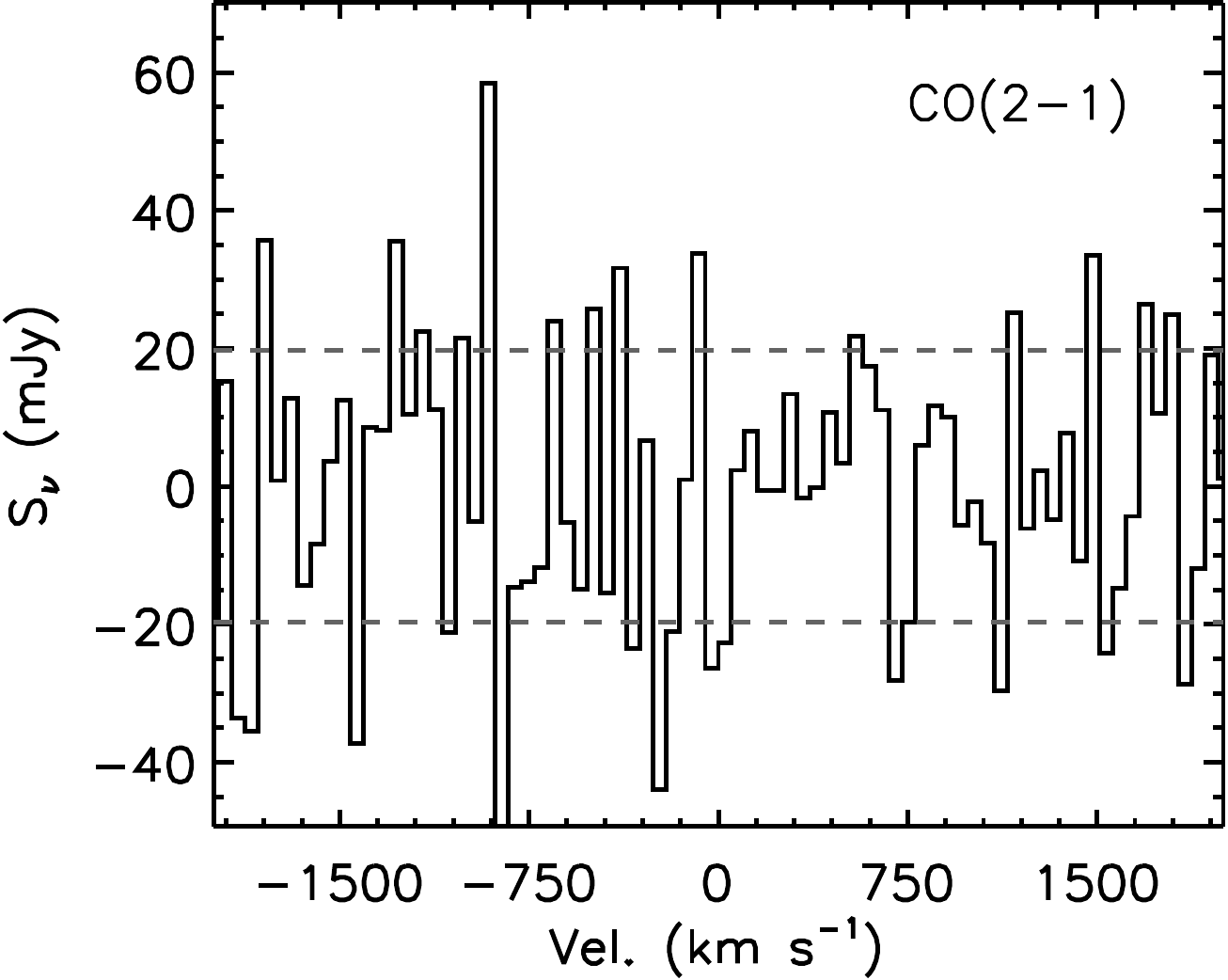}
\caption{NGC0057}
\end{subfigure}
\begin{subfigure}[b]{0.48\textwidth}
\includegraphics[width=0.45\textwidth,angle=0,clip,trim=0cm 0cm 0cm 0.0cm]{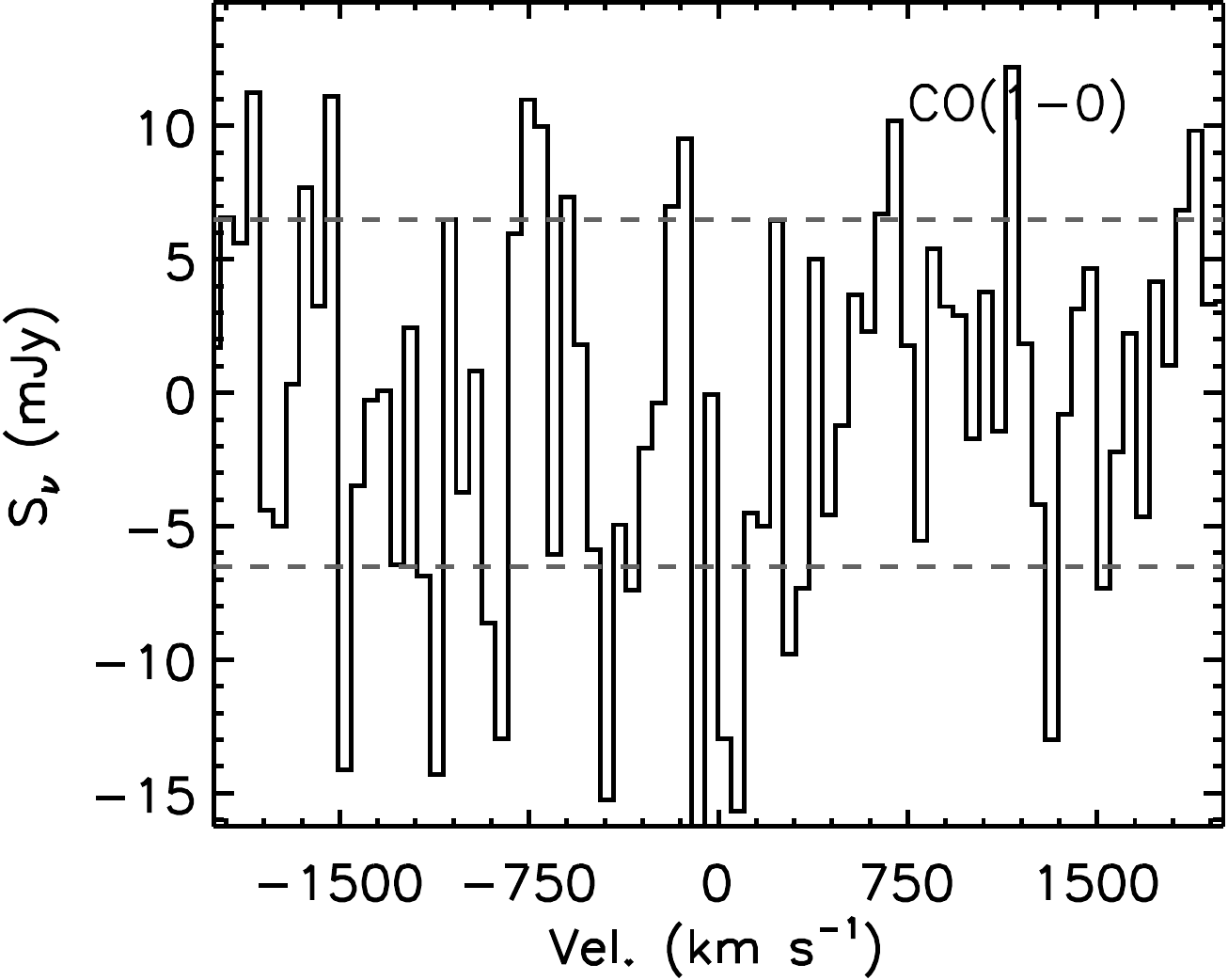}\hspace{0.25cm}
\includegraphics[width=0.45\textwidth,angle=0,clip,trim=0cm 0cm 0cm 0.0cm]{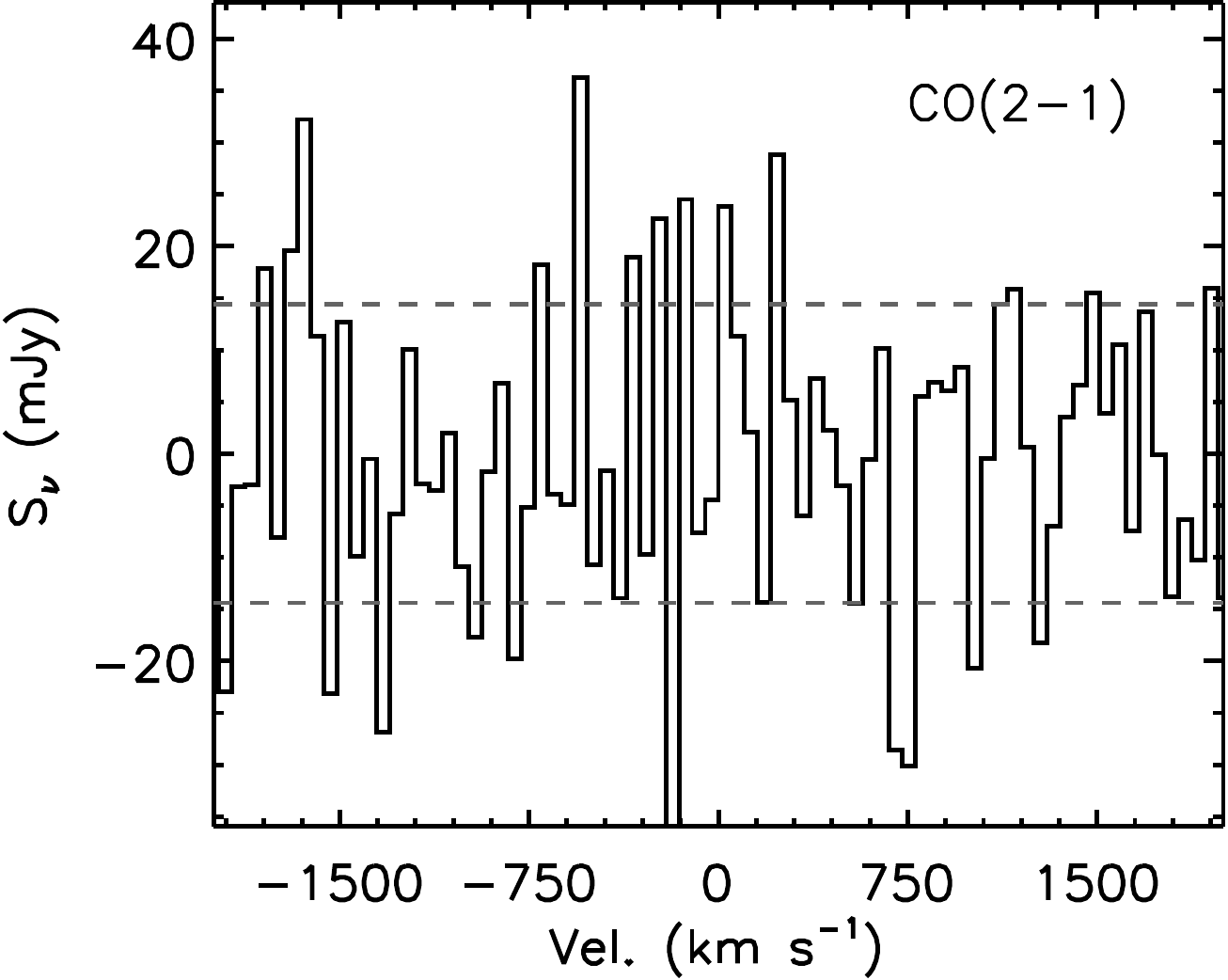}
\caption{NGC0227}
\end{subfigure}\vspace{0.5cm}
\begin{subfigure}[b]{0.48\textwidth}
\includegraphics[width=0.45\textwidth,angle=0,clip,trim=0cm 0cm 0cm 0.0cm]{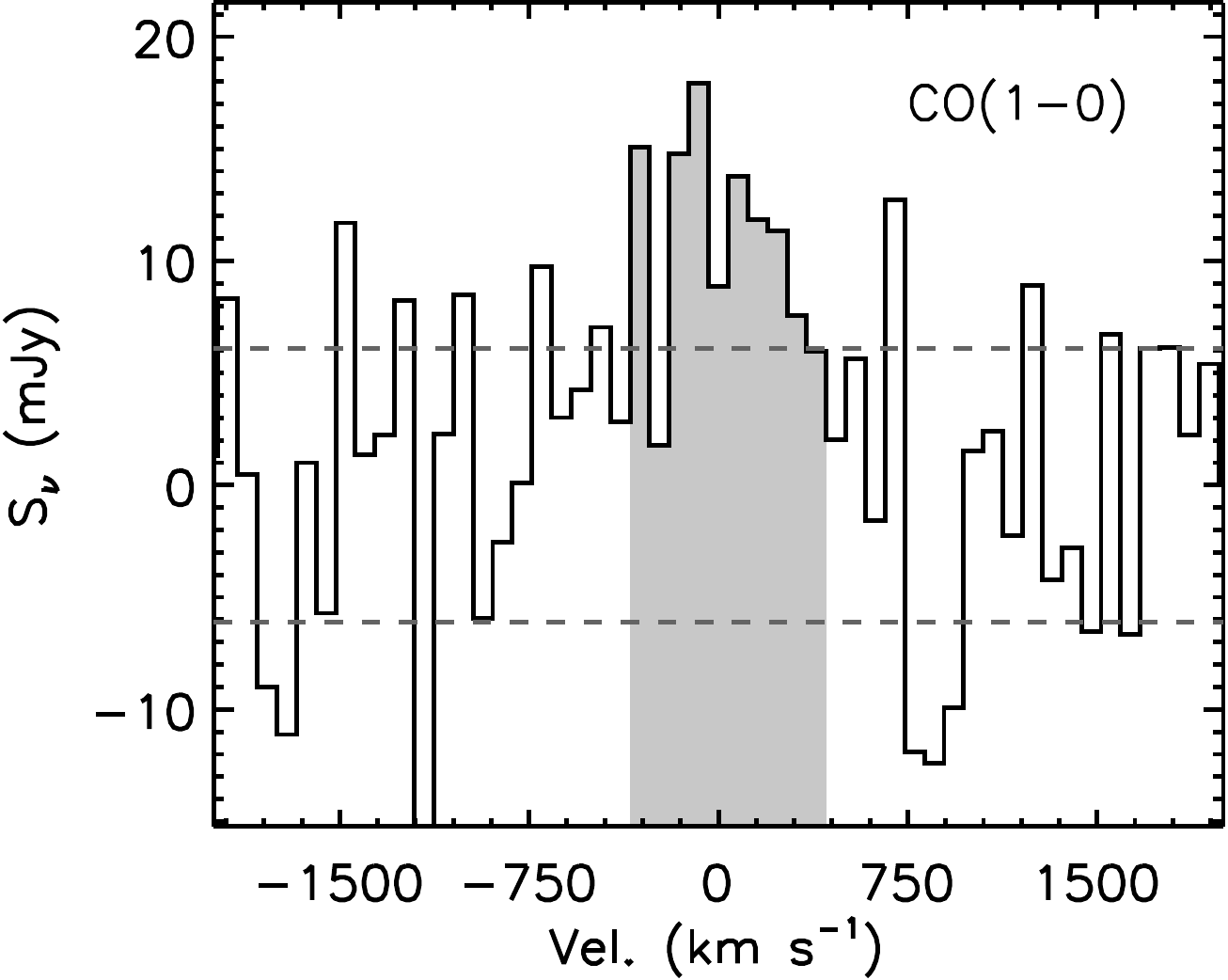}\hspace{0.25cm}
\includegraphics[width=0.45\textwidth,angle=0,clip,trim=0cm 0cm 0cm 0.0cm]{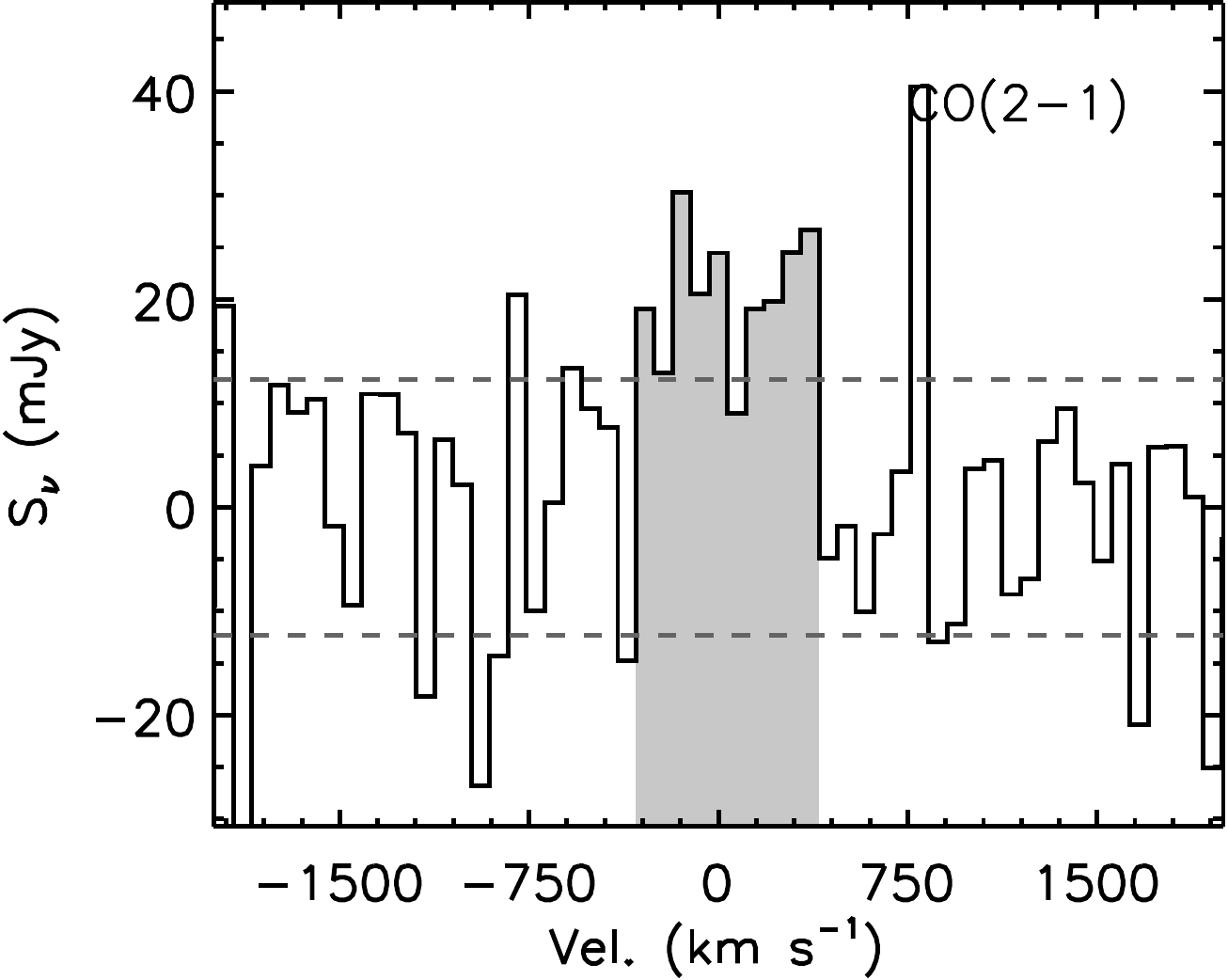}
\caption{NGC0467}
\end{subfigure}
\begin{subfigure}[b]{0.48\textwidth}
\includegraphics[width=0.45\textwidth,angle=0,clip,trim=0cm 0cm 0cm 0.0cm]{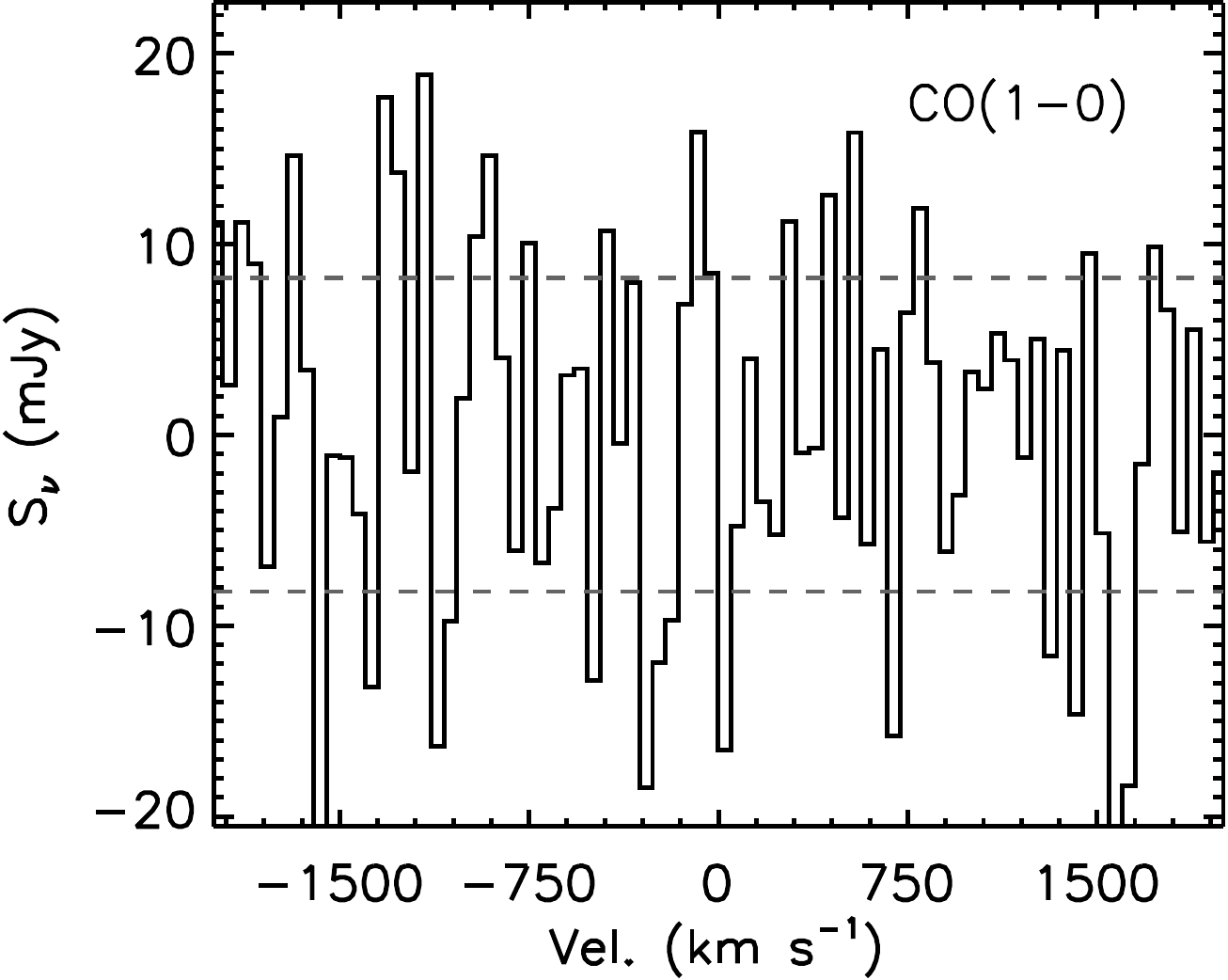}\hspace{0.25cm}
\includegraphics[width=0.45\textwidth,angle=0,clip,trim=0cm 0cm 0cm 0.0cm]{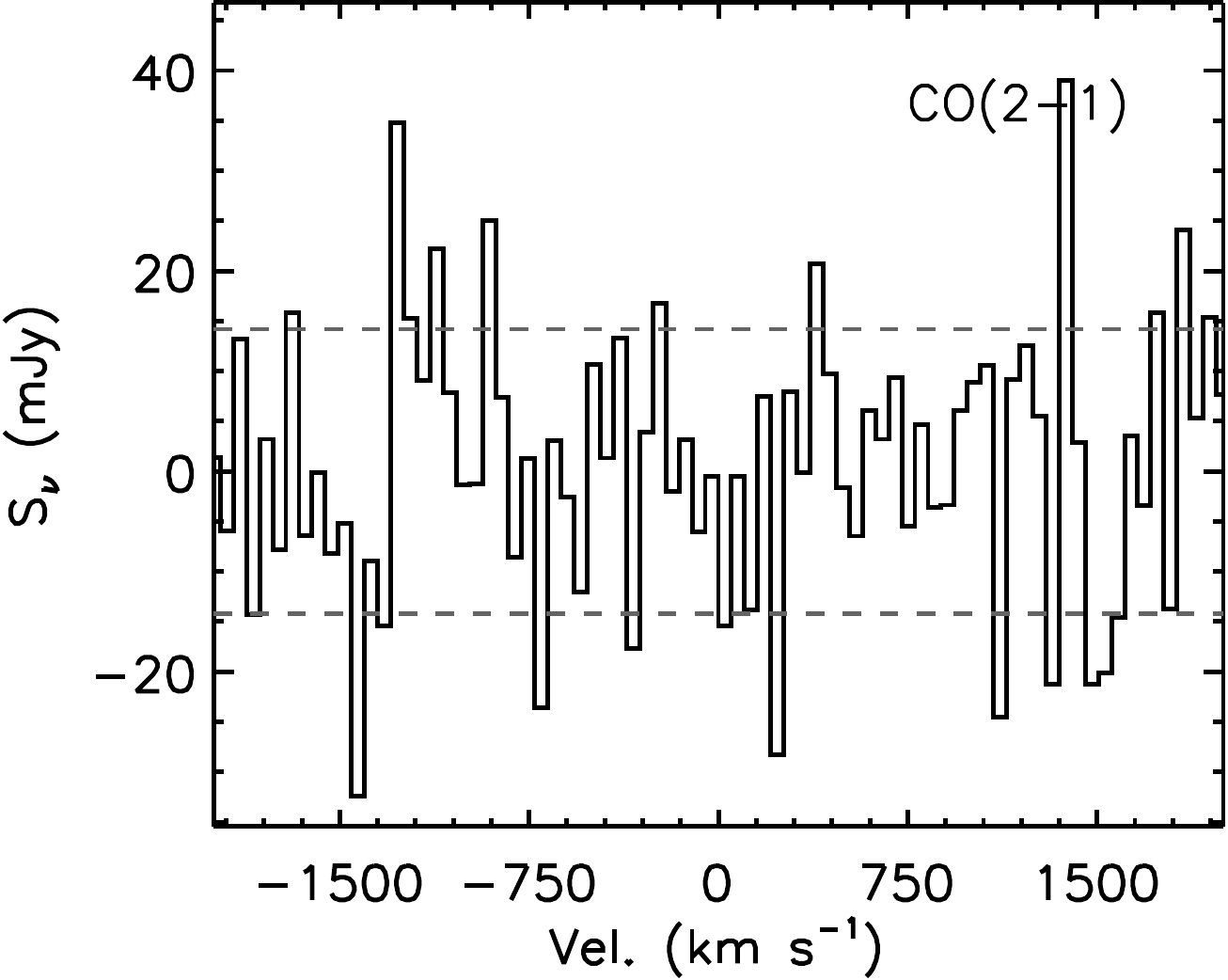}
\caption{NGC0499}
\end{subfigure}\vspace{0.5cm}
\begin{subfigure}[b]{0.48\textwidth}
\includegraphics[width=0.45\textwidth,angle=0,clip,trim=0cm 0cm 0cm 0.0cm]{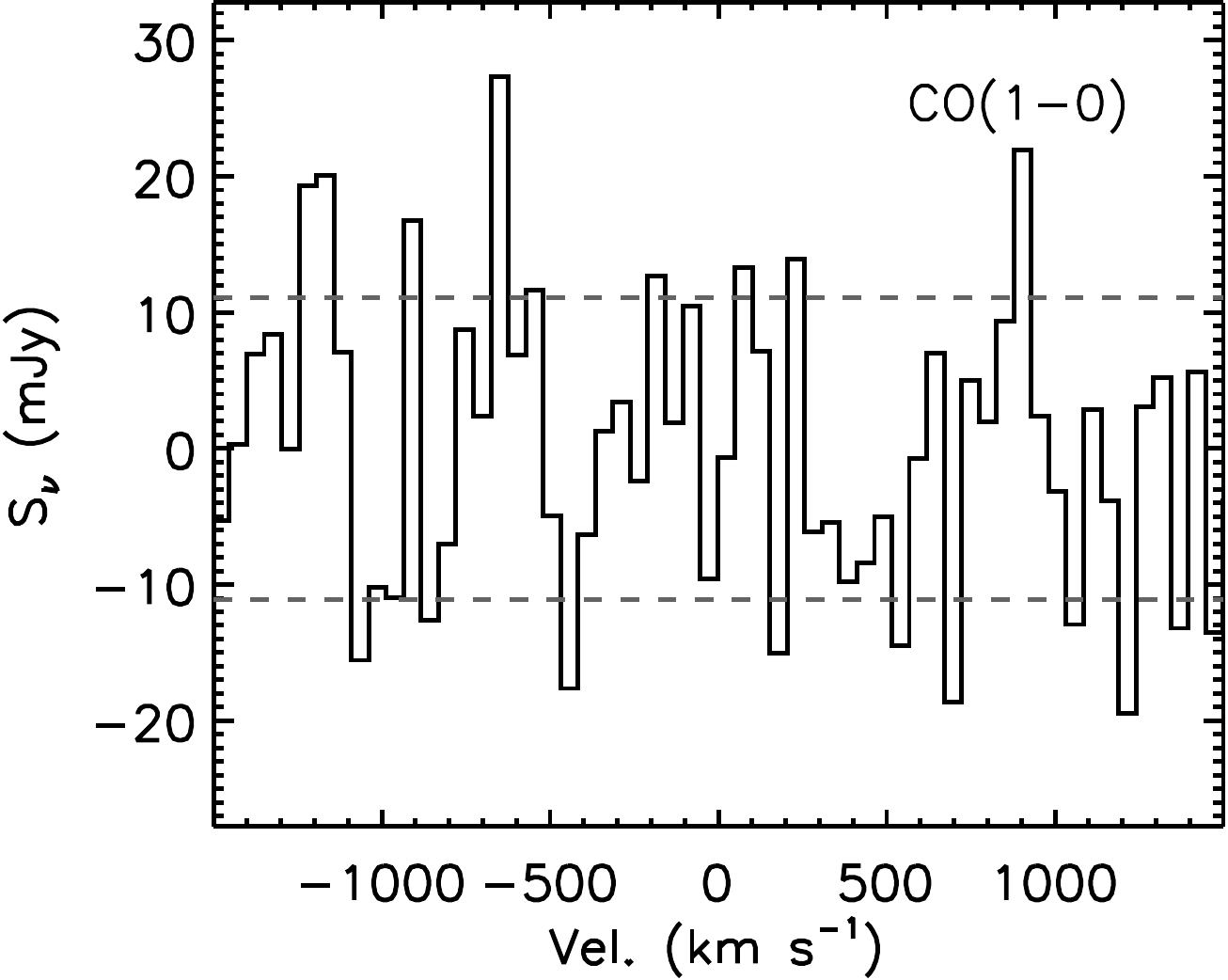}\hspace{0.25cm}
\includegraphics[width=0.45\textwidth,angle=0,clip,trim=0cm 0cm 0cm 0.0cm]{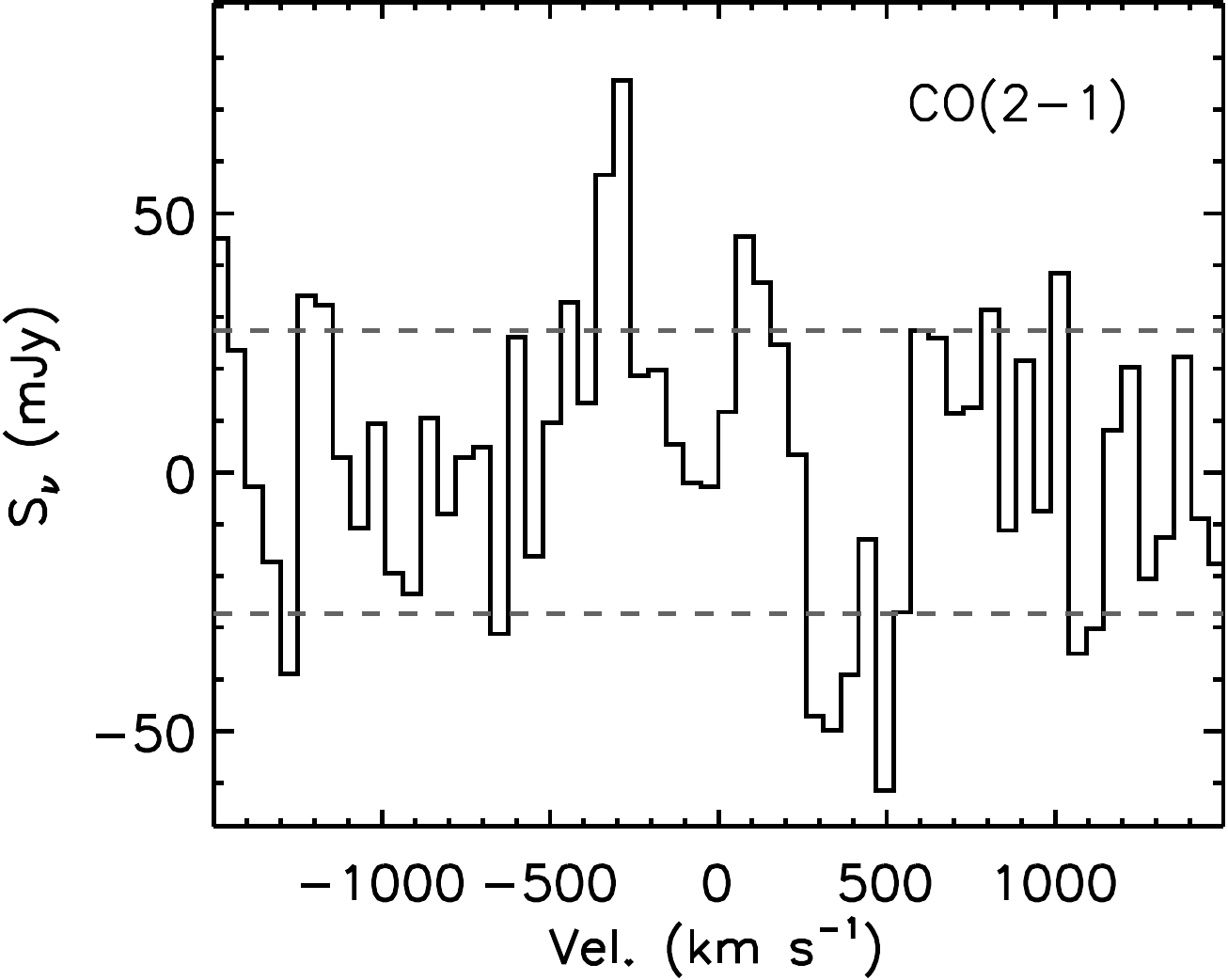}
\caption{NGC0507}
\end{subfigure}
\begin{subfigure}[b]{0.48\textwidth}
\includegraphics[width=0.45\textwidth,angle=0,clip,trim=0cm 0cm 0cm 0.0cm]{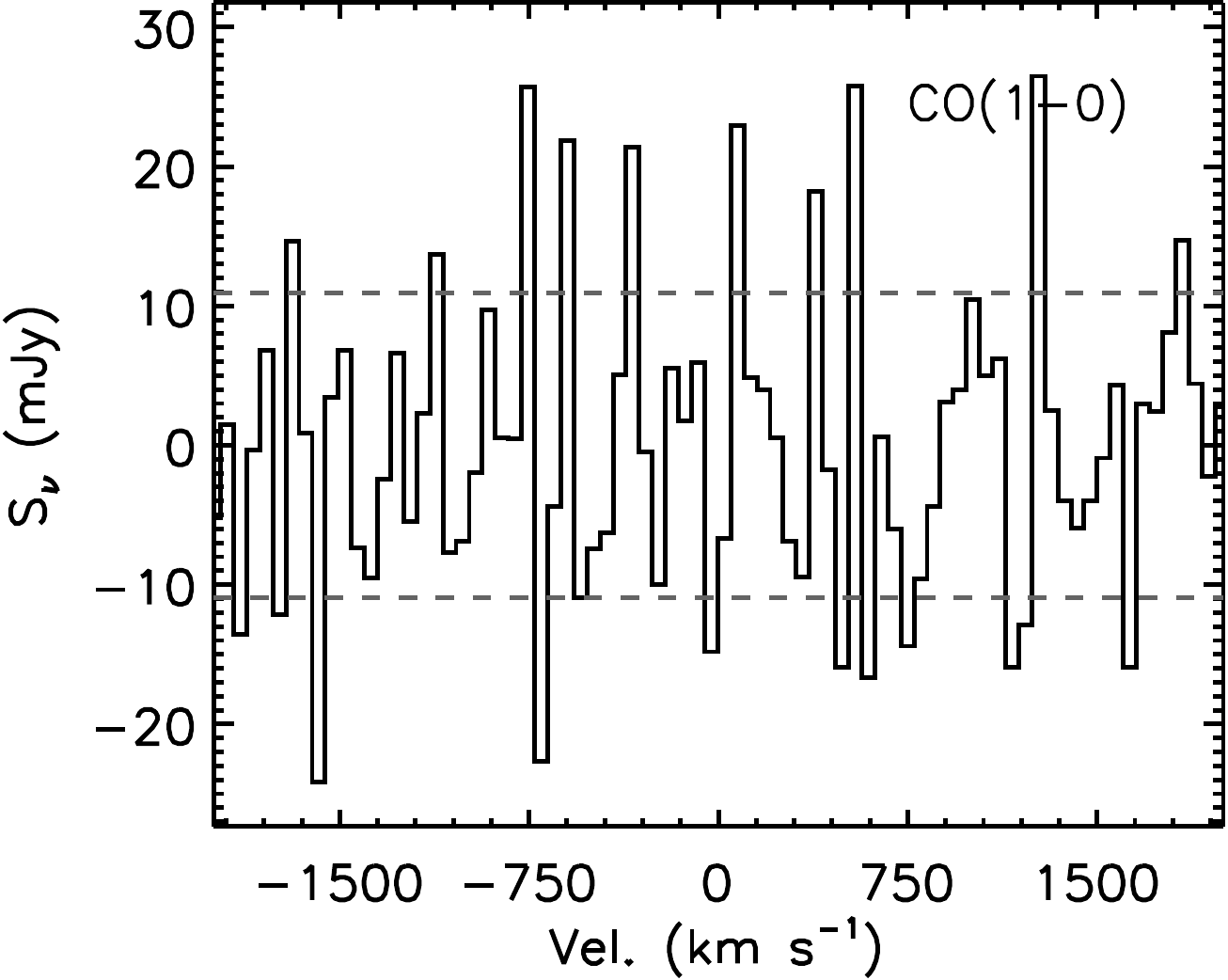}\hspace{0.25cm}
\includegraphics[width=0.45\textwidth,angle=0,clip,trim=0cm 0cm 0cm 0.0cm]{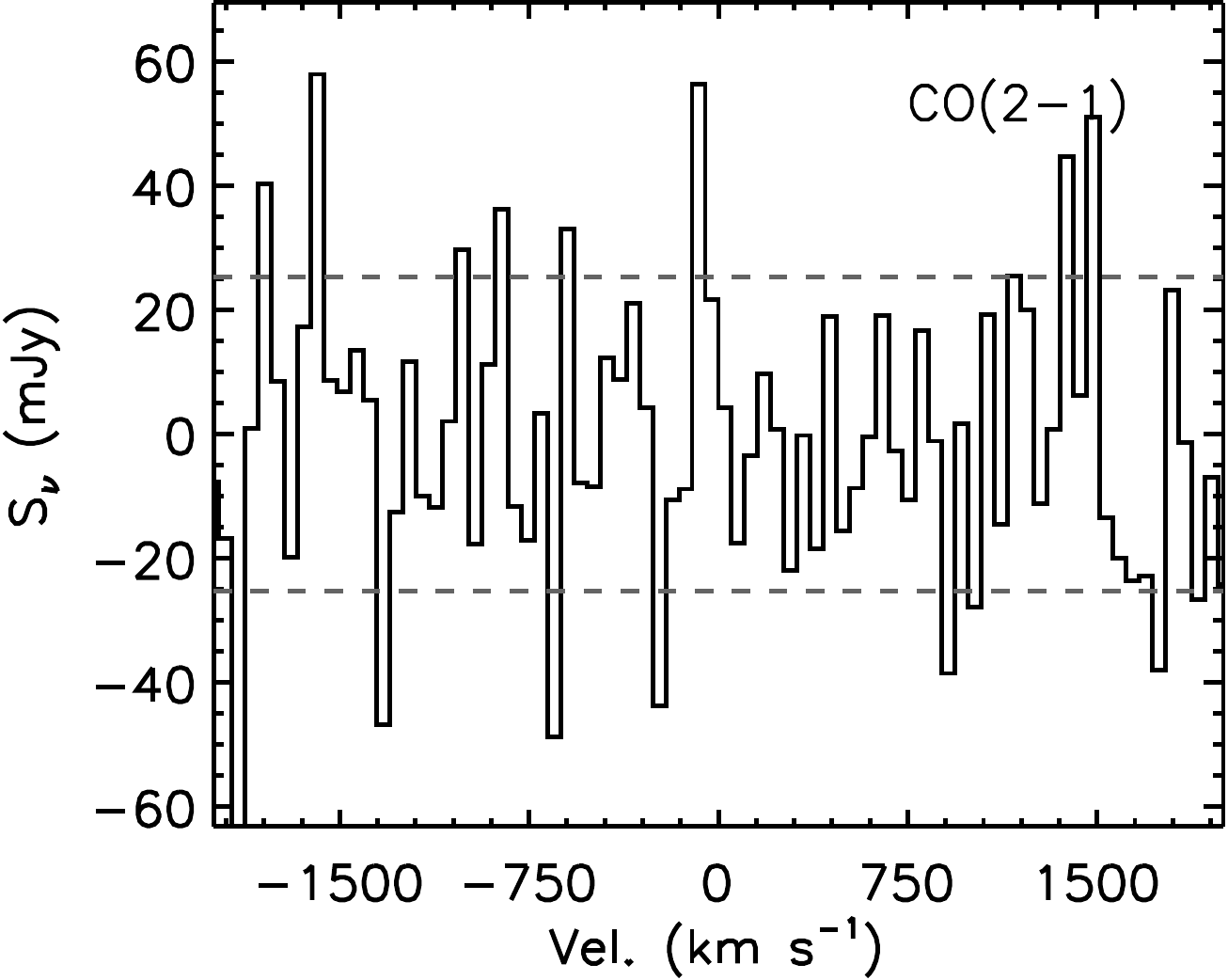}
\caption{NGC0533}
\end{subfigure}\vspace{0.5cm}
\begin{subfigure}[b]{0.48\textwidth}
\includegraphics[width=0.45\textwidth,angle=0,clip,trim=0cm 0cm 0cm 0.0cm]{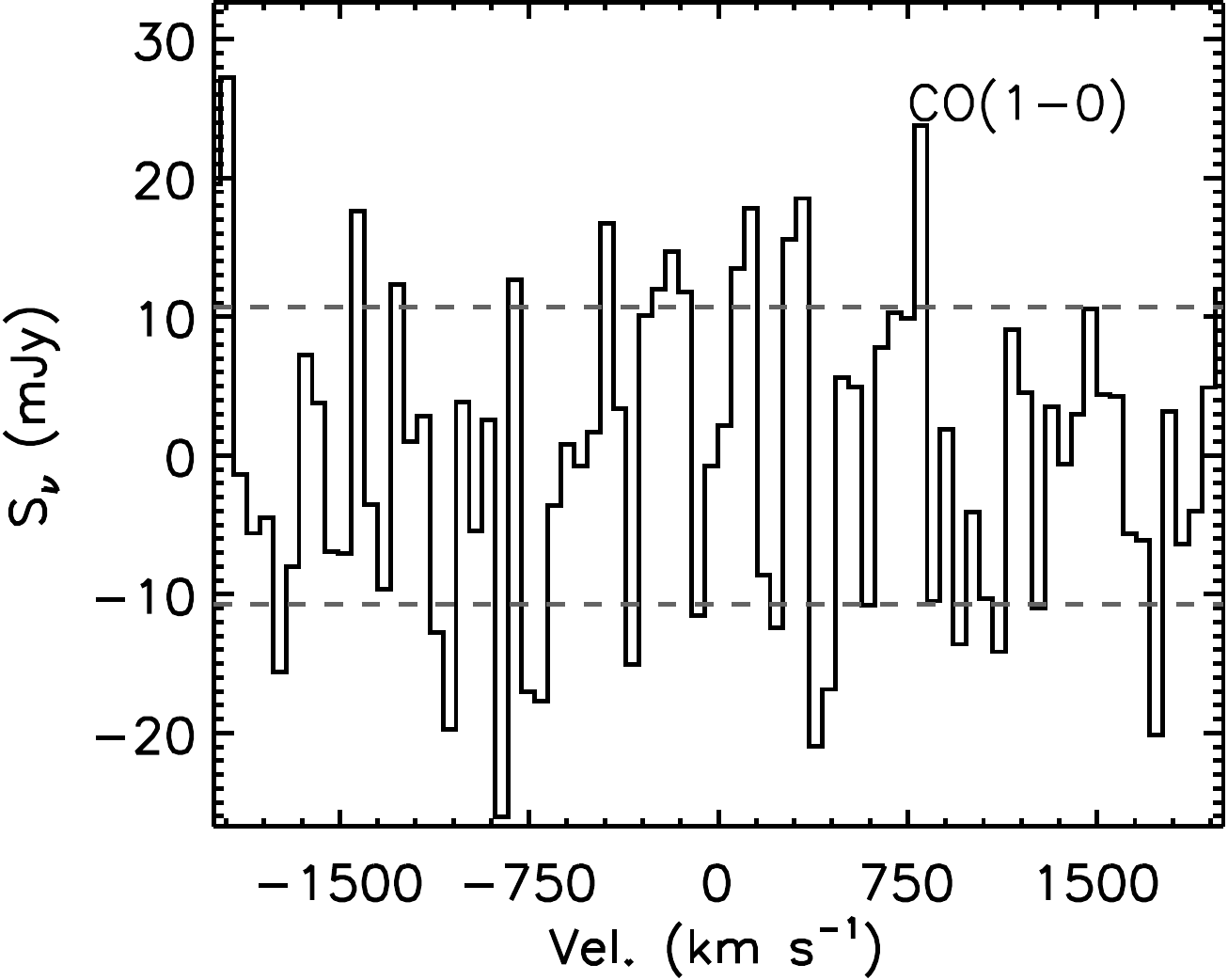}\hspace{0.25cm}
\includegraphics[width=0.45\textwidth,angle=0,clip,trim=0cm 0cm 0cm 0.0cm]{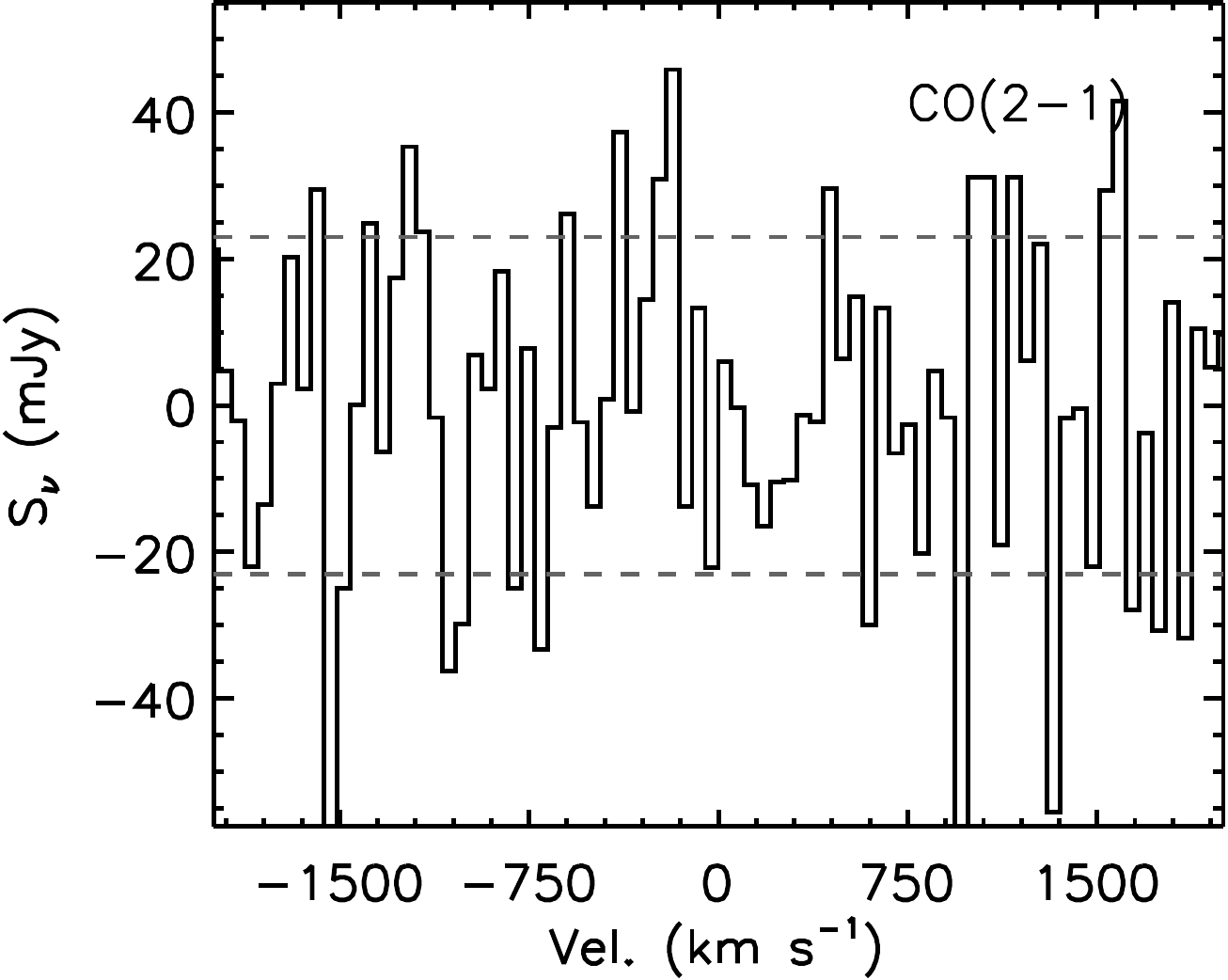}
\caption{NGC0547}
\end{subfigure}
\begin{subfigure}[b]{0.48\textwidth}
\includegraphics[width=0.45\textwidth,angle=0,clip,trim=0cm 0cm 0cm 0.0cm]{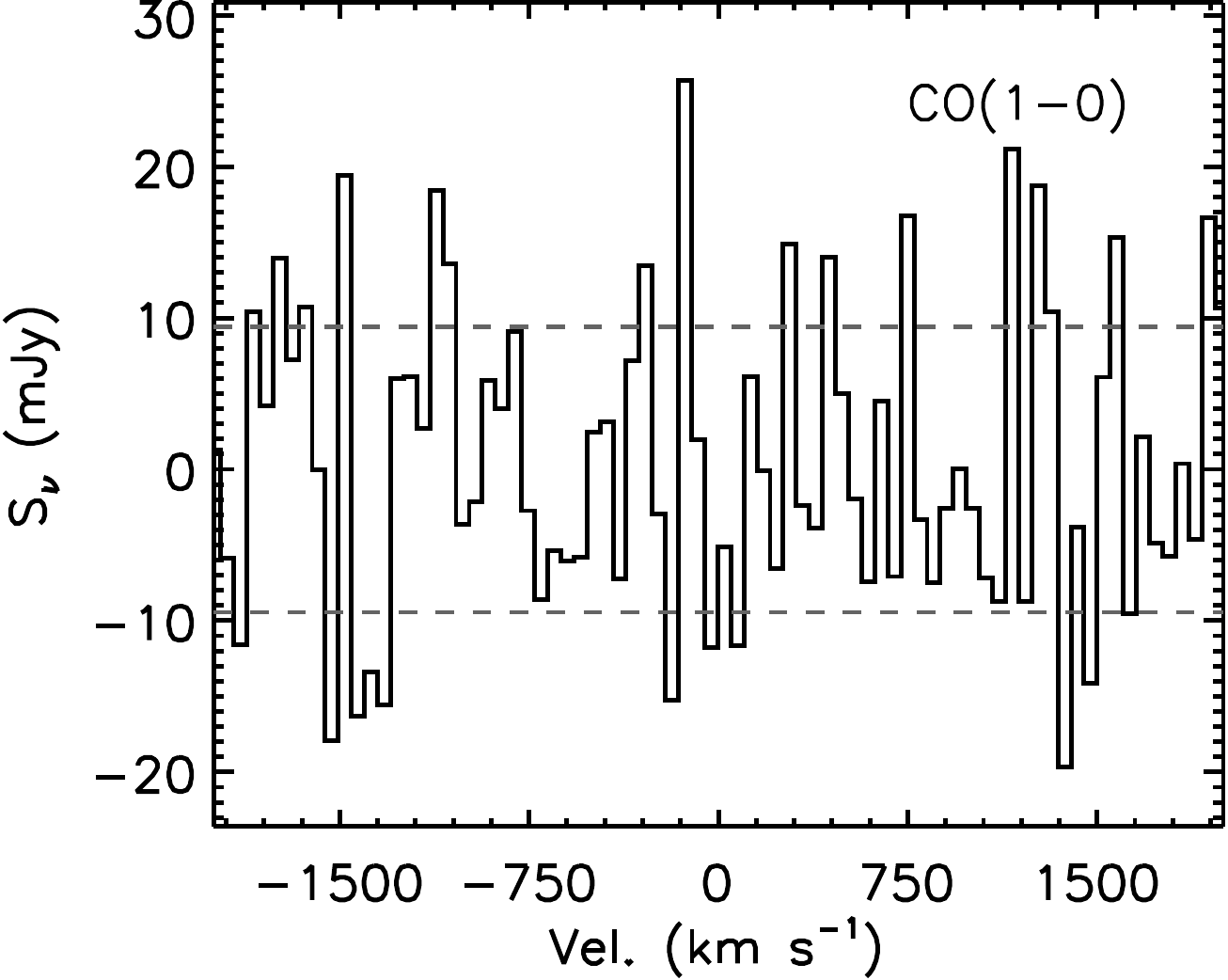}\hspace{0.25cm}
\includegraphics[width=0.45\textwidth,angle=0,clip,trim=0cm 0cm 0cm 0.0cm]{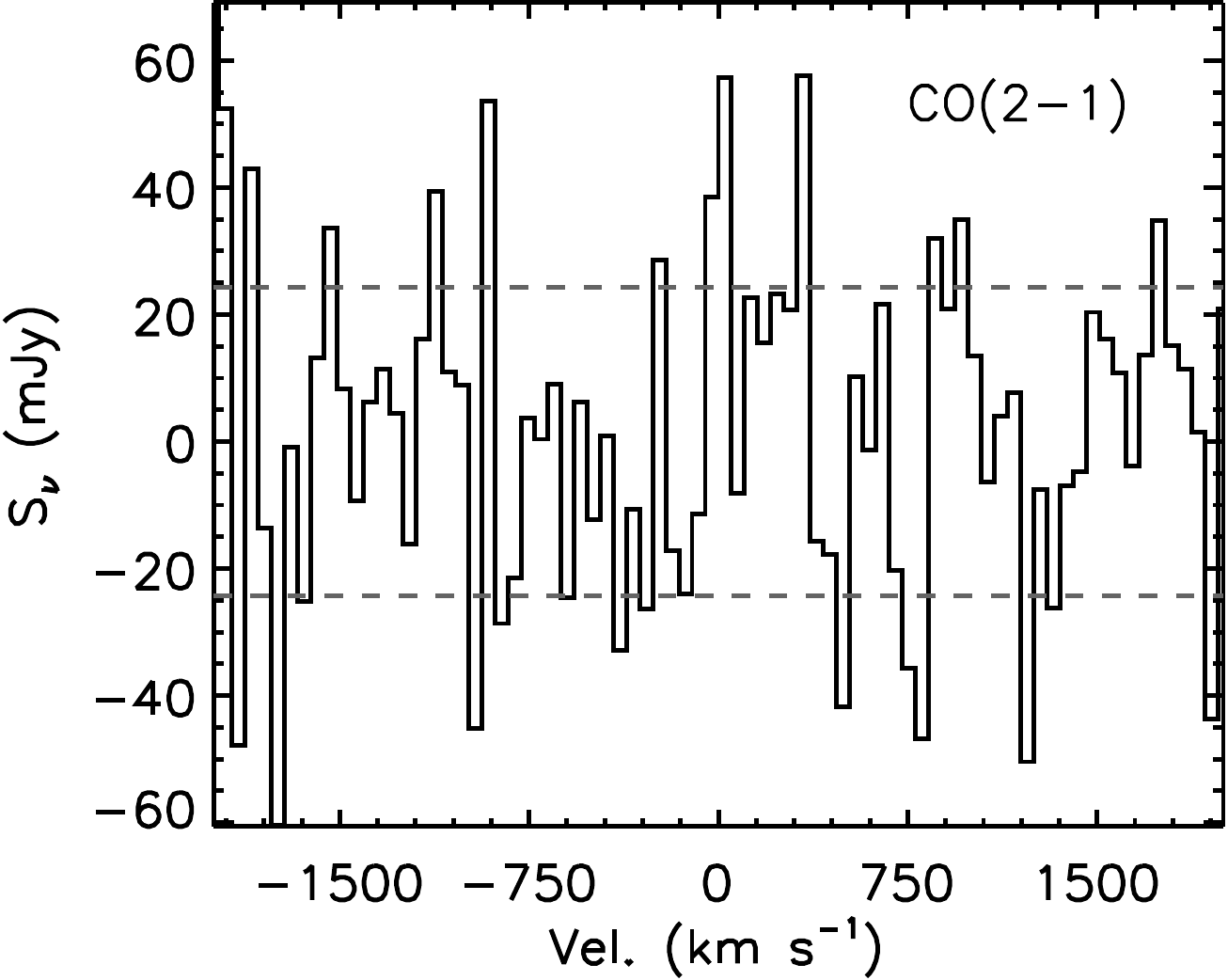}
\caption{NGC0741}
\end{subfigure}\vspace{0.5cm}
\begin{subfigure}[b]{0.48\textwidth}
\includegraphics[width=0.45\textwidth,angle=0,clip,trim=0cm 0cm 0cm 0.0cm]{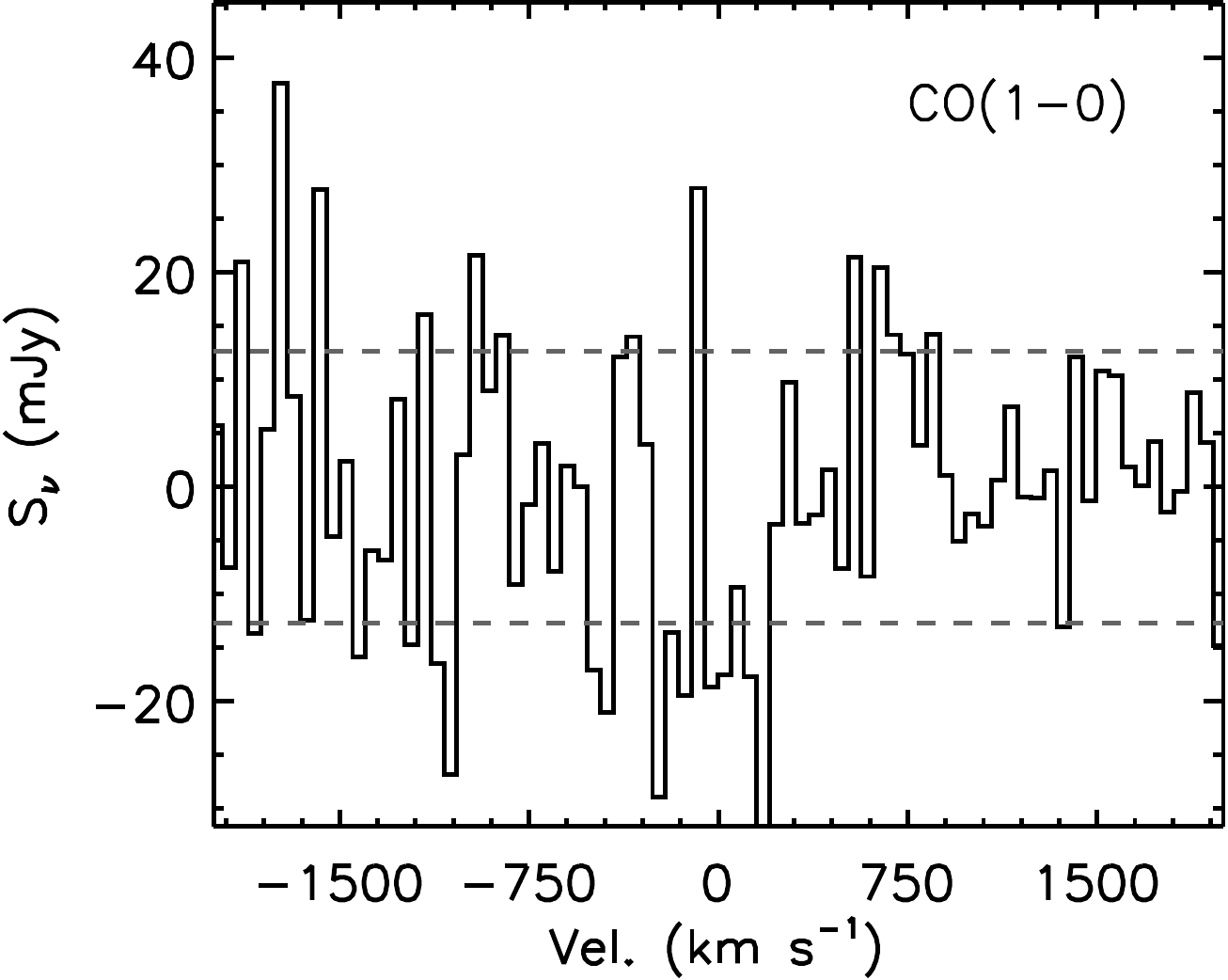}\hspace{0.25cm}
\includegraphics[width=0.45\textwidth,angle=0,clip,trim=0cm 0cm 0cm 0.0cm]{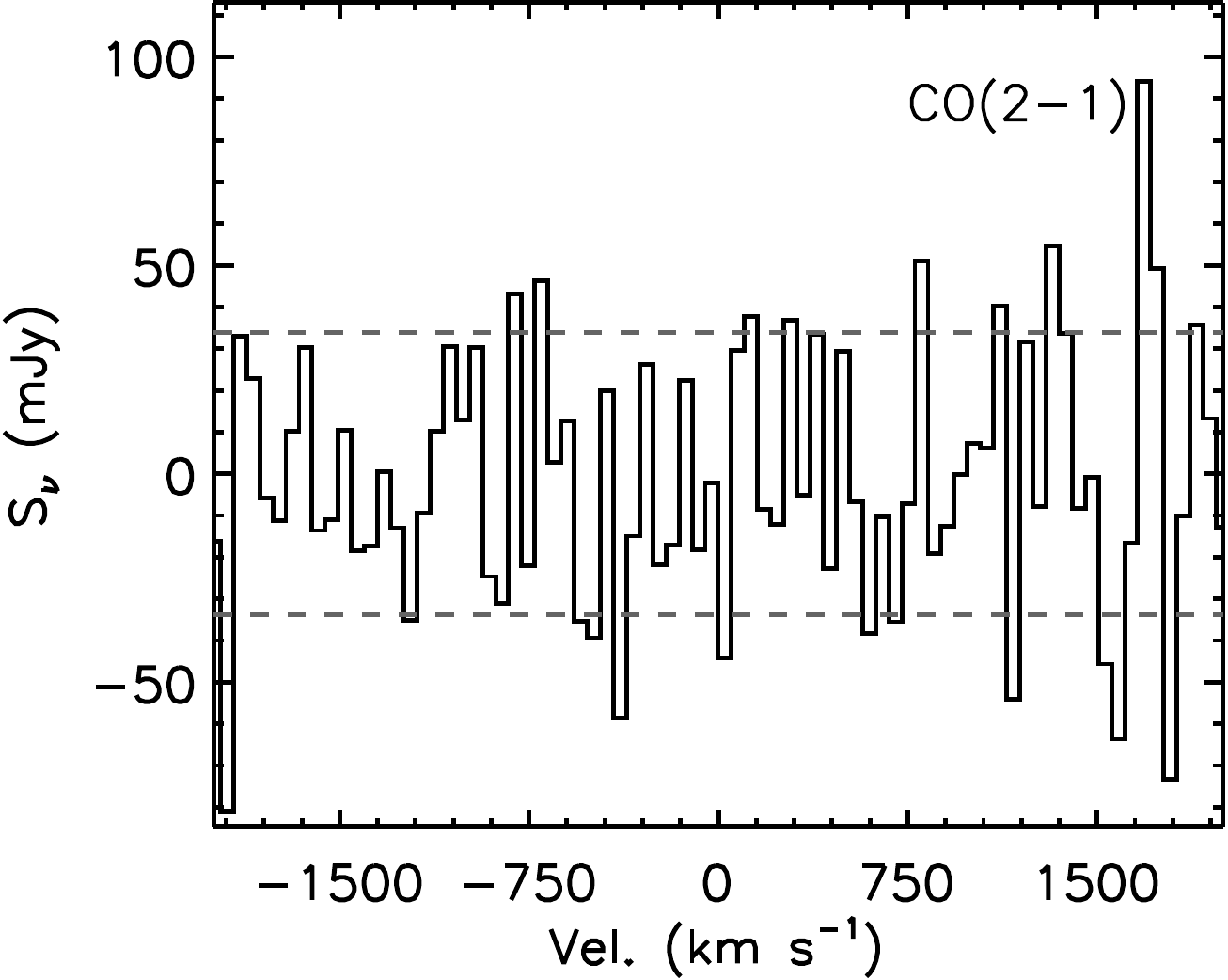}
\caption{NGC0890}
\end{subfigure}
\begin{subfigure}[b]{0.48\textwidth}
\includegraphics[width=0.45\textwidth,angle=0,clip,trim=0cm 0cm 0cm 0.0cm]{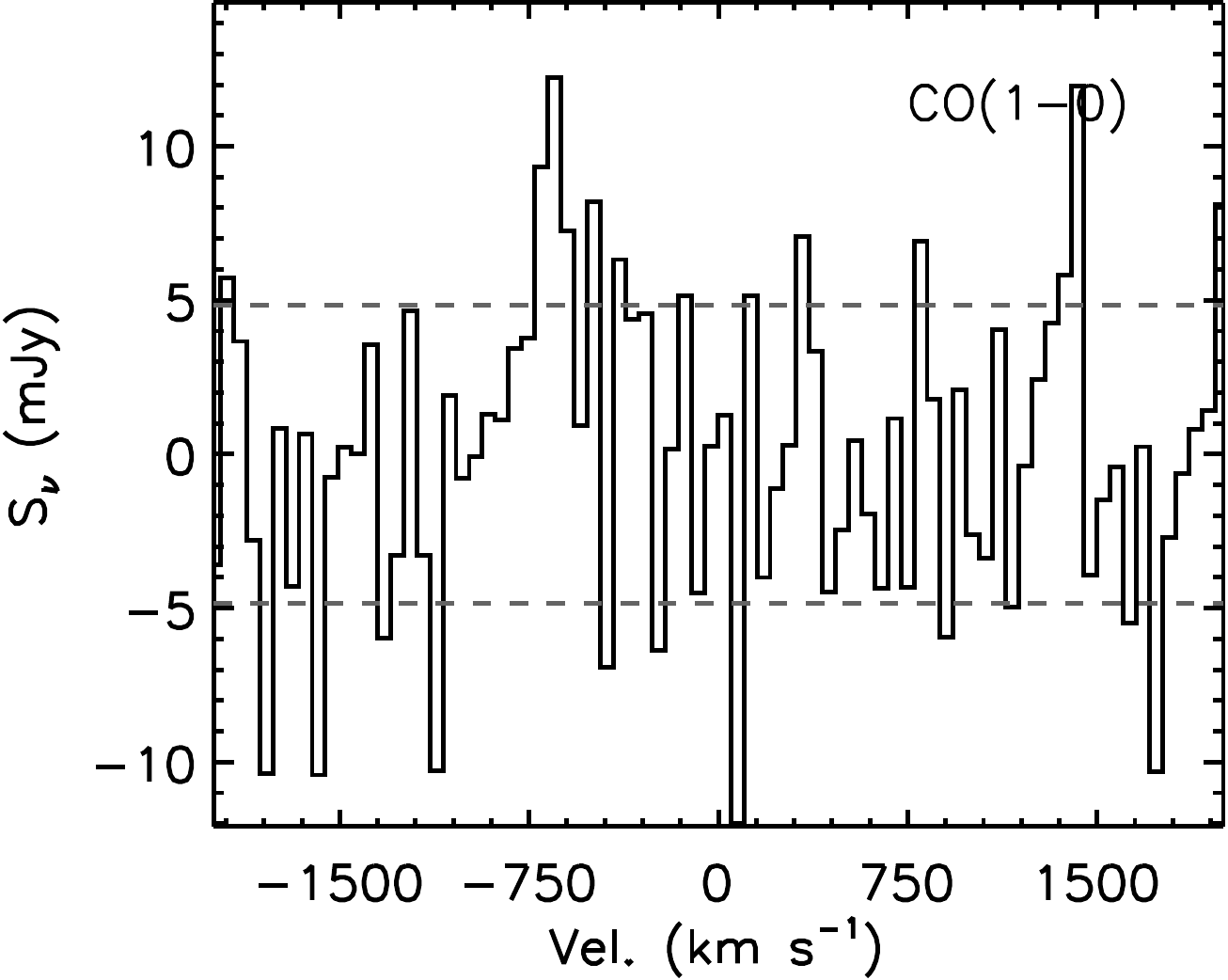}\hspace{0.25cm}
\includegraphics[width=0.45\textwidth,angle=0,clip,trim=0cm 0cm 0cm 0.0cm]{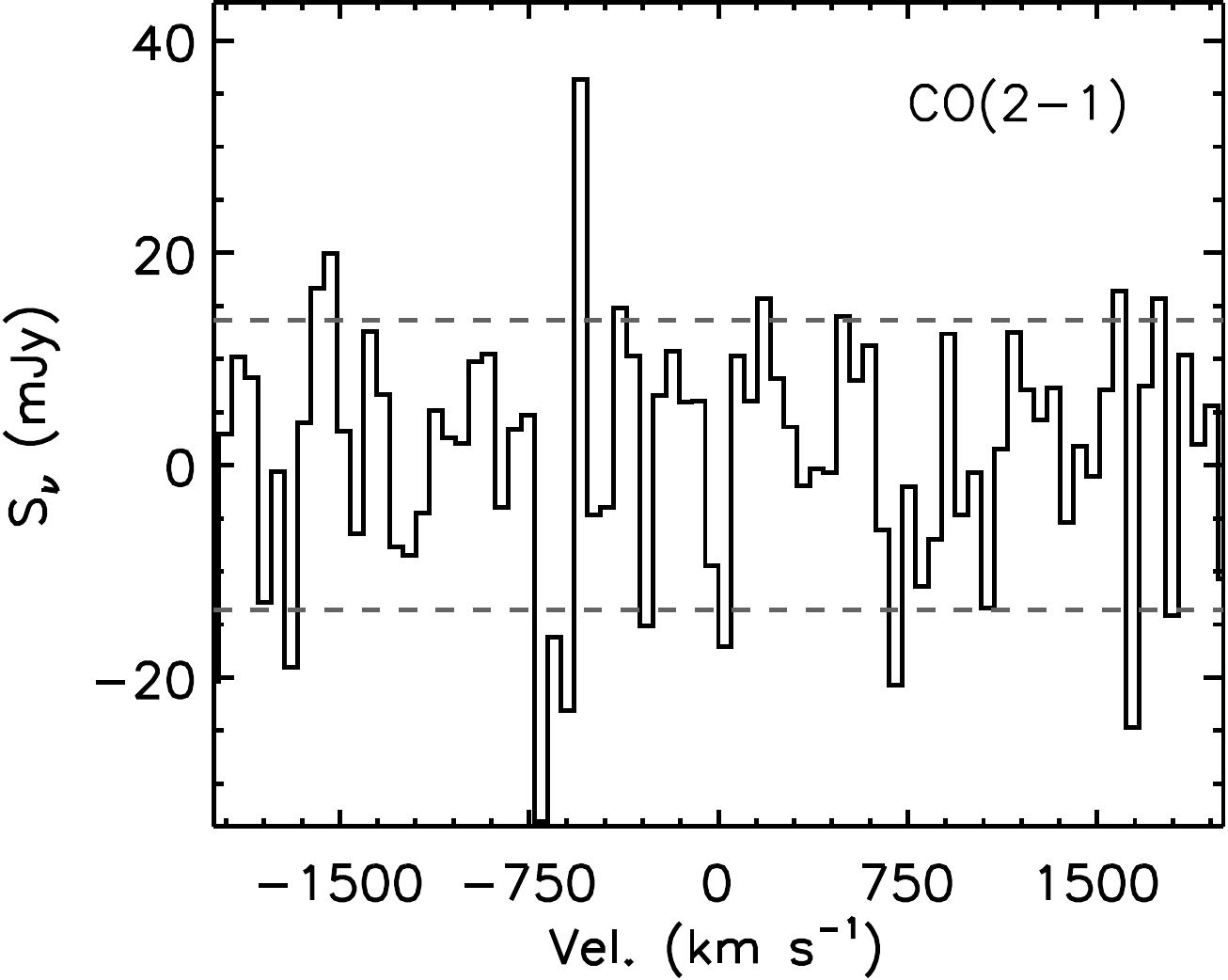}
\caption{NGC0910}
\end{subfigure}
\caption{CO(1-0) and CO(2-1) spectra of our sample objects. The velocity on the x-axis is
plotted over $\pm$2000 \kms\ with respect to the rest-frequency of that line, unless otherwise indicated. The grey shaded region on the spectra denotes the detected line, and the dashed
lines show the $\pm$1$\sigma$ RMS level. }
\label{codetsfig2}
 \end{center}
 \end{figure*}
 \begin{figure*} 
\begin{center} 
\begin{subfigure}[b]{0.48\textwidth}
\includegraphics[width=0.45\textwidth,angle=0,clip,trim=0cm 0cm 0cm 0.0cm]{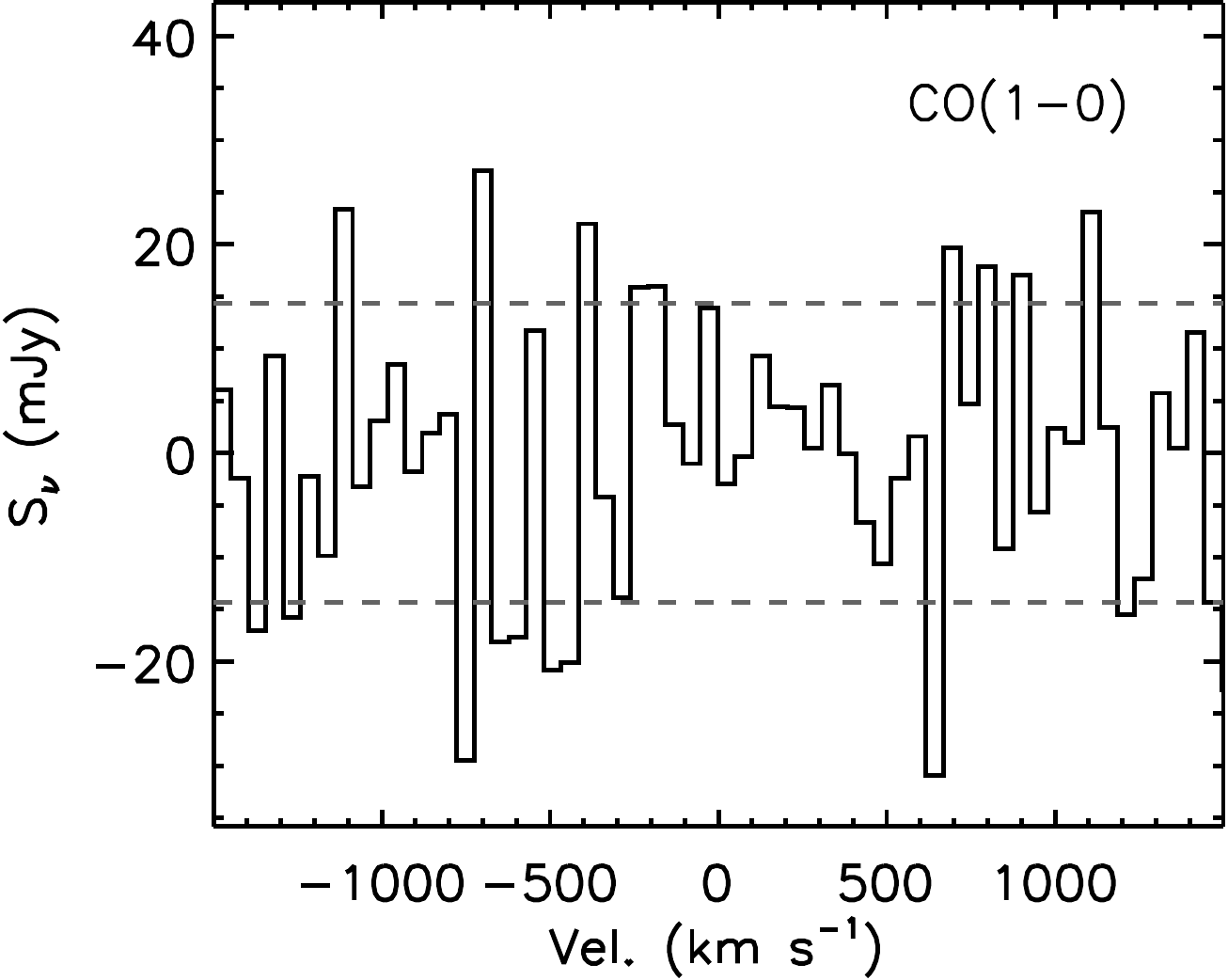}\hspace{0.25cm}
\includegraphics[width=0.45\textwidth,angle=0,clip,trim=0cm 0cm 0cm 0.0cm]{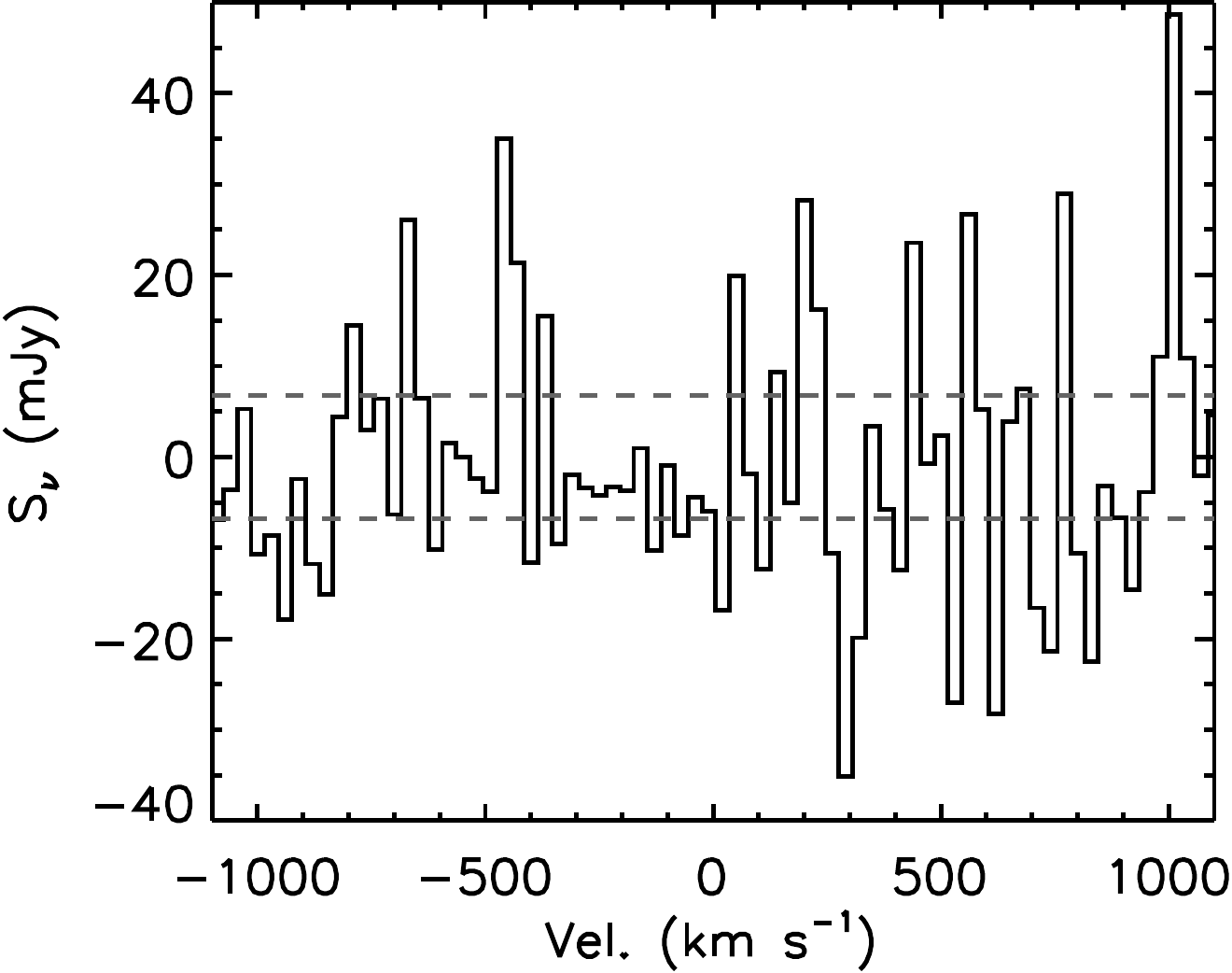}
\caption{NGC1453}
\end{subfigure}
\begin{subfigure}[b]{0.48\textwidth}
\includegraphics[width=0.45\textwidth,angle=0,clip,trim=0cm 0cm 0cm 0.0cm]{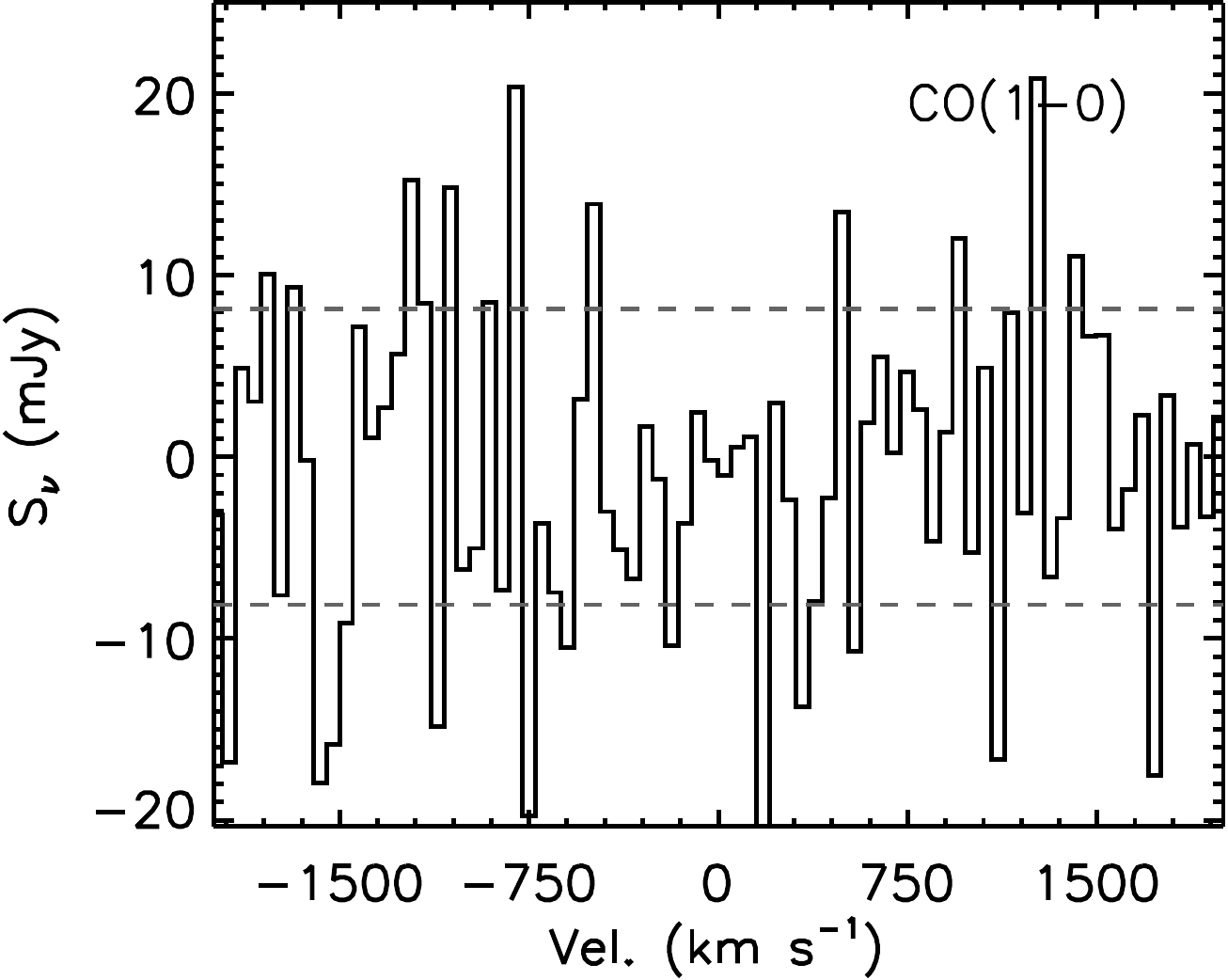}\hspace{0.25cm}
\includegraphics[width=0.45\textwidth,angle=0,clip,trim=0cm 0cm 0cm 0.0cm]{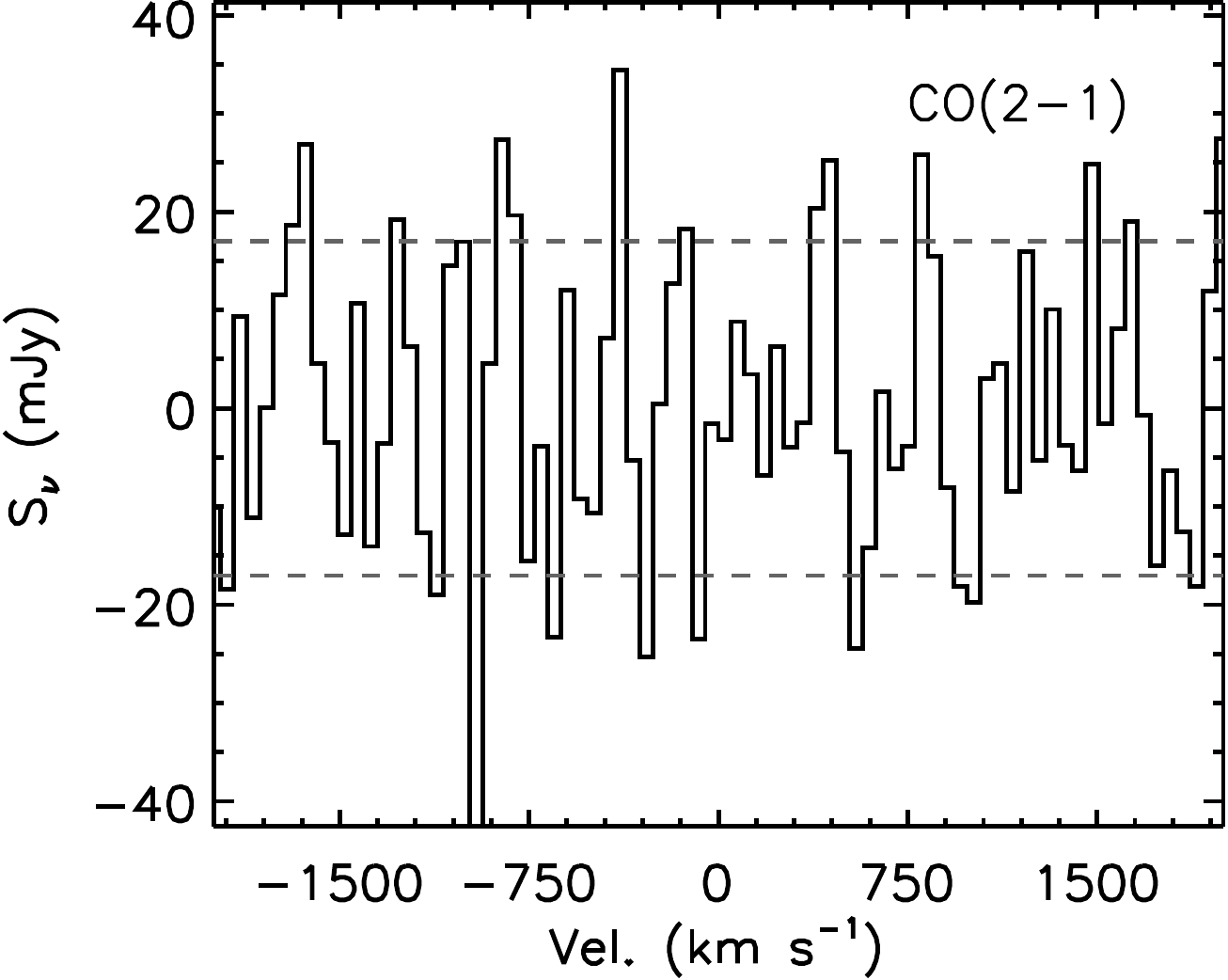}
\caption{NGC1573}
\end{subfigure}\vspace{0.5cm}
\begin{subfigure}[b]{0.48\textwidth}
\includegraphics[width=0.45\textwidth,angle=0,clip,trim=0cm 0cm 0cm 0.0cm]{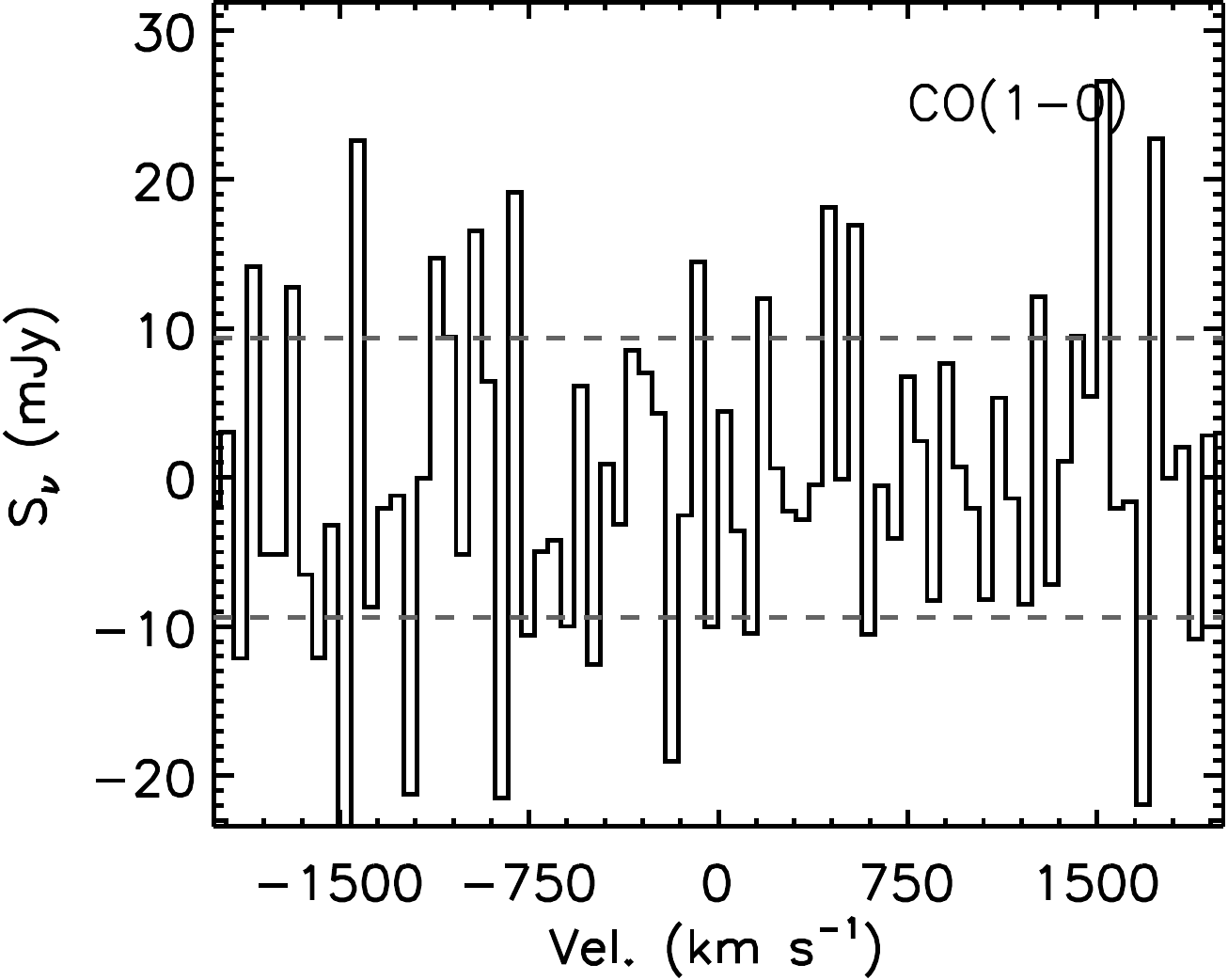}\hspace{0.25cm}
\includegraphics[width=0.45\textwidth,angle=0,clip,trim=0cm 0cm 0cm 0.0cm]{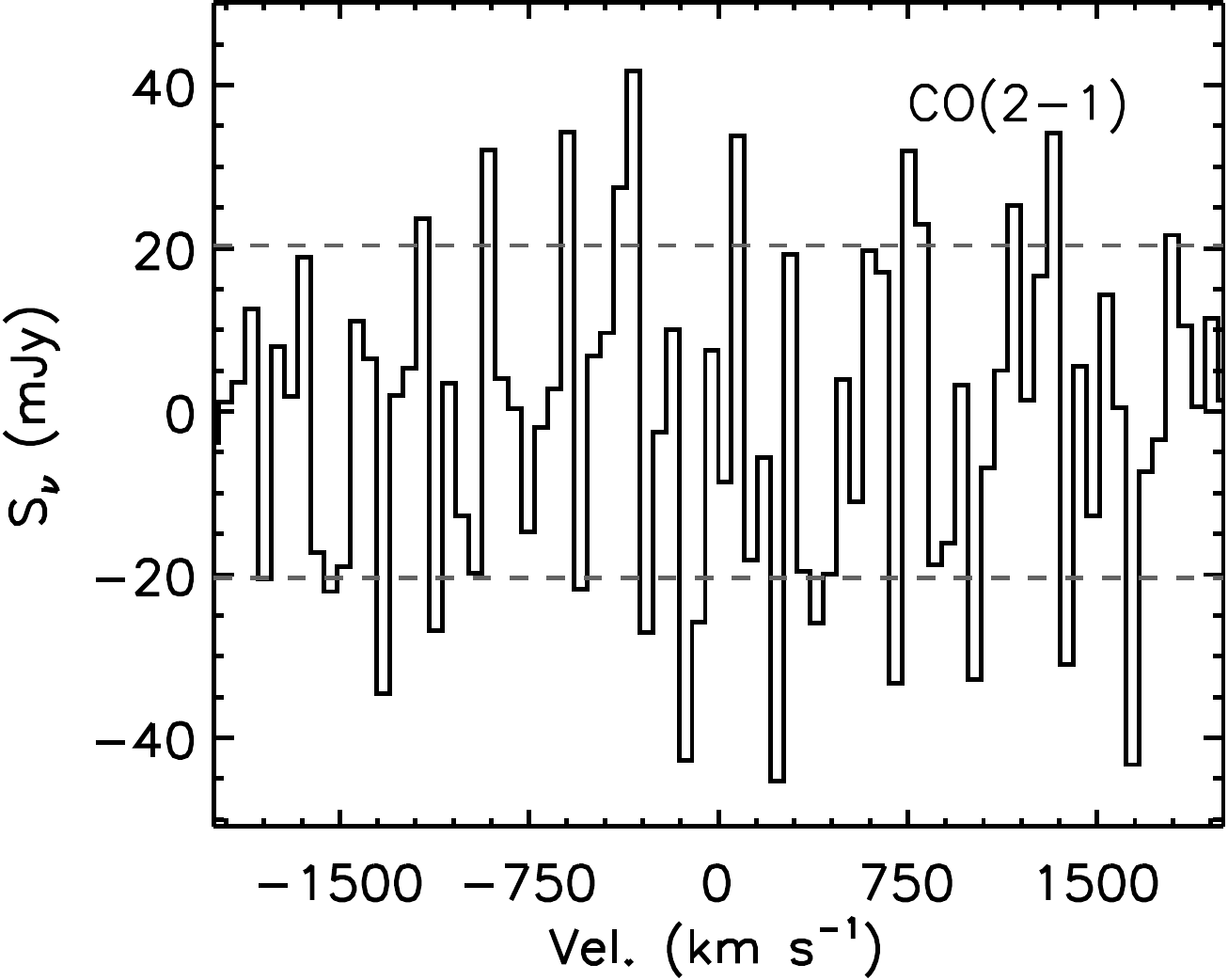}
\caption{NGC2256}
\end{subfigure}
\begin{subfigure}[b]{0.48\textwidth}
\includegraphics[width=0.45\textwidth,angle=0,clip,trim=0cm 0cm 0cm 0.0cm]{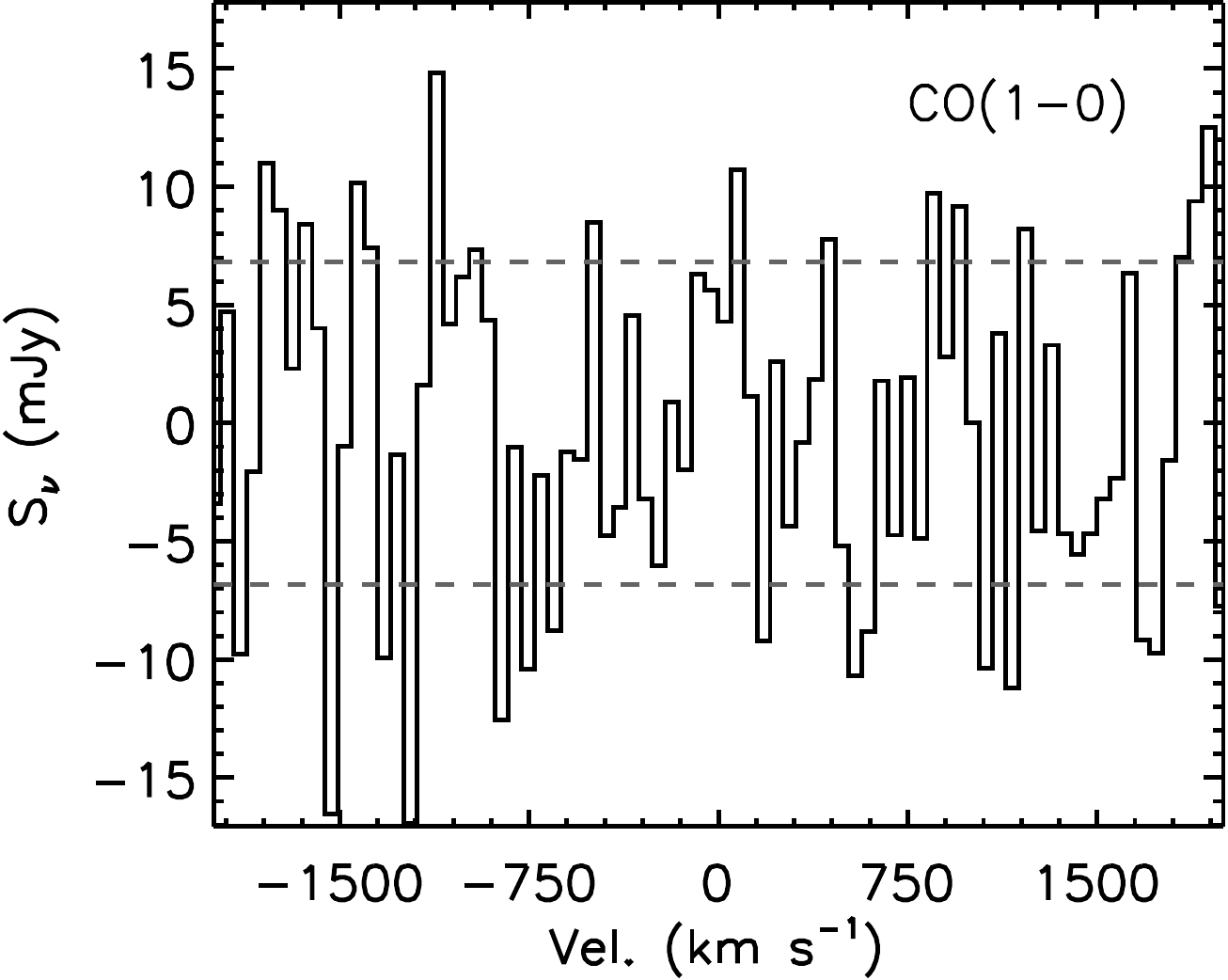}\hspace{0.25cm}
\includegraphics[width=0.45\textwidth,angle=0,clip,trim=0cm 0cm 0cm 0.0cm]{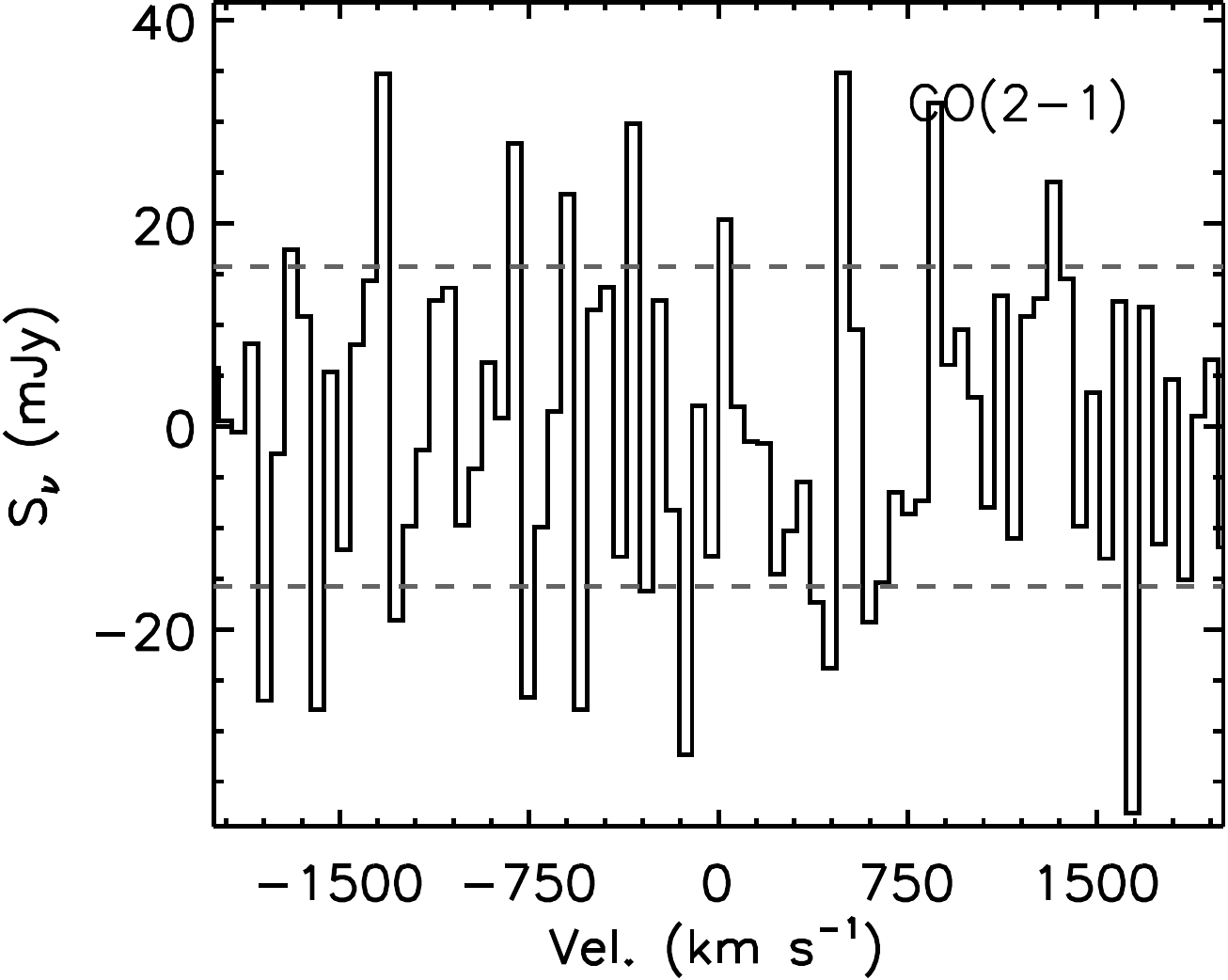}
\caption{NGC2274}
\end{subfigure}\vspace{0.5cm}
\begin{subfigure}[b]{0.48\textwidth}
\includegraphics[width=0.45\textwidth,angle=0,clip,trim=0cm 0cm 0cm 0.0cm]{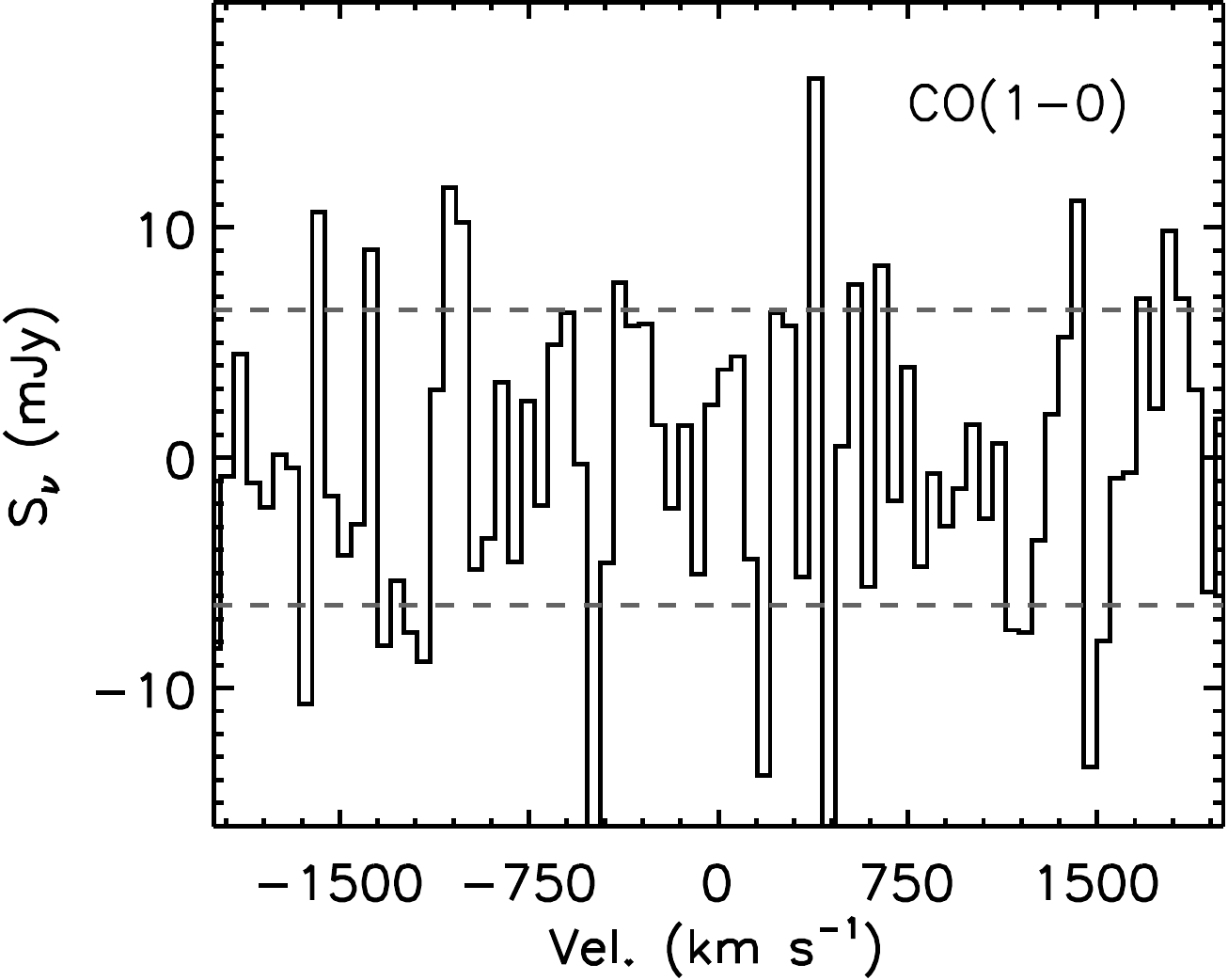}\hspace{0.25cm}
\includegraphics[width=0.45\textwidth,angle=0,clip,trim=0cm 0cm 0cm 0.0cm]{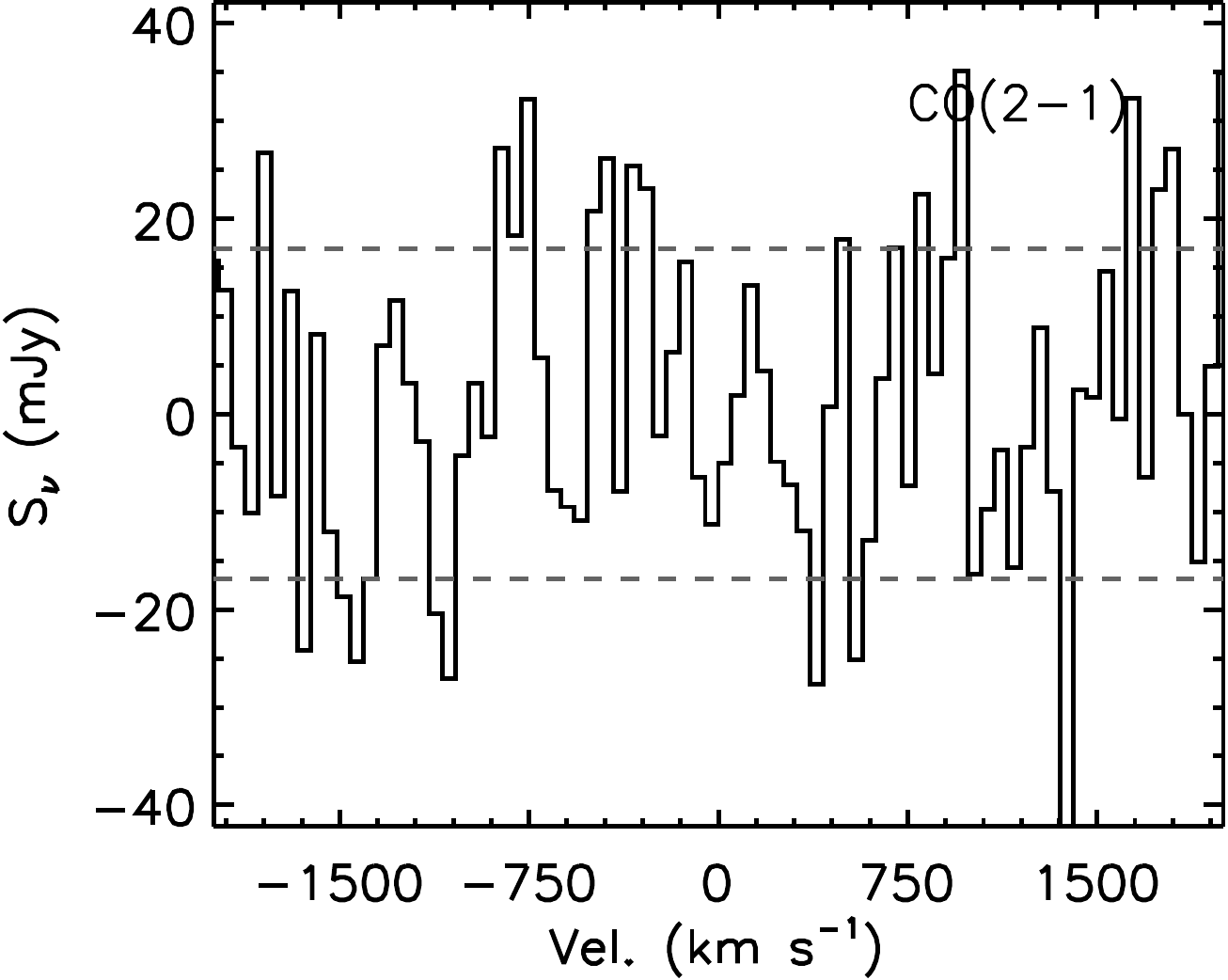}
\caption{NGC2513}
\end{subfigure}
\begin{subfigure}[b]{0.48\textwidth}
\includegraphics[width=0.45\textwidth,angle=0,clip,trim=0cm 0cm 0cm 0.0cm]{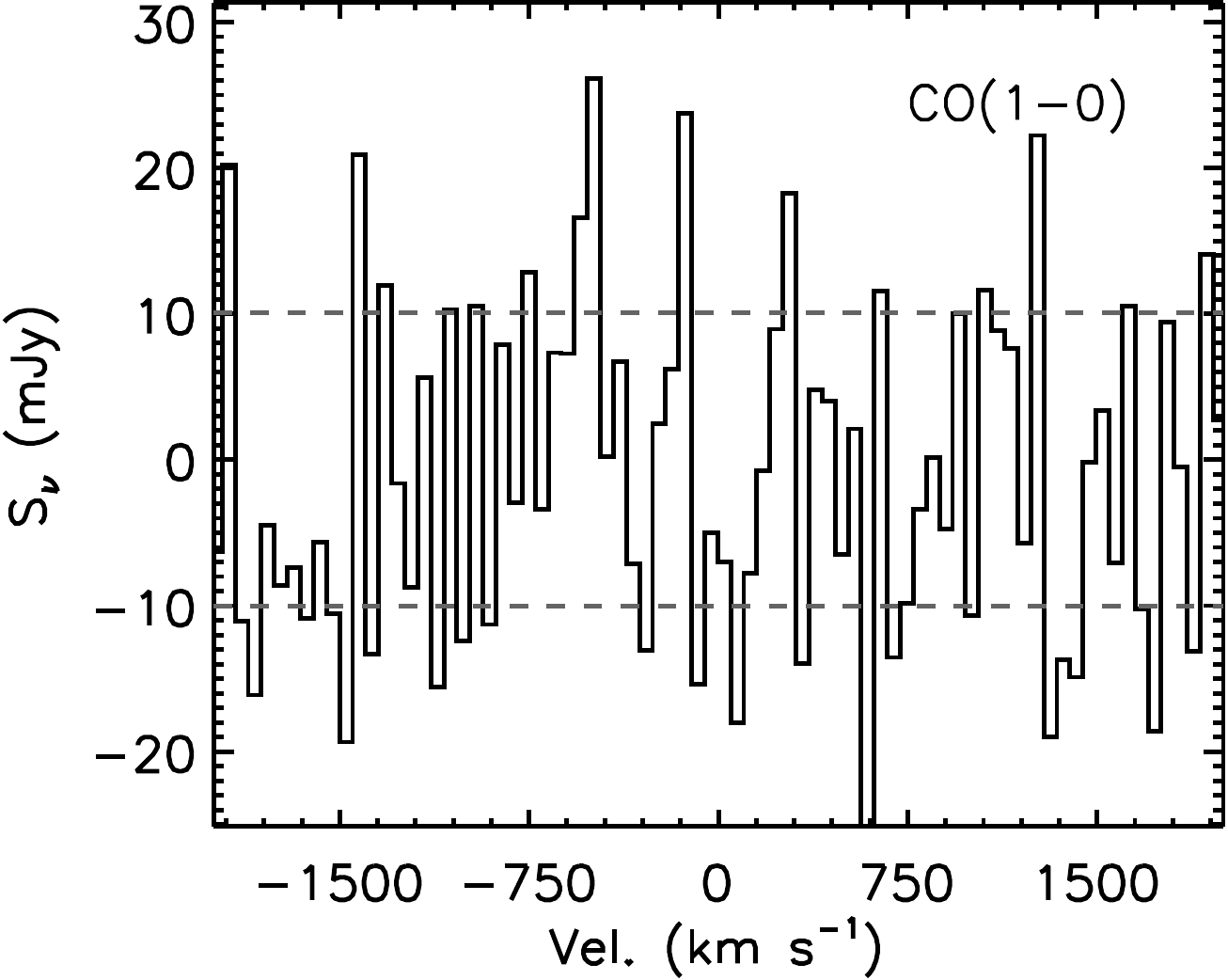}\hspace{0.25cm}
\includegraphics[width=0.45\textwidth,angle=0,clip,trim=0cm 0cm 0cm 0.0cm]{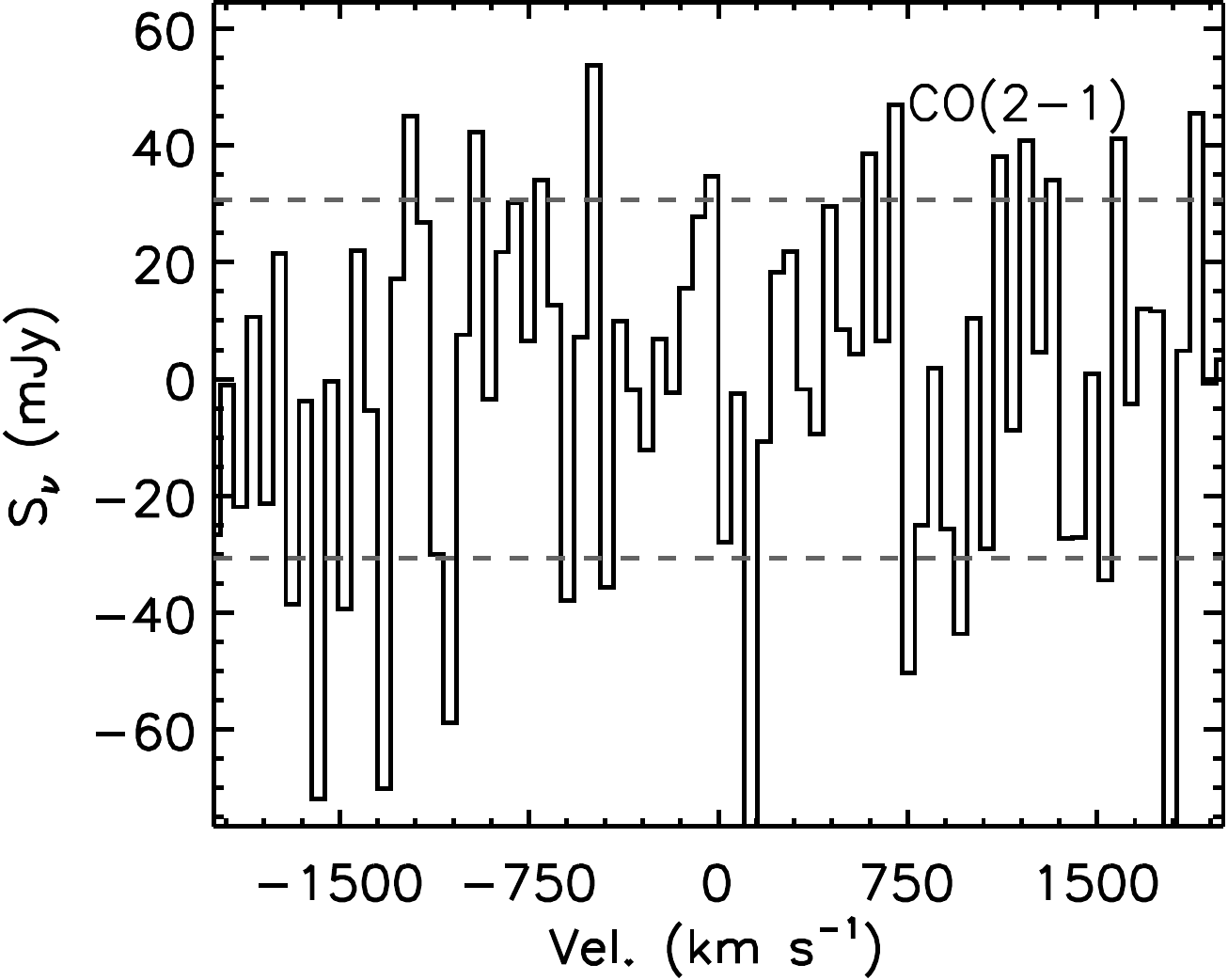}
\caption{NGC2672}
\end{subfigure}\vspace{0.5cm}
\begin{subfigure}[b]{0.48\textwidth}
\includegraphics[width=0.45\textwidth,angle=0,clip,trim=0cm 0cm 0cm 0.0cm]{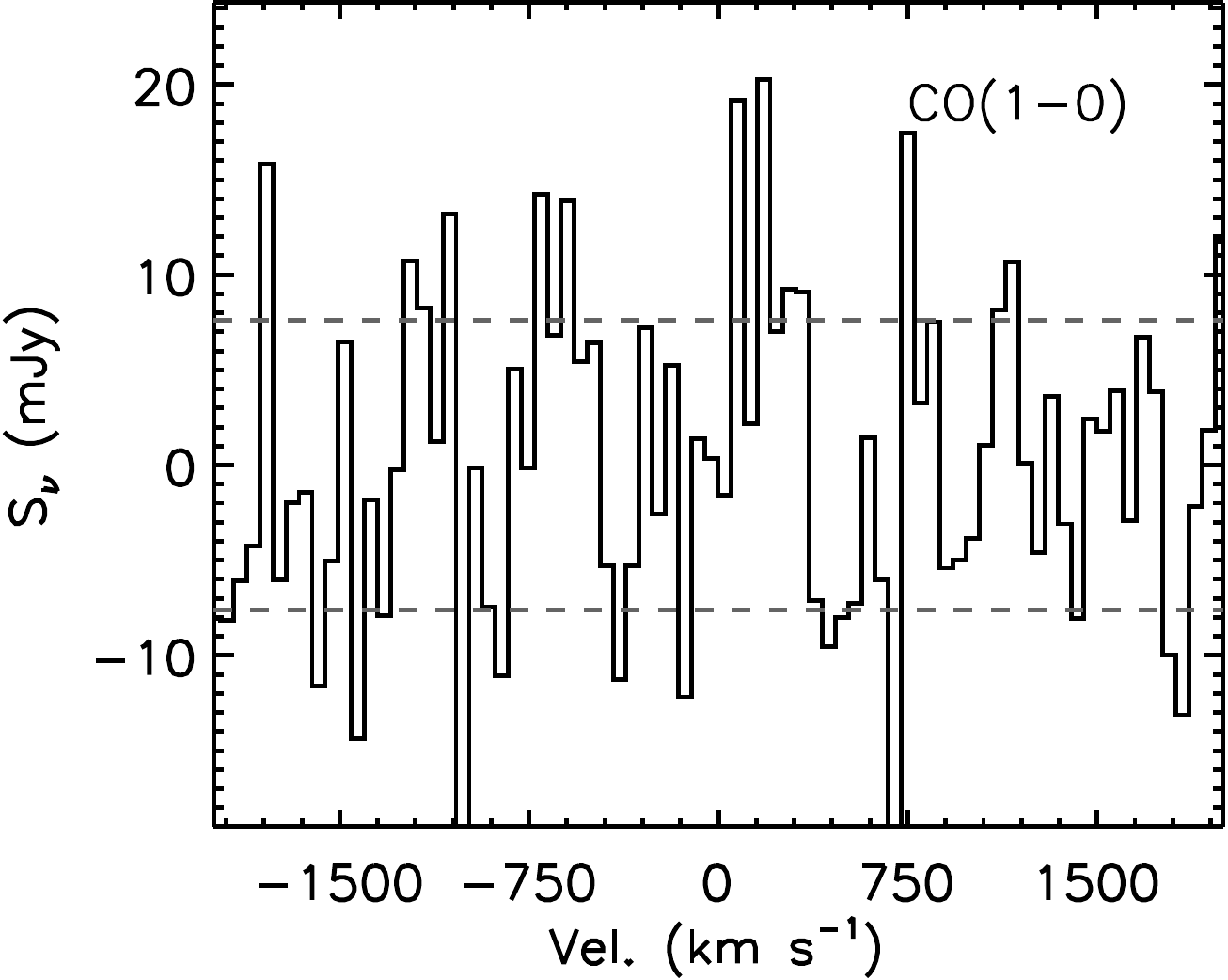}\hspace{0.25cm}
\includegraphics[width=0.45\textwidth,angle=0,clip,trim=0cm 0cm 0cm 0.0cm]{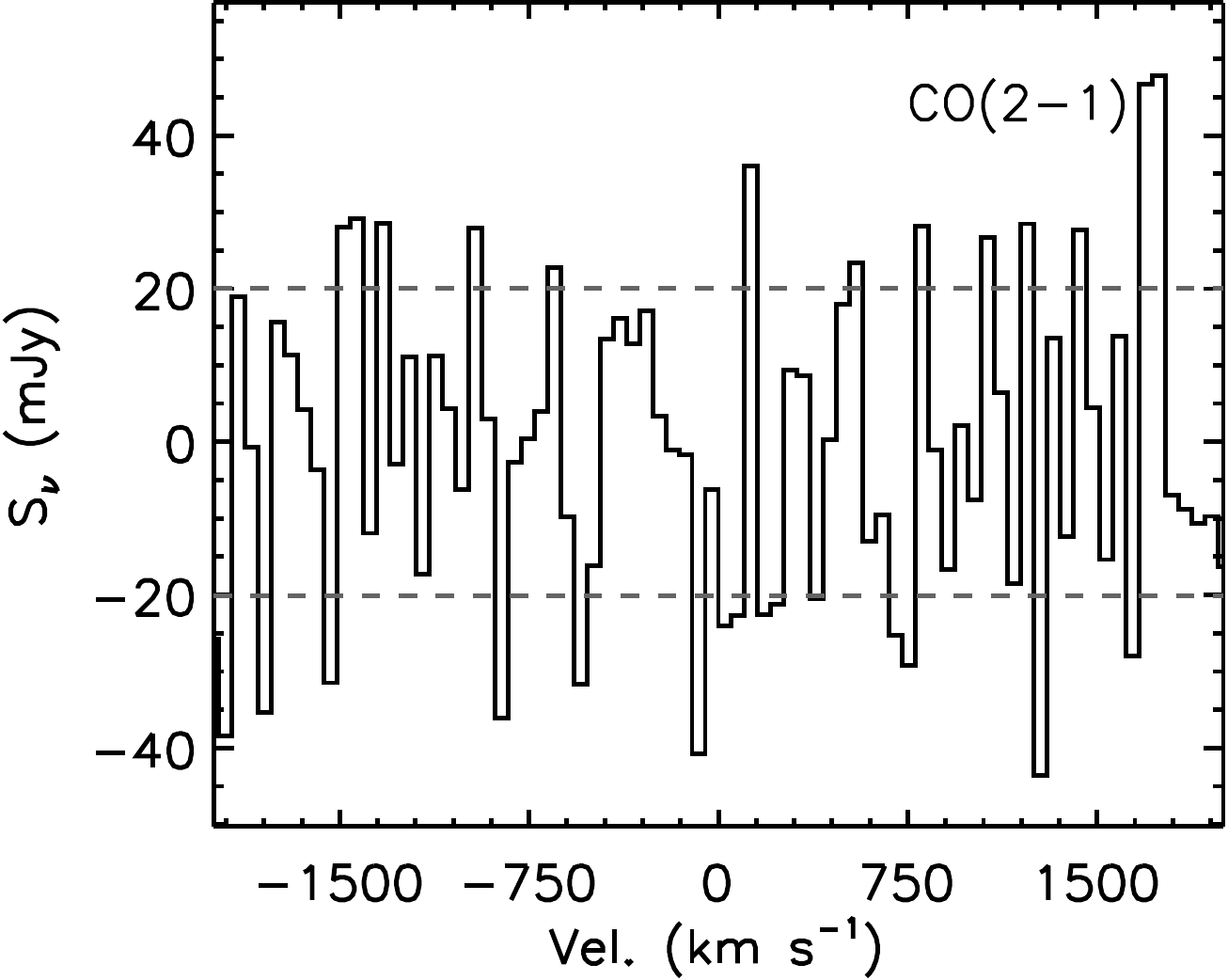}
\caption{NGC2693}
\end{subfigure}
\begin{subfigure}[b]{0.48\textwidth}
\includegraphics[width=0.45\textwidth,angle=0,clip,trim=0cm 0cm 0cm 0.0cm]{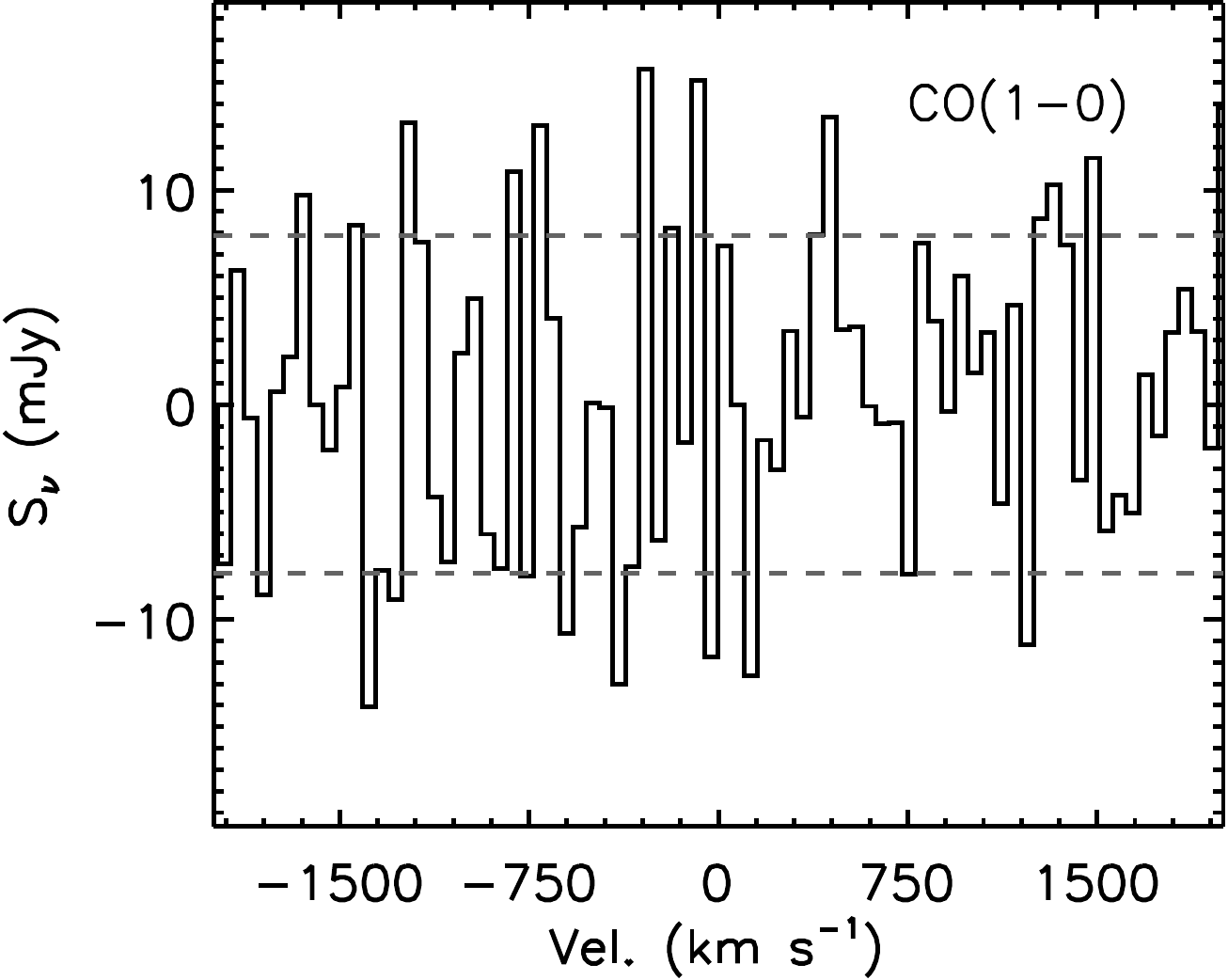}\hspace{0.25cm}
\includegraphics[width=0.45\textwidth,angle=0,clip,trim=0cm 0cm 0cm 0.0cm]{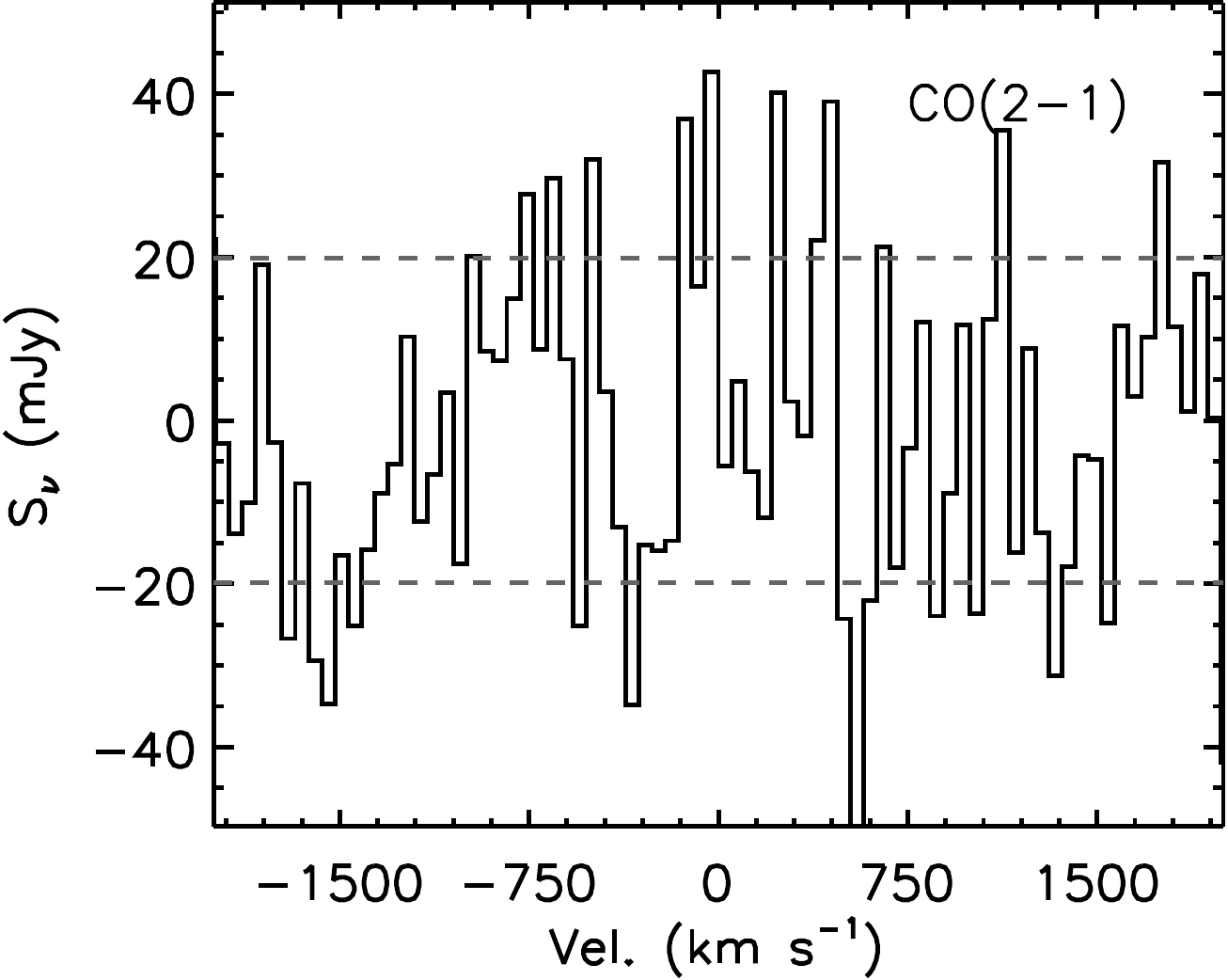}
\caption{NGC2832}
\end{subfigure}\vspace{0.5cm}
\begin{subfigure}[b]{0.48\textwidth}
\includegraphics[width=0.45\textwidth,angle=0,clip,trim=0cm 0cm 0cm 0.0cm]{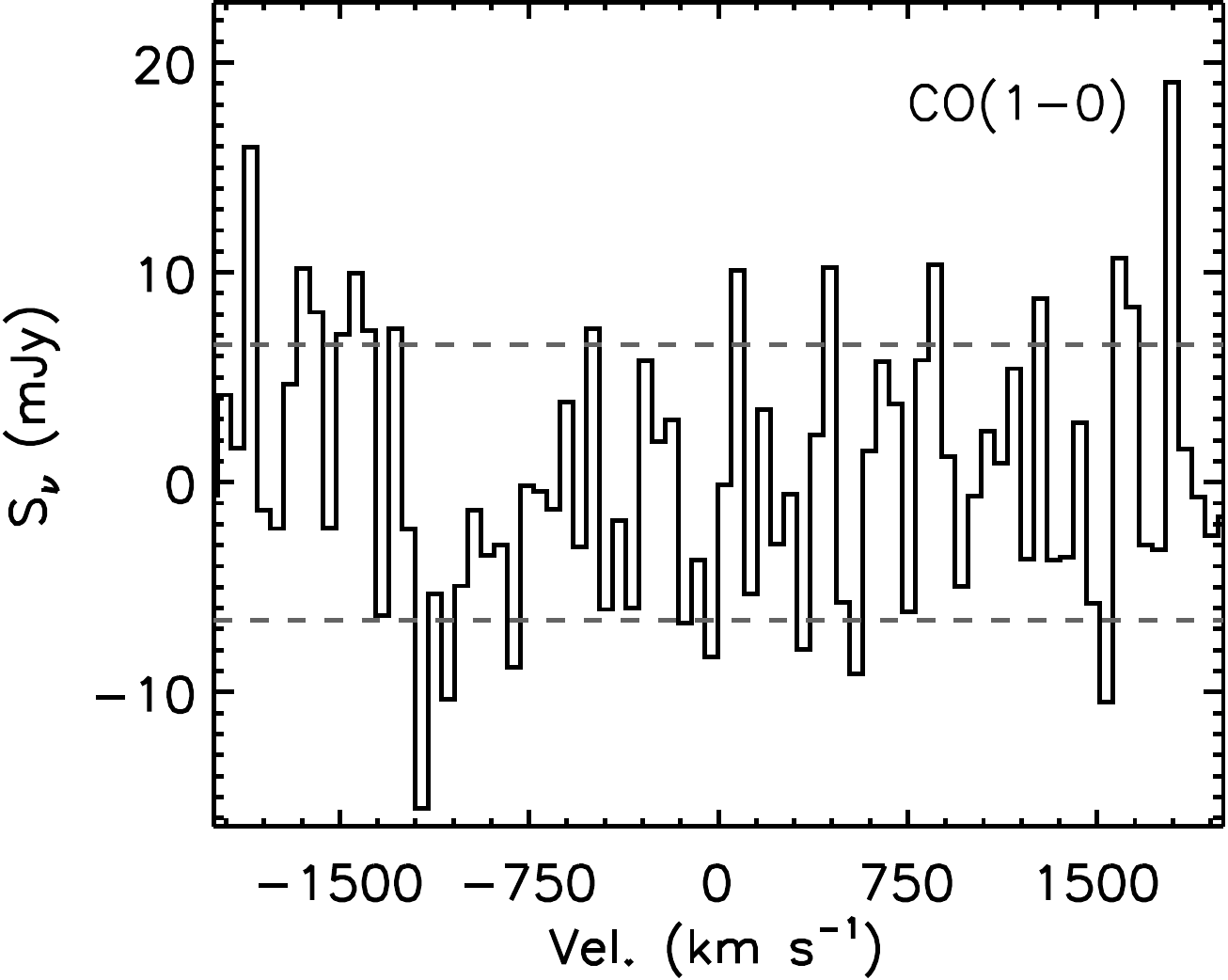}\hspace{0.25cm}
\includegraphics[width=0.45\textwidth,angle=0,clip,trim=0cm 0cm 0cm 0.0cm]{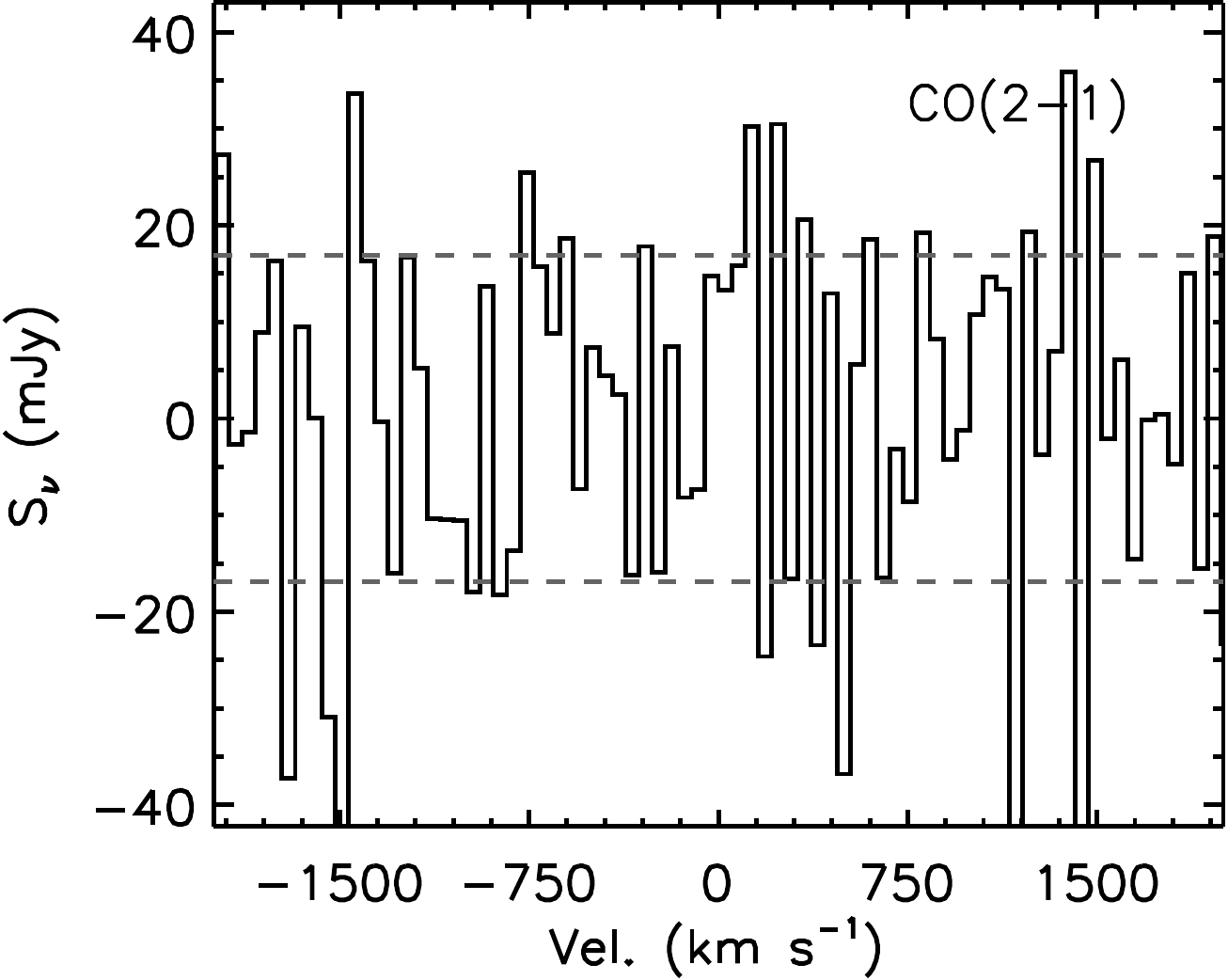}
\caption{NGC3158}
\end{subfigure}
\begin{subfigure}[b]{0.48\textwidth}
\includegraphics[width=0.45\textwidth,angle=0,clip,trim=0cm 0cm 0cm 0.0cm]{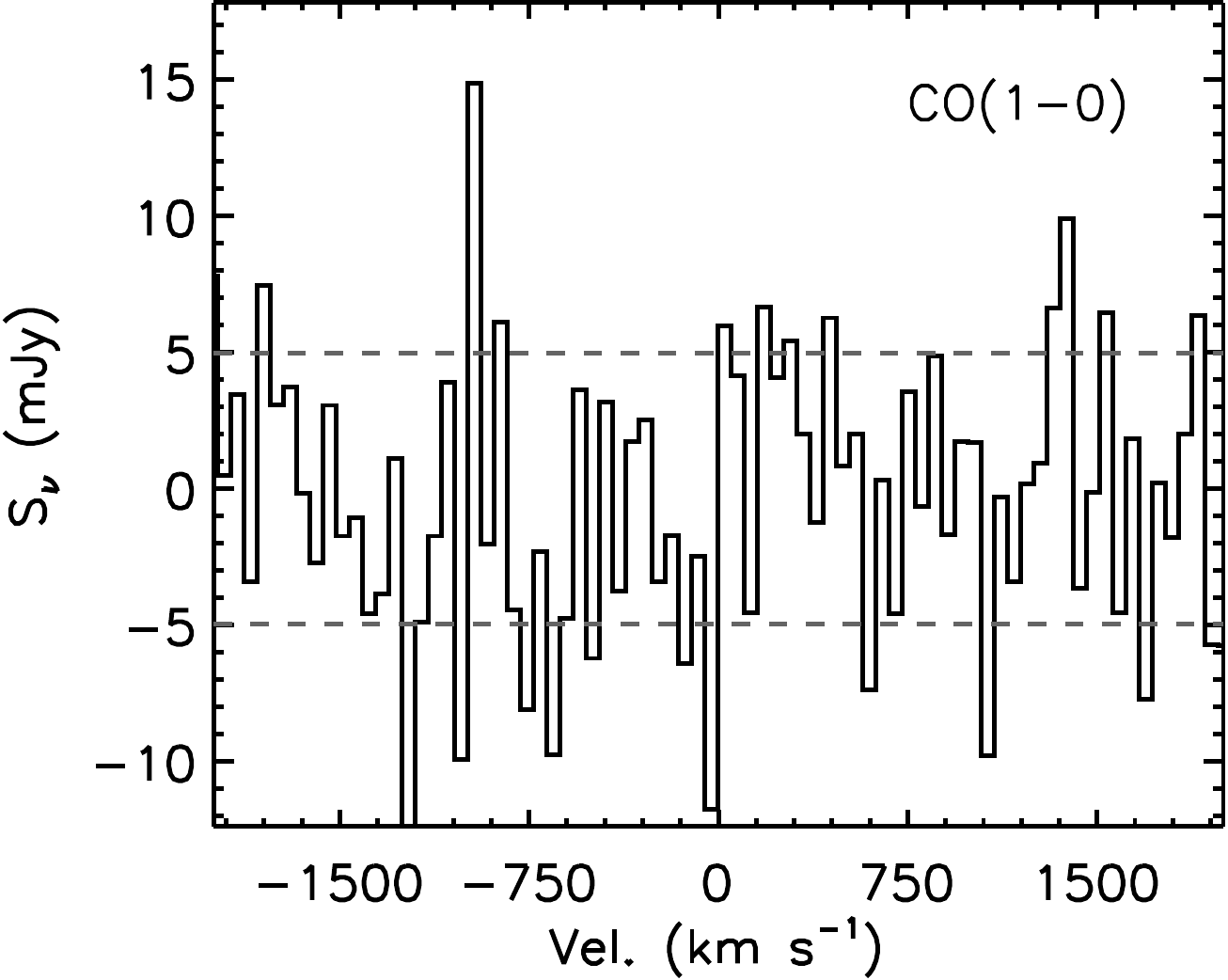}\hspace{0.25cm}
\includegraphics[width=0.45\textwidth,angle=0,clip,trim=0cm 0cm 0cm 0.0cm]{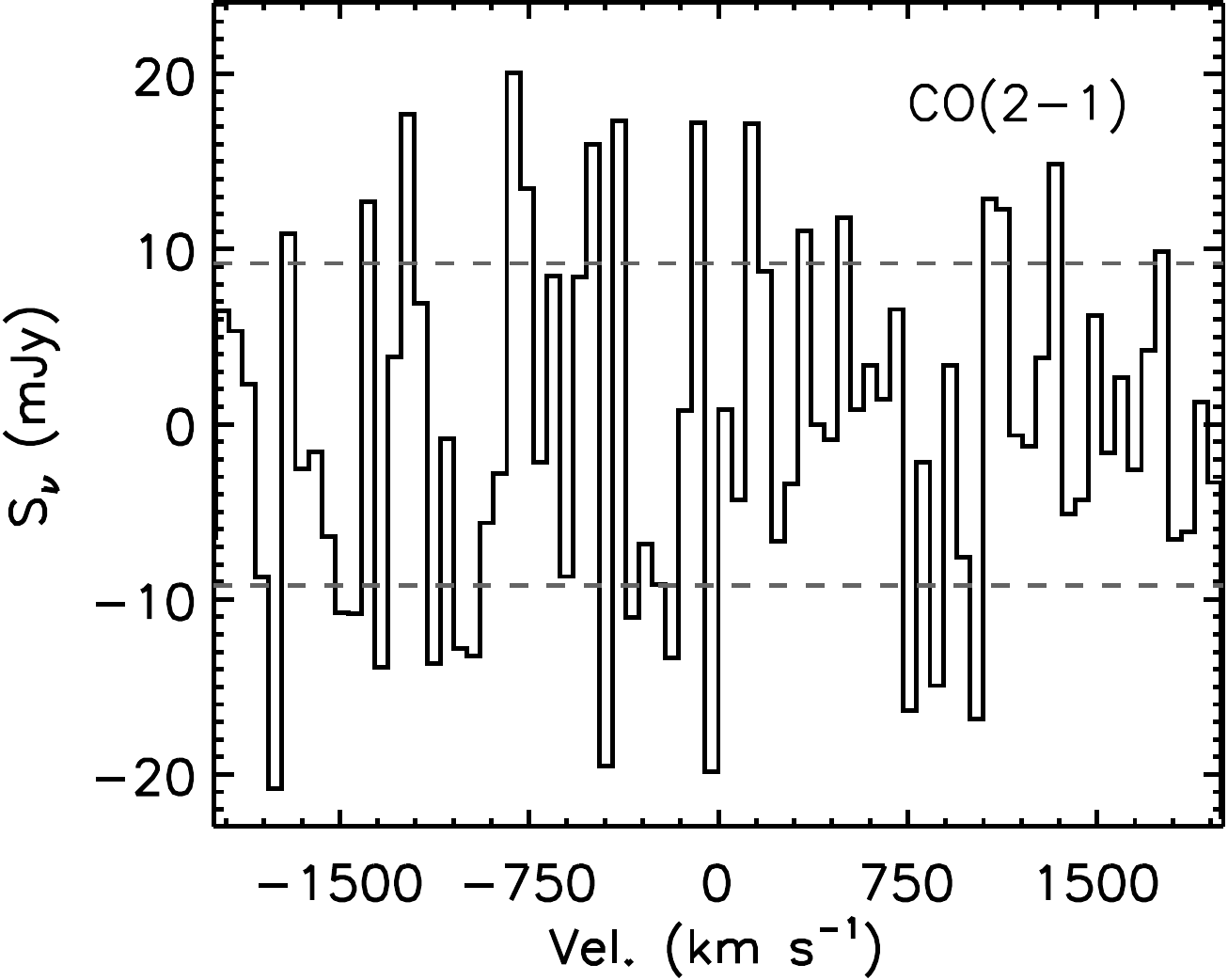}
\caption{NGC3816}
\end{subfigure}
\contcaption{ }
\label{codetsfig2}
 \end{center}
 \end{figure*}
 \begin{figure*} 
\begin{center} 
\begin{subfigure}[b]{0.48\textwidth}
\includegraphics[width=0.45\textwidth,angle=0,clip,trim=0cm 0cm 0cm 0.0cm]{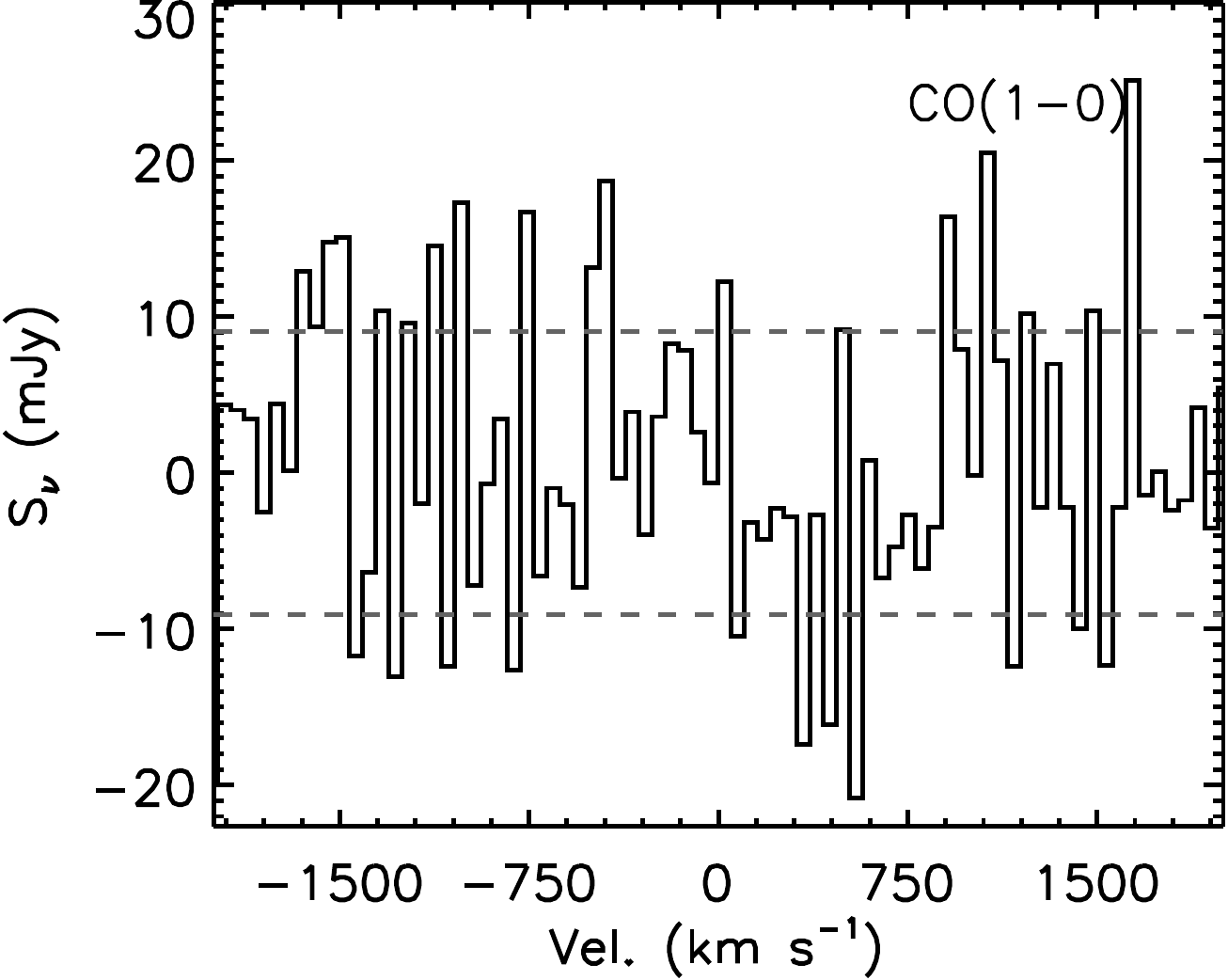}\hspace{0.25cm}
\includegraphics[width=0.45\textwidth,angle=0,clip,trim=0cm 0cm 0cm 0.0cm]{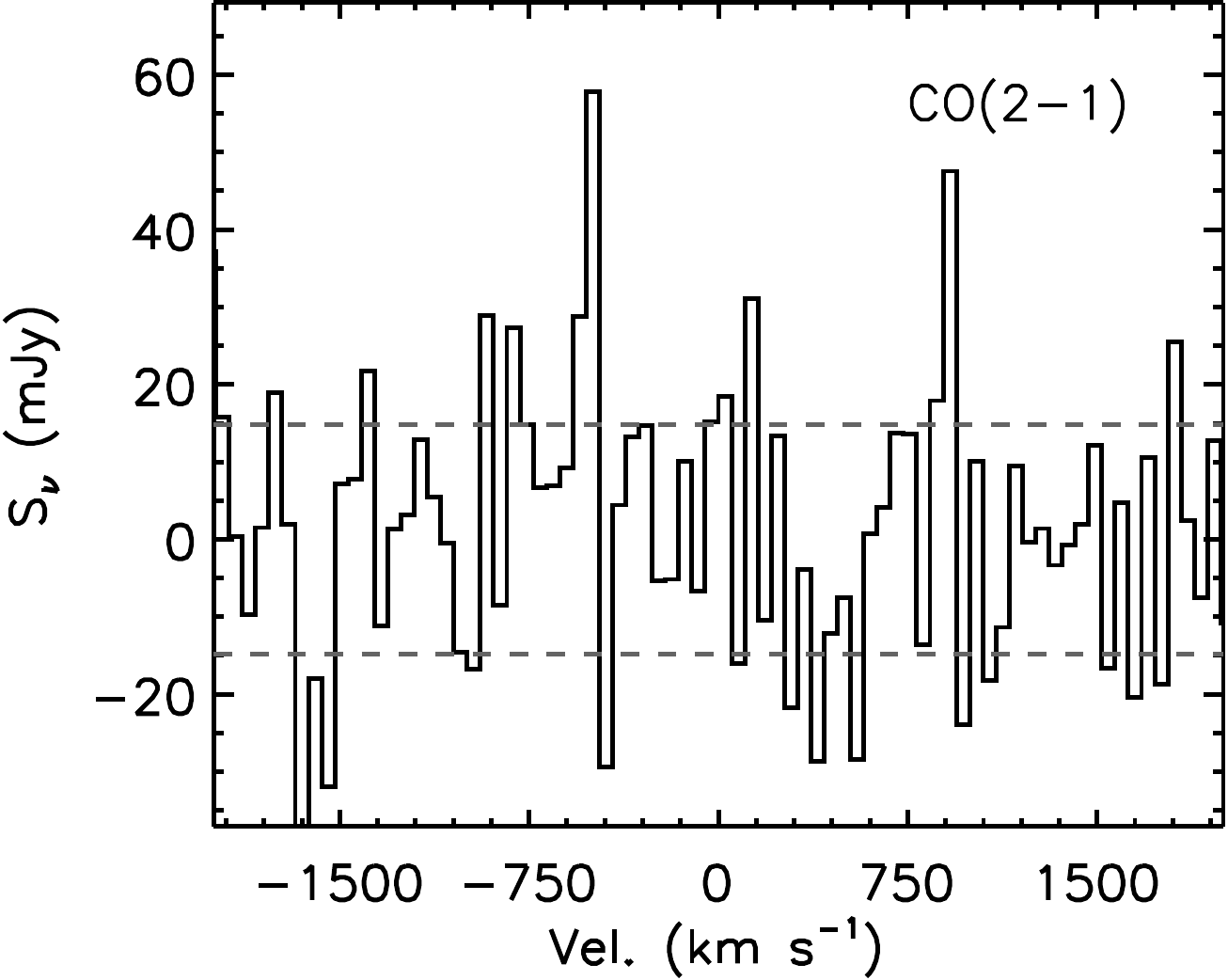}
\caption{NGC4073}
\end{subfigure}
\begin{subfigure}[b]{0.48\textwidth}
\includegraphics[width=0.45\textwidth,angle=0,clip,trim=0cm 0cm 0cm 0.0cm]{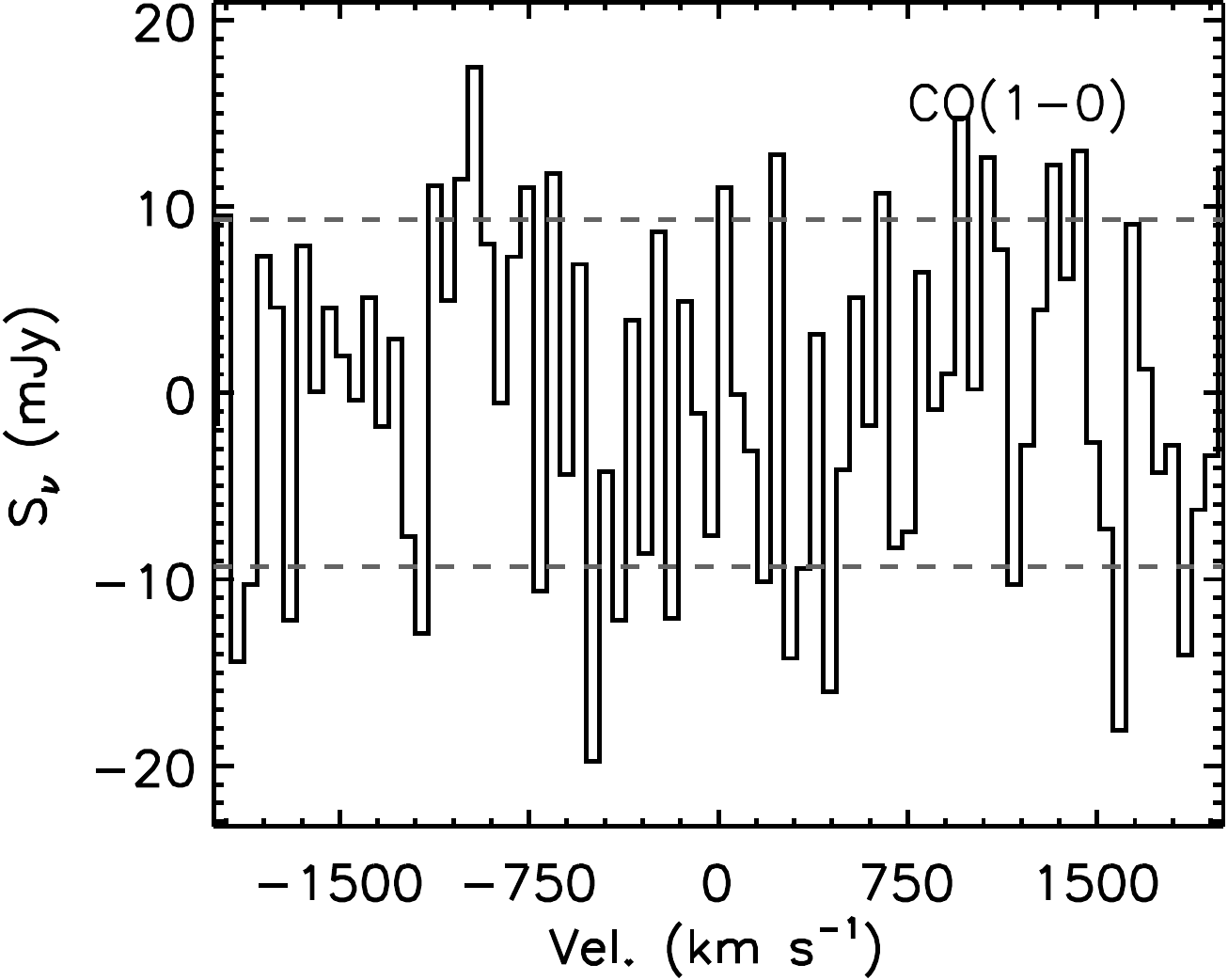}\hspace{0.25cm}
\includegraphics[width=0.45\textwidth,angle=0,clip,trim=0cm 0cm 0cm 0.0cm]{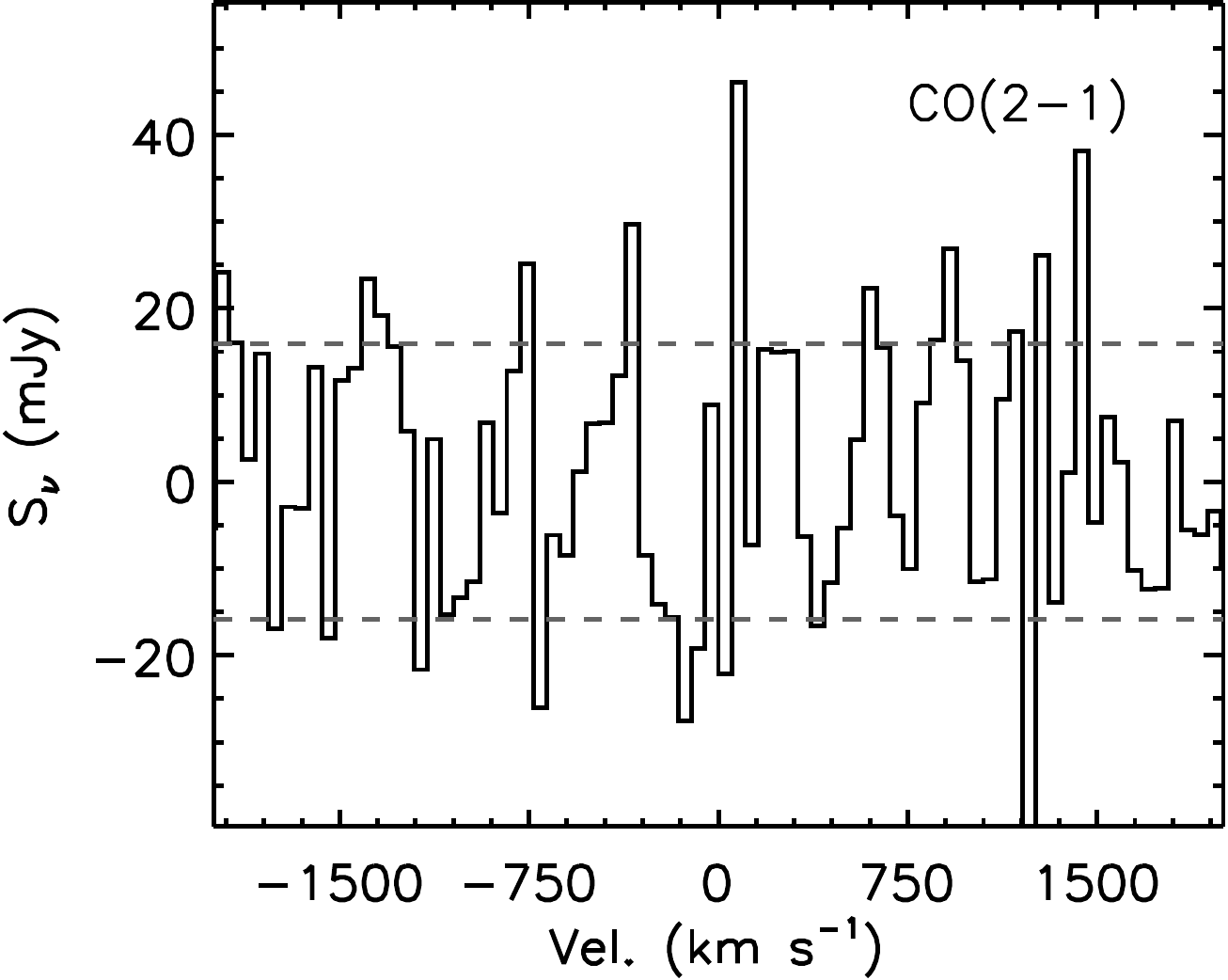}
\caption{NGC4889}
\end{subfigure}\vspace{0.5cm}
\begin{subfigure}[b]{0.48\textwidth}
\includegraphics[width=0.45\textwidth,angle=0,clip,trim=0cm 0cm 0cm 0.0cm]{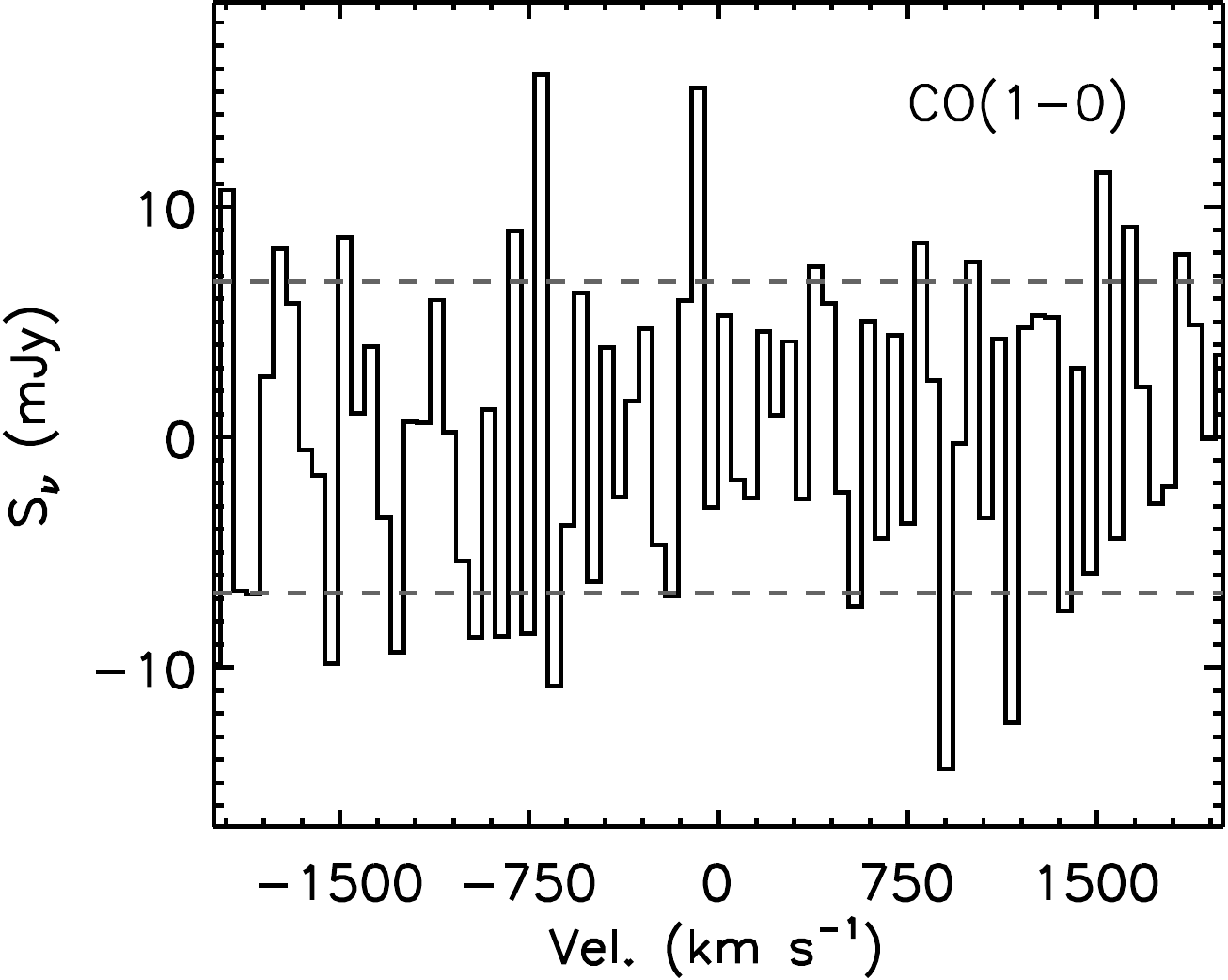}\hspace{0.25cm}
\includegraphics[width=0.45\textwidth,angle=0,clip,trim=0cm 0cm 0cm 0.0cm]{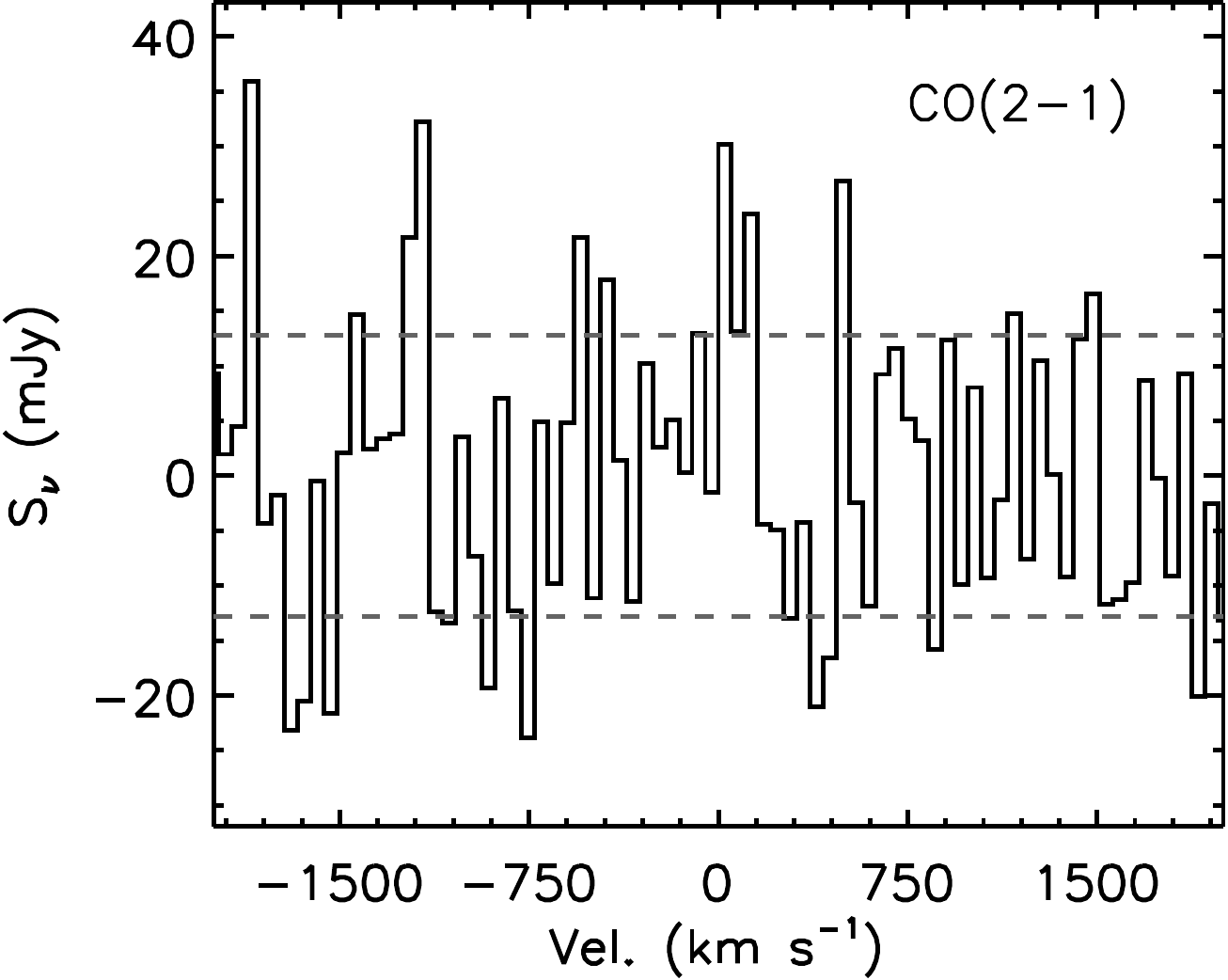}
\caption{NGC4914}
\end{subfigure}
\begin{subfigure}[b]{0.48\textwidth}
\includegraphics[width=0.45\textwidth,angle=0,clip,trim=0cm 0cm 0cm 0.0cm]{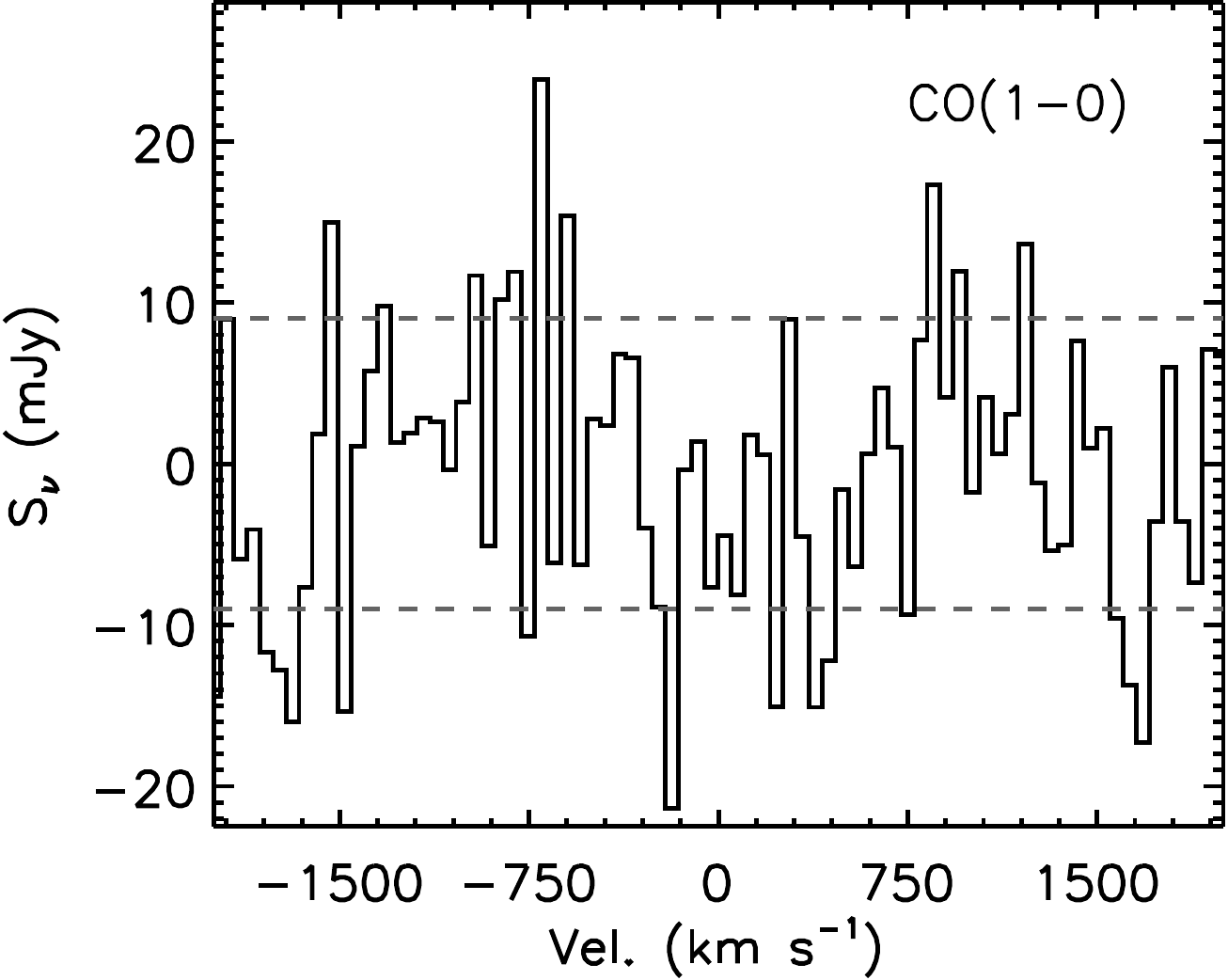}\hspace{0.25cm}
\includegraphics[width=0.45\textwidth,angle=0,clip,trim=0cm 0cm 0cm 0.0cm]{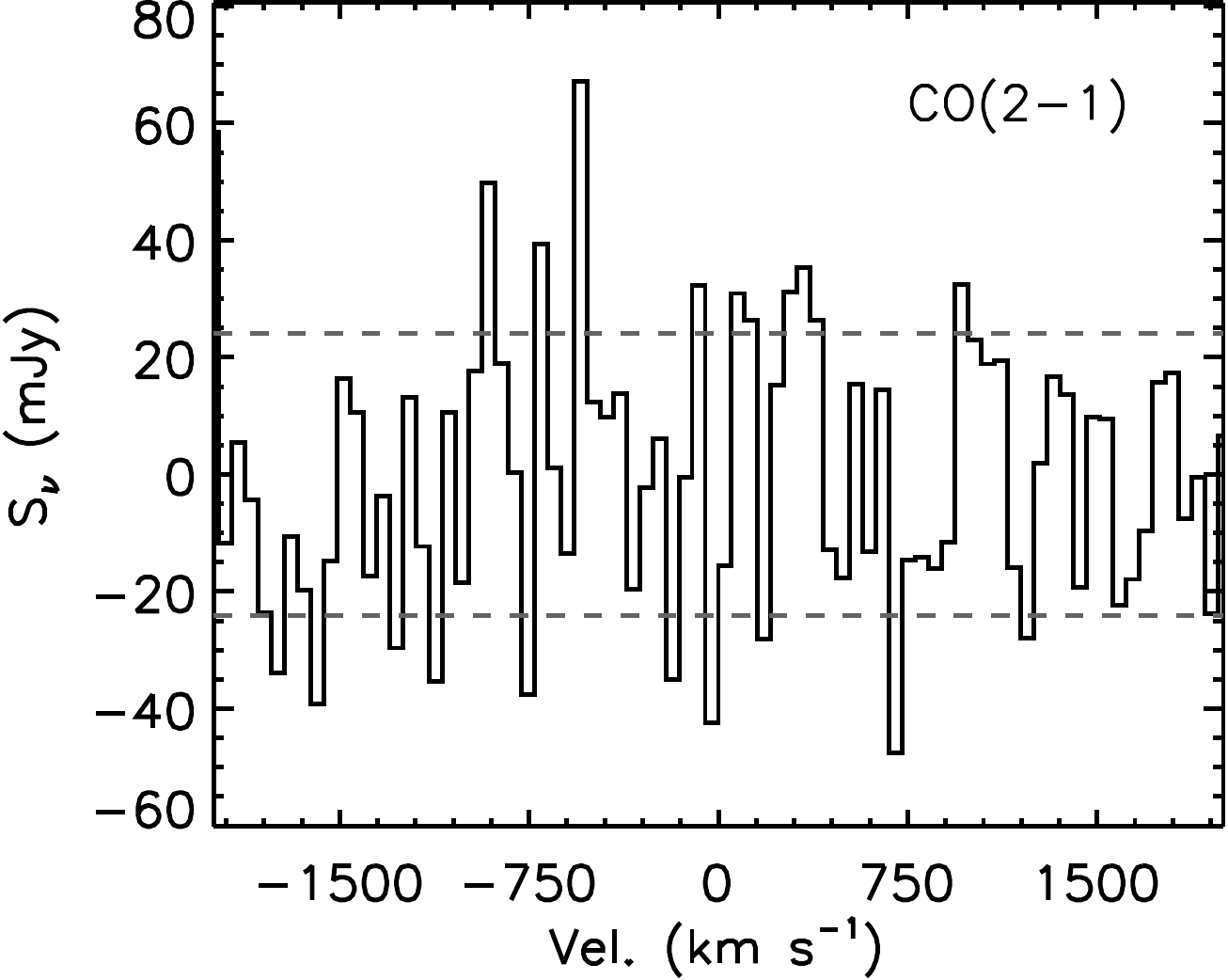}
\caption{NGC7265}
\end{subfigure}\vspace{0.5cm}
\begin{subfigure}[b]{0.48\textwidth}
\includegraphics[width=0.45\textwidth,angle=0,clip,trim=0cm 0cm 0cm 0.0cm]{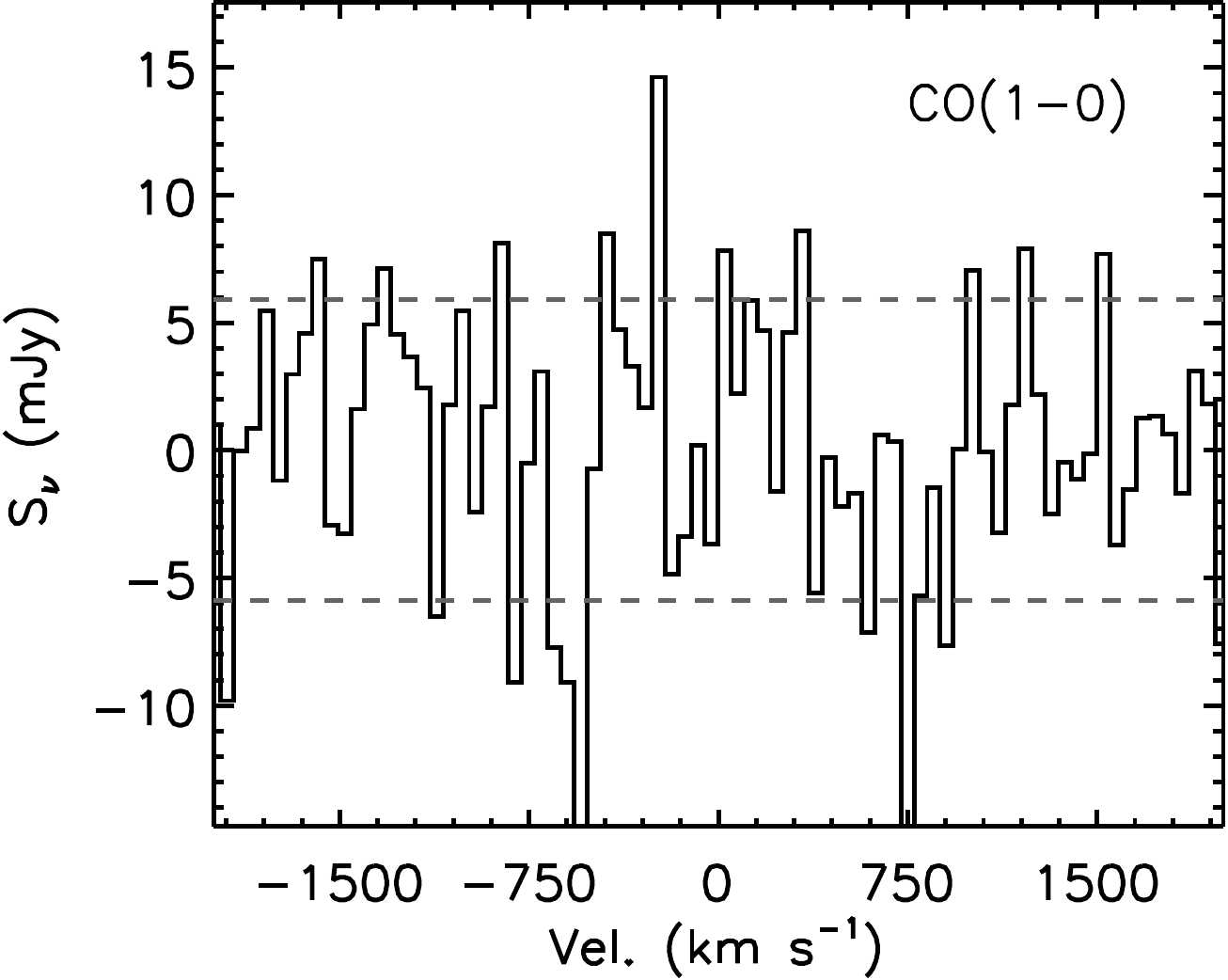}\hspace{0.25cm}
\includegraphics[width=0.45\textwidth,angle=0,clip,trim=0cm 0cm 0cm 0.0cm]{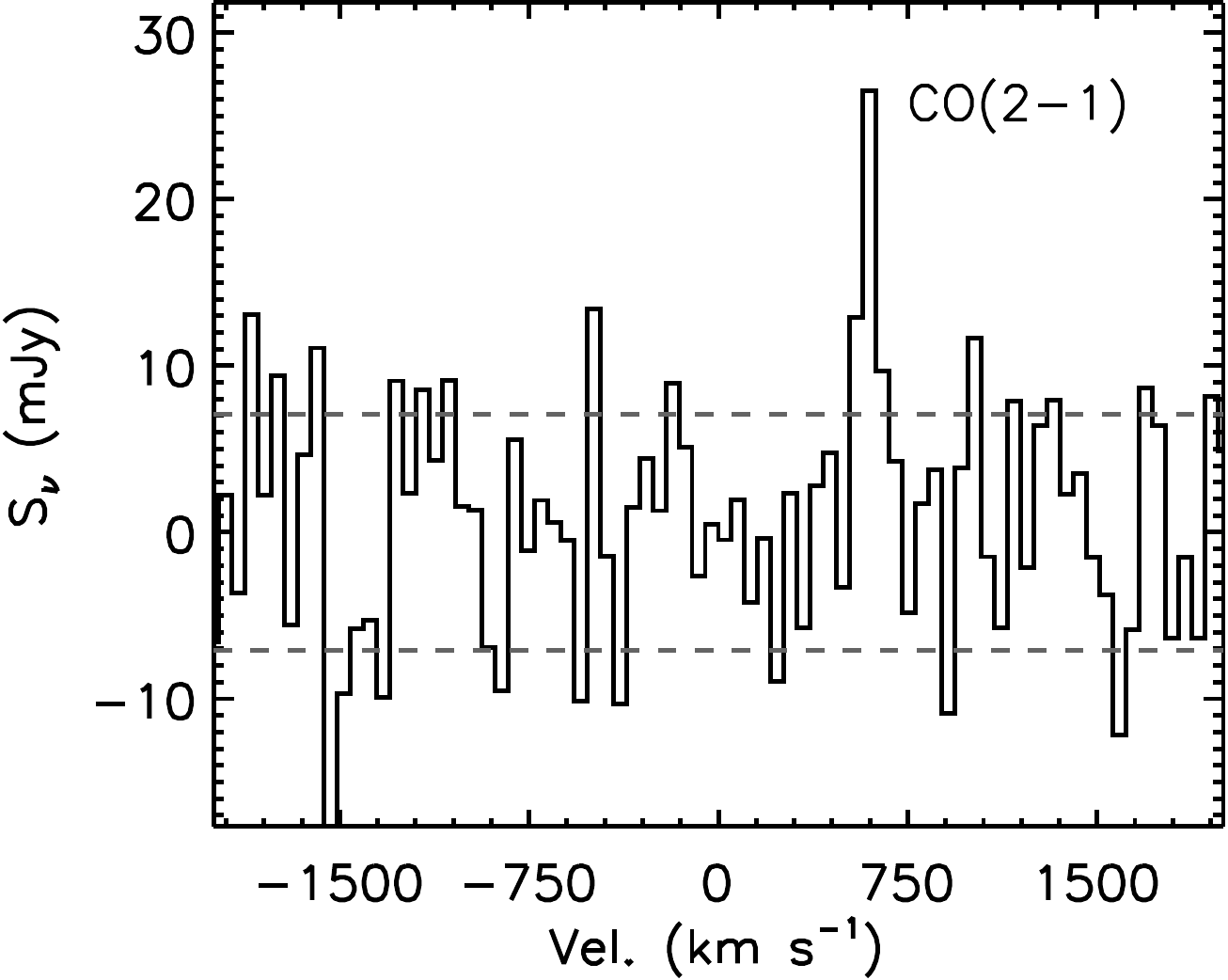}
\caption{NGC7274}
\end{subfigure}
\begin{subfigure}[b]{0.48\textwidth}
\includegraphics[width=0.45\textwidth,angle=0,clip,trim=0cm 0cm 0cm 0.0cm]{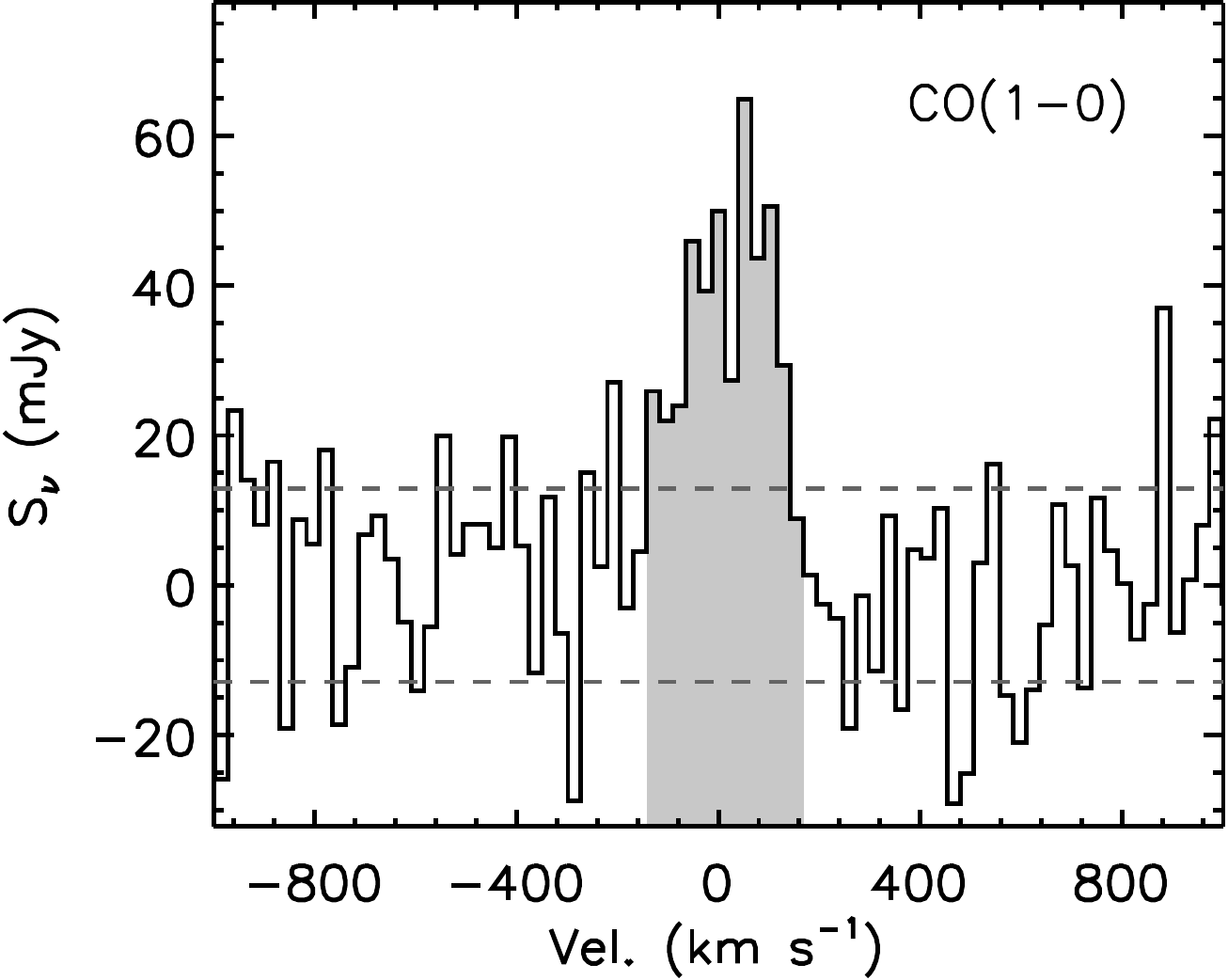}\hspace{0.25cm}
\includegraphics[width=0.45\textwidth,angle=0,clip,trim=0cm 0cm 0cm 0.0cm]{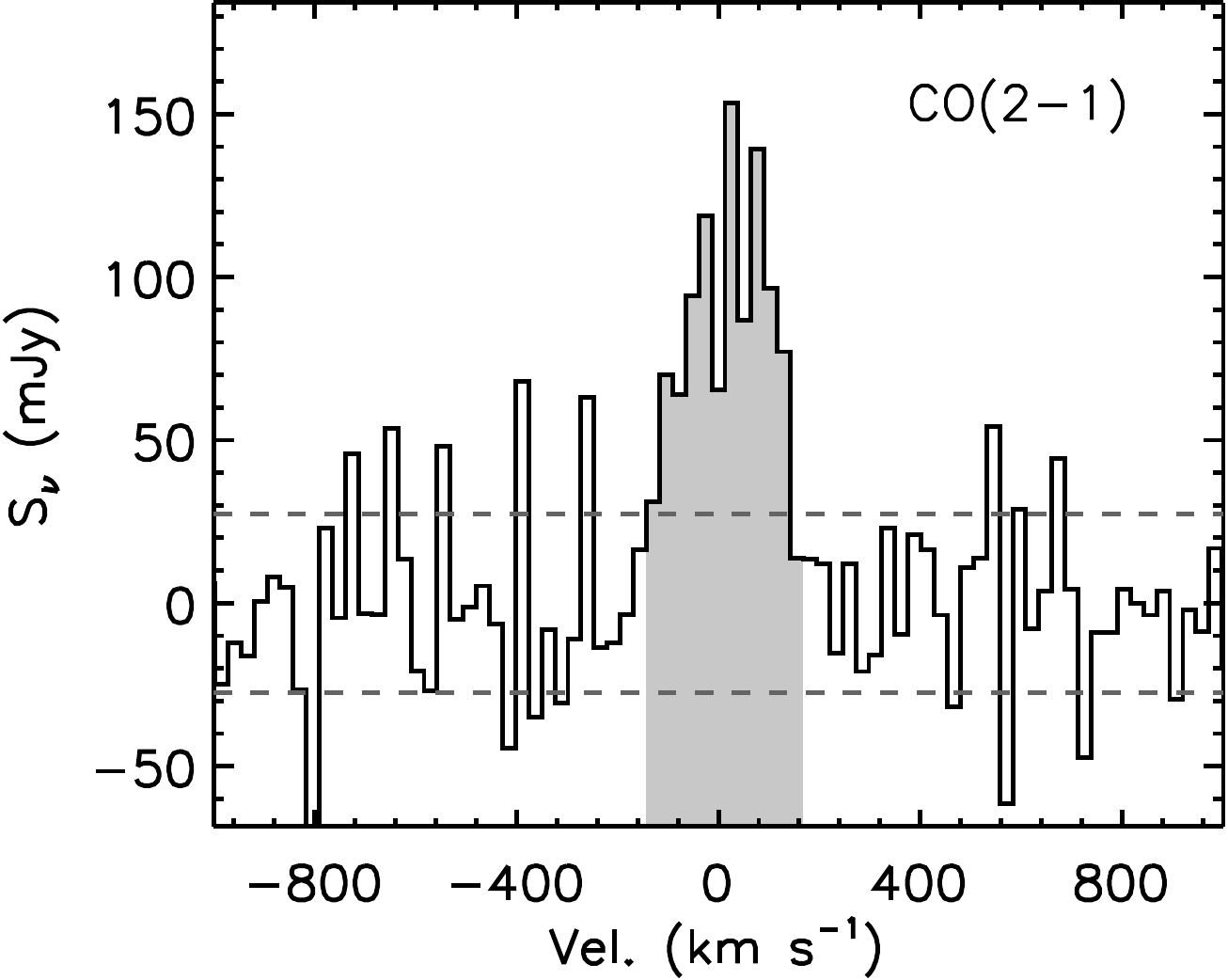}
\caption{NGC7550}
\end{subfigure}\vspace{0.5cm}
\begin{subfigure}[b]{0.48\textwidth}
\includegraphics[width=0.45\textwidth,angle=0,clip,trim=0cm 0cm 0cm 0.0cm]{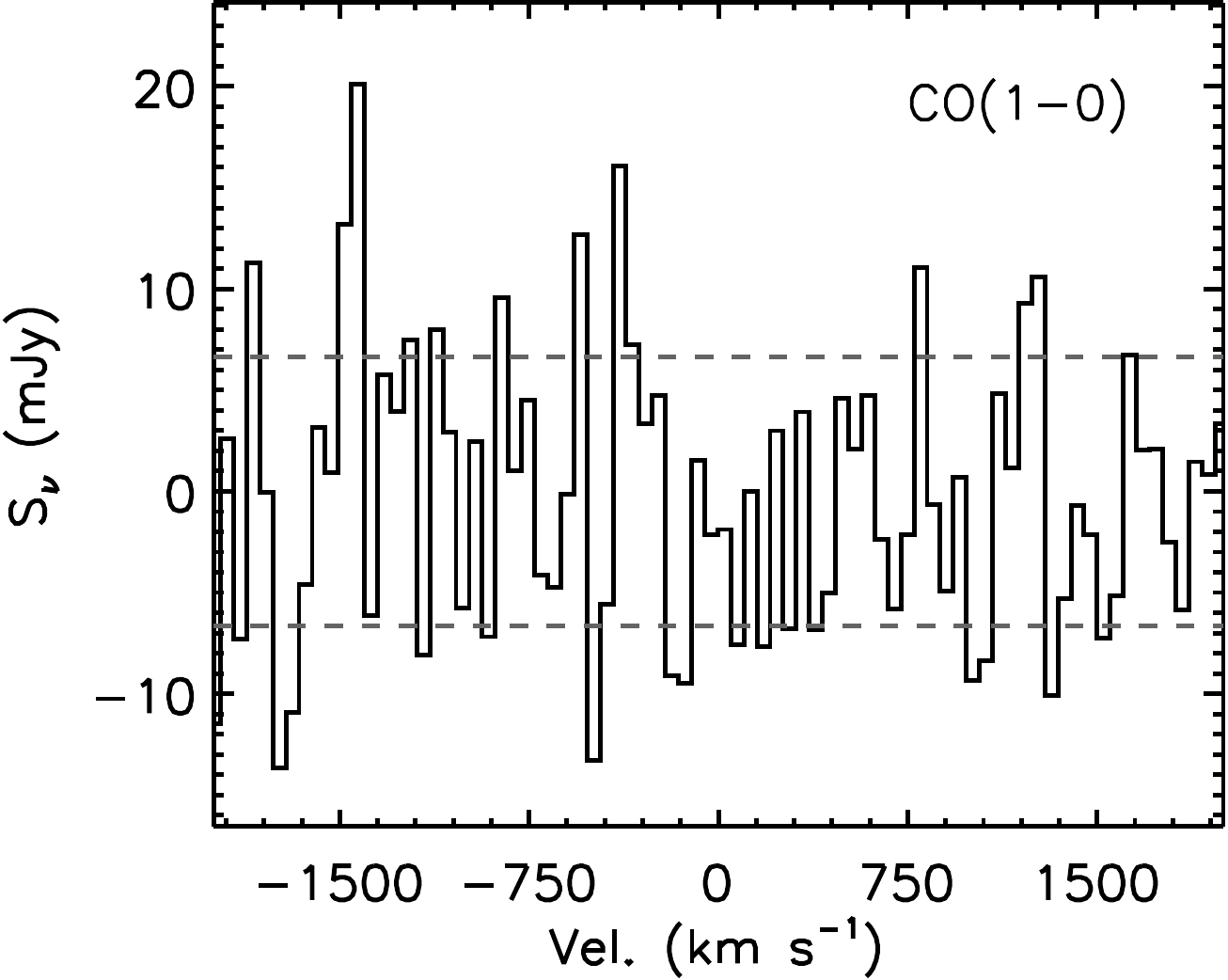}\hspace{0.25cm}
\includegraphics[width=0.45\textwidth,angle=0,clip,trim=0cm 0cm 0cm 0.0cm]{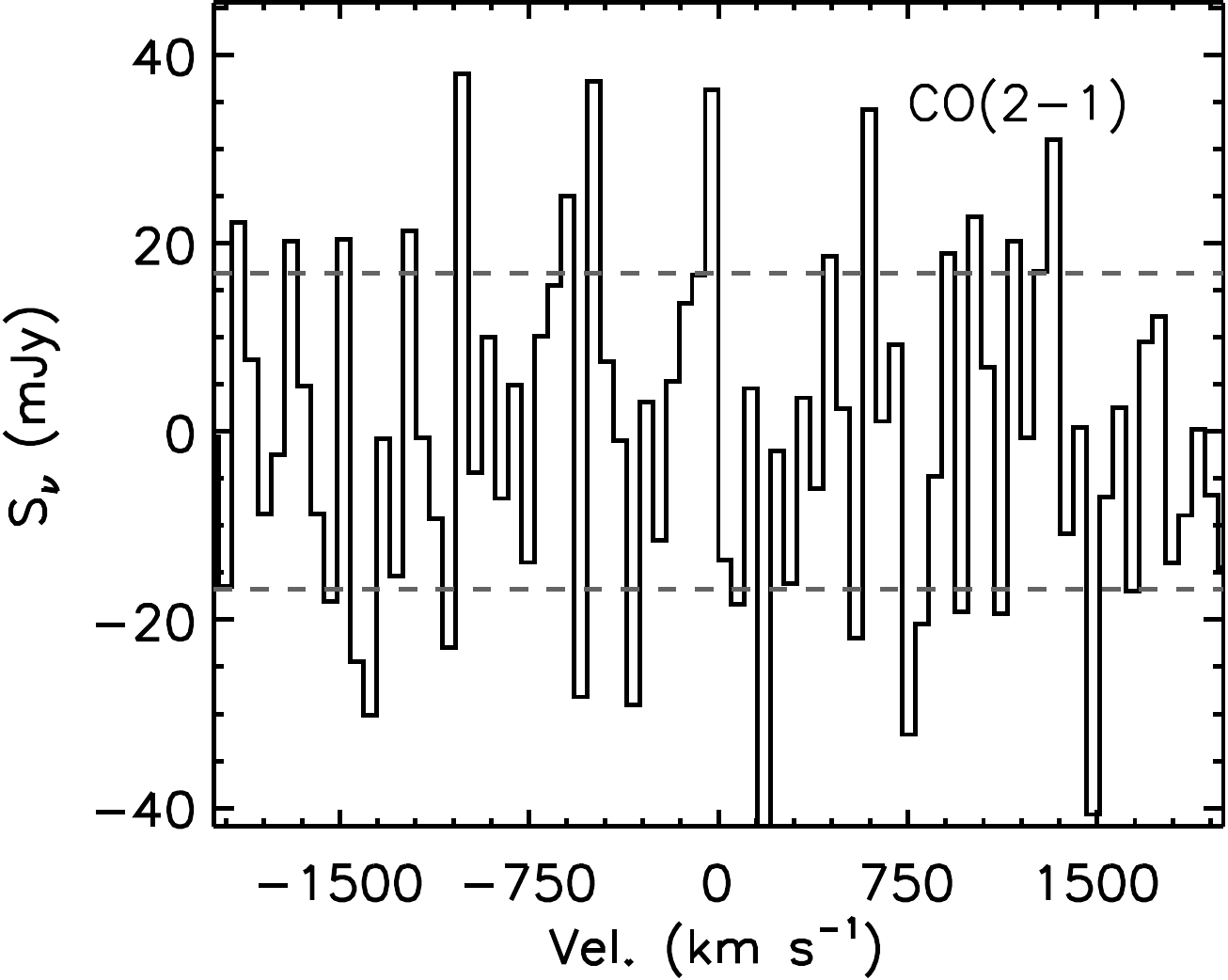}
\caption{NGC7618}
\end{subfigure}
\contcaption{ }
\label{codetsfig2}
 \end{center}
 \end{figure*}

\section{X-rays and Ionised gas}

\begin{figure*} 
\begin{center}
\includegraphics[width=0.75\textwidth,angle=0,clip,trim=-0.57cm 1.6cm 0cm 0.0cm]{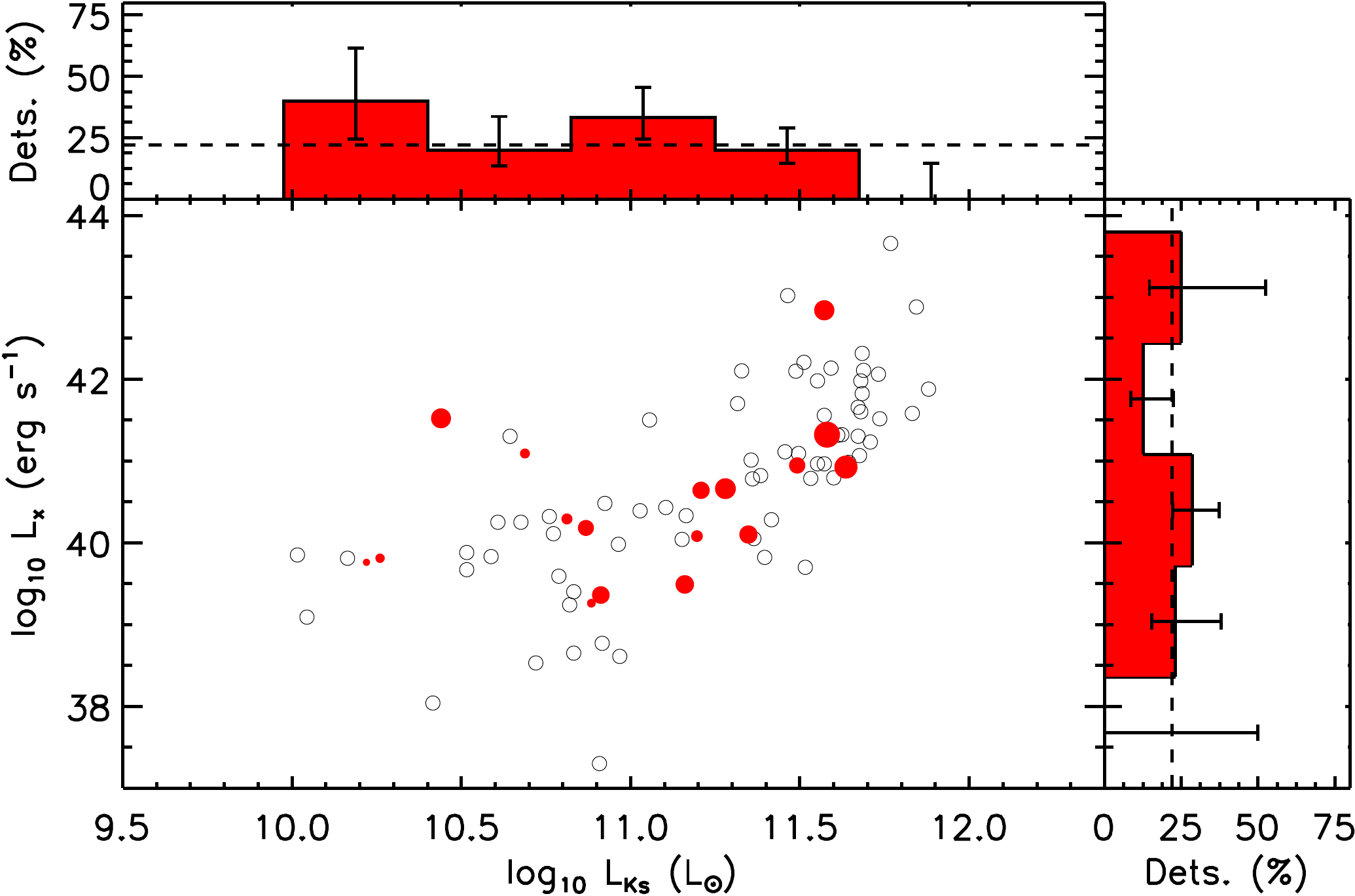}
\includegraphics[width=0.75\textwidth,angle=0,clip,trim=0cm 0cm 0cm 0.0cm]{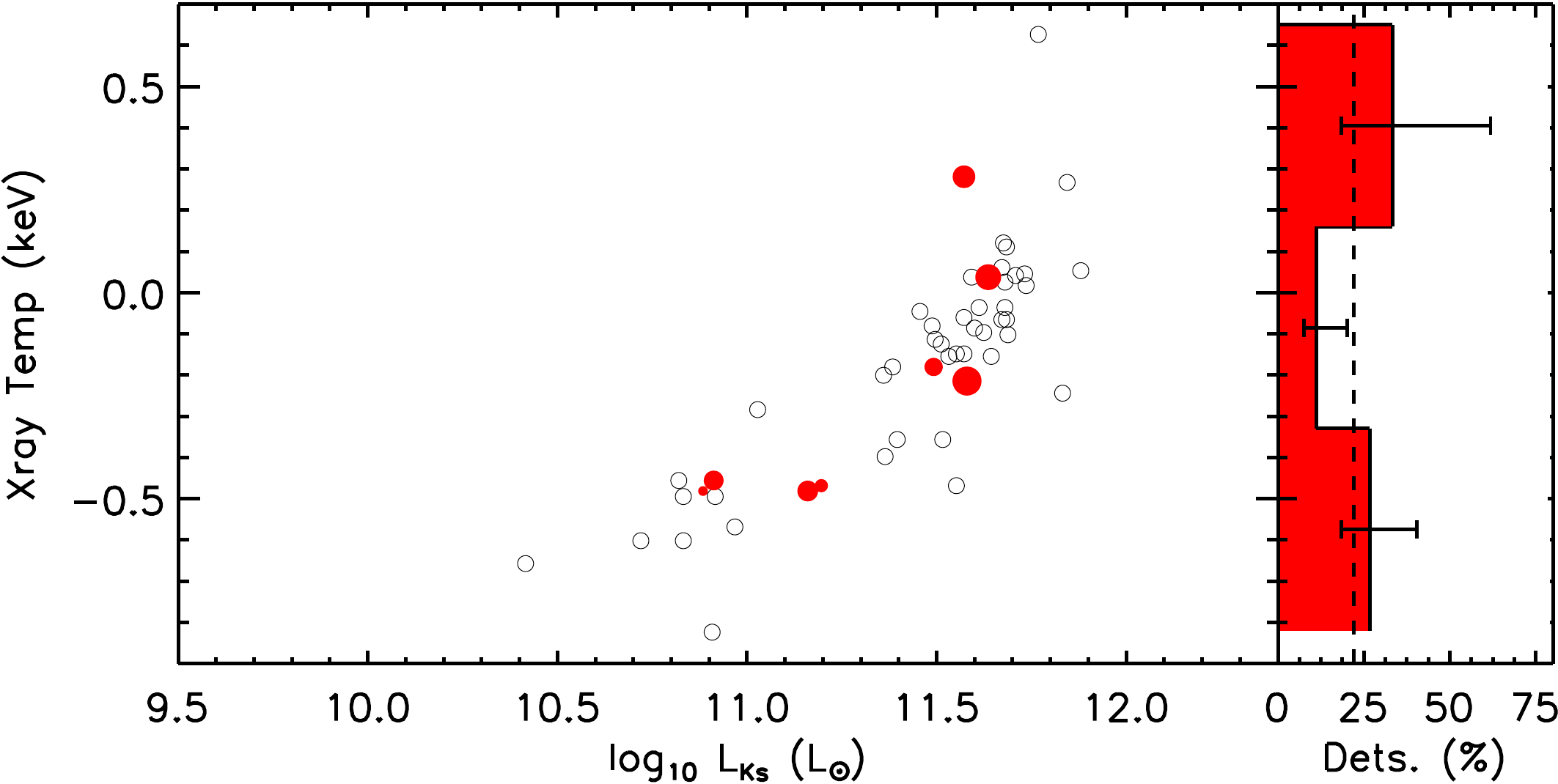}
\caption{As Figure \ref{fig:fundplane}, but showing the $K_s$-band luminosity of the MASSIVE and \atlas\ galaxies against their X-ray luminosity (top panel) and X-ray temperature (bottom panel), taken from {\protect \cite{2013MNRAS.432.1845S} and \protect \cite{2016ApJ...826..167G} where available}. As shown in the histogram on the right panel there is no clear correlation between these X-ray properties and the molecular gas detection rate in this subsample of galaxies.  }
\label{fig:lk_xray}
 \end{center}
 \end{figure*}

\end{document}